\documentclass[secnumarabic, reprint, superscriptaddress, tightenlines, amsmath,amssymb, aps, prl, nobibnotes,
]{revtex4-1}

\usepackage{float}
\usepackage{natbib}
\usepackage{graphicx,caption,subcaption,xcolor}
\usepackage[colorlinks=true,citecolor=blue,
linkcolor=blue,urlcolor=blue]{hyperref}
\usepackage[utf8]{inputenc} 
\usepackage[english]{babel}
\usepackage[OT2, T1]{fontenc}
\usepackage{upquote}
\usepackage{mathtools}
\usepackage [autostyle, english = american]{csquotes}
\usepackage{bigints}

\begin{document}

\title{Universal scaling relations in electron-phonon superconductors}

\author{Joshuah T. Heath}
\email{Joshuah.T.Heath@su.se}

\author{Rufus Boyack}
\affiliation{Department of Physics and Astronomy, Dartmouth College, Hanover, New Hampshire 03755, USA}
\email{Rufus.Boyack@dartmouth.edu}

\date{\today}

\begin{abstract}
\noindent 
We study linear scaling relations in electron-phonon superconductors. By combining numerical and analytical techniques, we find linear Homes scaling relations between the zero-temperature superfluid density and the normal-state DC conductivity. {This phenomenon arises via either} a large impurity scattering rate or inelastic scattering of electrons and Einstein phonons at large electron-phonon coupling. Our work thus shows that Homes scaling is more universal than either cuprate or BCS-like physics, and is instead a fundamental result in a wide class of superconductors. 
\end{abstract}

\pacs{1}

\maketitle

\indent {\it Introduction.--}Unconventional superconductors are often characterized by exotic normal phases above a characteristically high critical temperature $T_c$~\cite{Sachdev2003Jul,Stewart2017Apr}. One notable example is the cuprates~\cite{Bednorz1986,Pickett1989Apr, Wu1993,Schilling1993May,RevModPhys.72.969,Hwang2021Jun}, which exhibit a non-Fermi-liquid normal state at optimal doping known as the "strange metal"~\cite{Phillips2022Jul,Chowdhury2022Sep,Li2024Jun}. The unique electronic transport observed in the strange-metal phase~\cite{Gurvitch1987Sep,Martin1990Jan,Takagi1992Nov,Marel2003Sep} has been attributed to a linear scaling relation between observables in the $T=0$ ground state and the $T=T_c$ normal state~\cite{Zaanen2004Jul,Homes2004Jul}. The near-universal linear scaling behavior seen in clean high-$T_c$ superconductors was formally believed to be a hallmark of these quantum critical compounds. Nevertheless, linear scaling relationships have been generally considered within a wide array of physical phenomena, such as in quantum Hall physics~\cite{Nelson1977Nov}, weak localization~\cite{Smith1992}, and dirty BCS superconductors~\cite{Kogan2013}.

This letter concerns universal scaling relations beyond both high-$T_c$ and BCS-like superconductors. In regard to the latter, such scaling relationships are confined to the dirty limit~\cite{DeGennes1966, PhysRevLett.10.486,Nam1967,PhysRev.156.487,Kogan2013,Kogan2013b,Tao2017}, where there is a linear relationship between the $T=0$ superfluid density and the normal state electrical conductivity just above $T=T_c$. As articulated by de Gennes~\cite{DeGennes1966}, this linear relationship is a fundamental result for superconducting matter in the BCS limit, provided i) there is a diffuse scattering mechanism, and ii) the theory is gauge-invariant. {Our work builds upon de Gennes' criteria by softening condition i) to the less restrictive constraints of non-Galilean invariance and momentum relaxation.} 

Within the context of high-$T_c$ superconductors, similar linear scaling laws have been studied in the hope of identifying a universal fingerprint for these materials~\cite{Basov2011}. The first attempt to formulate such a relation was {provided} by Pimenov {\it et al.}~\cite{Pimenov1999Oct}, who suggested linear scaling between the zero-temperature normalized superfluid density $n_s(\tau)/n \equiv n_s(\tau, T=0)/n$ and $\sigma(\tau)\cdot\tau^{-1}$, where $\sigma(\tau)\equiv \sigma(\tau, T={T_c}^+)$ is the normal-state DC conductivity at $T={T_c}^+\equiv T_c+0^+$ and $\tau$ is the scattering time. This "Pimenov scaling" relation (which was partially motivated by the earlier "Uemura scaling" relation between $n_s(\tau)/n$ and $T_c$~\cite{Uemura1988Jul, Uemura1989May, PhysRevLett.66.2665,Tanner1998Jan}) failed to serve as a universal hallmark for high-$T_c$ physics, since heavily doped samples of certain YBaCuO species violated the proposed scaling law~\cite{Ulm1995Apr,Brorson1996Sep}. The work of both Uemura~{\it et al.} and Pimenov~{\it et al.} led to the landmark result of Homes~{\it et al.}~\cite{Homes2004Jul}, who identified that the so-called "Homes scaling" relation between $n_s(\tau)/n$ and $\sigma(\tau)\cdot T_c$ was a more universal feature of high-$T_c$ superconductors.

Unlike Uemura and Pimenov scaling, Homes scaling is obeyed in a wide class of compounds regardless of doping and other sample details~\cite{Homes2012Oct, Pimenov1999Oct, Dressel1994Nov, Milbradt2013Aug, Drichko2002Sep,Homes2009Nov, Wu2010Dec,Klein1994, Pronin1998Jun, Homes2004Jul,Homes2005Oct,Dordevic2022Jun}. While it was suggested by Zaanen~\cite{Zaanen2004Jul} that Planckian dissipation in the normal state of the cuprates (and thus strange-metal physics itself) is fundamentally tied to Homes scaling, both Zaanen and Homes pointed out that Homes scaling is present in low-$T_c$ compounds such as Pb and Nb~\cite{Homes2004Jul,Zaanen2004Jul,Lide2009Jun}. 
{Therefore, the} wide range of applicability {of} such linear scaling relations may very well suggest certain universal physics underlying a broad class of superconductors. {Nevertheless, theoretical works which demonstrate linear scaling beyond the BCS limit are quite scarce~\cite{Erdmenger2012, Kim2016Oct, Niu2016, Tao2017, Kim2022}. }

In our work, we provide numerical evidence and theoretical justification for Homes scaling in a strongly-correlated model of superconductivity distinct from both BCS-type and high-$T_c$-like physics. Specifically, we study a general family of scaling relations given by
\begin{align}
\dfrac{n_{s}(\tau, \lambda)}{n}=
\eta(\tau, \lambda)\dfrac{\sigma(\tau, \lambda)\cdot \psi }{\omega_{p}^{2}/(8\pi^{2})},\label{eq1} 
\end{align}
where $\lambda$ quantifies the interaction strength, 
$\eta(\tau, \lambda)$ is a proportionality factor, {$\omega_p$ is the plasma frequency of the free-electron gas}, and $\psi=T_c$ ($\tau^{-1}$) for Homes (Pimenov) scaling. {(See Sec.~IV A of the Supplemental Material for a distinction between linear proportionality and linear scaling).} We argue that BCS physics cannot explain such scaling relations outside of a dirty, weak-coupling scenario. In this letter, we go beyond BCS theory and consider scaling relations of the form given in Eq.~\eqref{eq1} using the framework of Eliashberg theory~\cite{Eliashberg1, Eliashberg2, RickayzenBook, Parks1, Bardeen1973Jul, Karakozov1975, Allen1983Jan, Marsiglio1988, Combescot1990, RevModPhys.62.1027, Marsiglio1991, Karakozov1991, Marsiglio1992,Combescot1995May, Bennemann, AlexandrovBook, Cappelluti2007Sep, PhysRevB.101.064506, Chubukov2020Jun, Marsiglio2020Jun, Protter2021}. Electron-phonon interactions provide an additional parameter (besides $\tau$) with which to "tune" the normal-state conductivity and superfluid density. Likewise, strong electron-phonon coupling results in a violation of the Planckian bound~\cite{Hartnoll2022Nov}, making Eliashberg theory an ideal setting to investigate the universality of Homes scaling. 

We find linear scaling behavior in the electron-phonon system in both the clean and dirty limits~\cite{Tinkham1959,AGDBook,Lemberger1978Dec}, with a fundamental ingredient for such scaling behavior being Galilean non-invariance {and momentum relaxation}, either via elastic impurity scattering or inelastic scattering between electrons and Einstein phonons in an isotropic system~\cite{Anderson1959,PhysRev.131.563,Millis1988}. 
\\\\
\indent{\it Numerical calculations on the imaginary frequency axis.}--We consider the isotropic, single-band Eliashberg equations on the imaginary Matsubara frequency axis~\cite{Szczesniak2006,Marsiglio2020Jun}. Eliashberg theory goes beyond BCS theory by incorporating a dynamical electron-phonon interaction~\cite{Boyack2023Mar}, and thus the gap function $\Delta(i\omega_n)$ and renormalization function $Z(i\omega_n)$ depend upon the fermionic Matsubara frequencies $\omega_n=(2n+1)\pi T$. For simplicity, we assume an Einstein (or Holstein) phonon model~\cite{Holstein1959Nov, Bennemann, Marsiglio2020Jun} with  a dimensionless electron-phonon coupling $\lambda$ and an Einstein phonon frequency $\omega_E$.

\begin{figure}[t!]
\hspace{-10mm} \includegraphics[width=1\columnwidth]{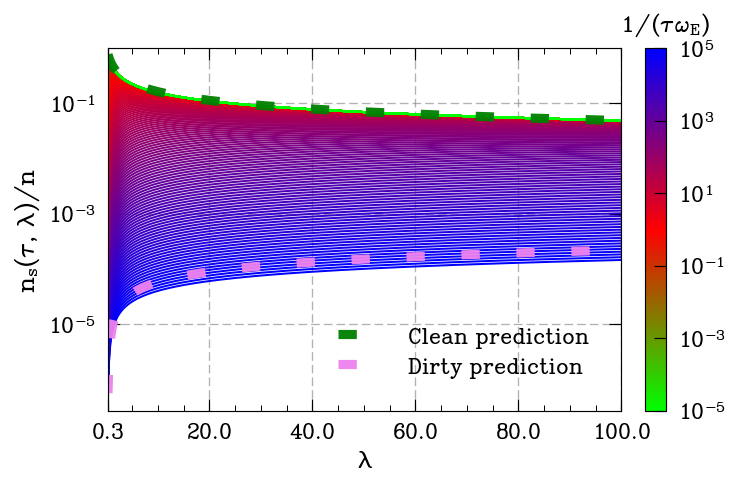}
\vspace{0mm} \caption{ { Superfluid density $n_s(\tau, \lambda)/n$ versus $\lambda$ for various scattering rates. In the clean limit, $n_s(\tau, \lambda)/n\sim 1/Z_0$ (green dashed curve), while $n_s(\tau, \lambda)/n\sim \pi \tau \Delta_0$ (violet dashed curve) in the dirty limit.}}
\label{Fig1}
\end{figure}

We iteratively solve the Eliashberg equations for a fixed $\lambda \in [0.3,\,100]$, with convergence criteria of the Matsubara summation determined by an algorithm discussed in the Supplemental Material~\cite{JHRB_SM}. The gap and the renormalization functions follow a Lorentzian structure for all values of $\lambda$, as already noted for $\lambda\lessapprox 0.5$~\cite{Marsiglio2018Jul}. While Eliashberg theory remains valid for large coupling strengths as long as $\omega_E$ is much smaller than the Fermi energy $\epsilon_F$~\cite{Marsiglio1991, Chubukov2020Jun, Zhang2022Oct}, $\lambda$ is usually no more than $3.5-4$ in most present-day materials \cite{RevModPhys.62.1027,Errea2020Feb}. 
The motivation for considering the large-$\lambda$ limit follows from the formulation of asymptotically strong Eliashberg theory (ASETh)~\cite{Allen1975Aug,Marsiglio1988b,Marsiglio1991,Combescot1995May}, in which the Eliashberg equations reduce to a universal theory characterized by an Einstein phonon spectrum~\cite{Combescot1995May}. Our results for an Einstein phonon model with $\lambda\gg1$ should therefore remain appropriate for other strongly coupled models of Eliashberg superconductivity. {Although several arguments have been presented for a theoretical upper bound $\lambda_c$ on the electron-phonon coupling strength in realistic materials~\cite{PhysRevB.106.064502,Yuzbashyan2022Jul,PhysRevB.106.054518, Esterlis2019}, the motivation for considering arbitrary $\lambda$ (and, in particular, the ASETh limit) is to explore the universal behavior of the Homes slope, regardless of the precise numerical value for $\lambda_c$.} 

After numerically obtaining the gap and renormalization functions, we calculate the superfluid density for arbitrary $\tau$, $\lambda$, and $T$~\cite{Glover1957, Ferrell1958, Abrikosov1959, Karakozov1991, Leyronas1996, Berlisnky1993, Tao2017, Dutta2022}:

\begin{align}
\dfrac{n_{s}(\tau, \lambda, T)}{n}=
&\pi T\sum_{n=-\infty}^\infty\dfrac{\Delta^{2}(i\omega_{n})}{\omega_{n}^{2}+\Delta^{2}(i\omega_{n})}\nonumber\\
\quad&\times\dfrac{1}{Z(i\omega_{n})\sqrt{\omega_{n}^{2}+\Delta^{2}(i\omega_{n})}+1/(2\tau)}.
\label{eq:SupDens}
\end{align}

\noindent In Fig.~\ref{Fig1}, we plot the $T=0$ superfluid density $n_s(\tau, \lambda)/n$ versus $\lambda$. In the dirty limit $1/(\tau \omega_E)\gg 1$, severe suppression of $n_s(\tau, \lambda)/n$ occurs regardless of the interaction strength~\cite{Nam1967, Kogan2013, Mandal2020, Dutta2022}. In the clean limit, we find that $n_s(\tau, \lambda)/n$ goes as $1/Z_0$, where $Z_0\equiv \lim\limits_{T\rightarrow 0}Z(i\omega_0)$ is the $T=0$ limit of the renormalization function. This is in stark contrast to the clean BCS limit, where $n_s/n\rightarrow 1$ as $1/(\tau \omega_E)\rightarrow 0$ \cite{Kogan2013, Tao2017, Dutta2022,GRBook}. 

{The localized Einstein phonon acts as an impurity~\cite{Chubukov2020Jun}. As a consequence, the DC conductivity for the $T>T_c$ normal state is modified from the Drude result~\cite{Marsiglio1996,Klein1994,Marsiglio1994}. Extending previous work done on the Einstein-phonon model~\cite{Marsiglio1995}, we obtain}
$\sigma(\tau, \lambda, T)\equiv (\omega_p^2/(4\pi))\cdot \zeta(\tau, \lambda, T)$, where we define~\cite{Millis1988, Marsiglio1991, Klein1994, Marsiglio1994, Marsiglio1995, Marsiglio1996, Bennemann}:
\begin{widetext}
\begin{align}
\zeta(\tau, \lambda, T)&\equiv\dfrac{1}{2\pi \lambda T}\bigintss_{0}^{\infty}\,\frac{
\textrm{sech}^2\left(
\dfrac{\omega_E}{2T}x
\right)
}{
\textrm{coth} \left( \dfrac{\omega_E}{2T} \right)
-\dfrac{1}{2} \bigg\{
\textrm{tanh}\left[\dfrac{\omega_E}{2T} \left(1-x\right)\right]+\textrm{tanh}\left[\dfrac{\omega_E}{2T}\left(1+x\right)\right]
\bigg\} + \dfrac{1}{\pi \lambda \tau\omega_E}}dx.
\label{eqn3}
\end{align}
\end{widetext}

\noindent In the dirty limit, the above expression reduces to $\zeta(\tau, \lambda, T)=\tau$, reproducing the Drude result~\cite{Zimmerman1991}. In the clean limit, Eq.~\eqref{eqn3} reduces to $\zeta(\lambda, T)=[1/(2\pi\lambda \omega_E)]\cdot \sinh(\omega_E/T)$, yielding a finite DC conductivity independent of $\tau$. At $T=T_c$, we can simplify Eq.~\eqref{eqn3} further by recalling the semi-analytical formula for the Eliashberg critical temperature derived by Combescot~\cite{Combescot1989} for arbitrary $\lambda$ assuming an Einstein phonon model, given by $T_c=a\omega_E(e^{2/\lambda}-1)^{-1/2}$ where $a\approx 0.256$. 
As such, the DC conductivity $\sigma(\tau, \lambda)$ for the $T=T_c$ normal state can be cast as a function purely of $\tau$ and $\lambda$. 
\begin{figure*}
\hspace{3mm}\subfloat[Homes phase diagram\label{FigHomes}]{%
  \includegraphics[width=.39\linewidth]{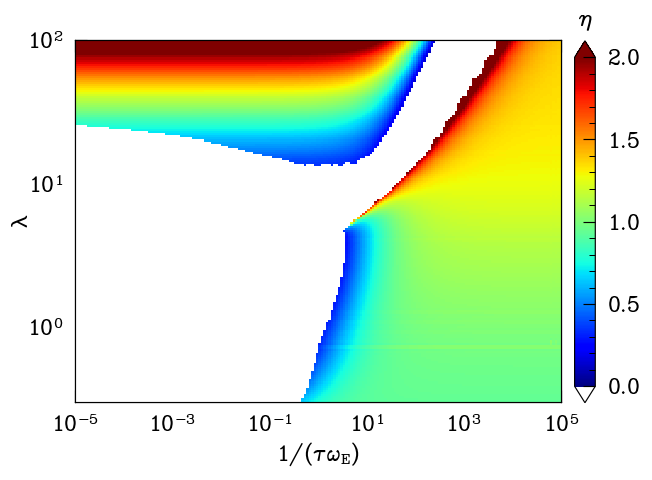}%
}\hspace{-1mm}
\subfloat[Pimenov phase diagram\label{FigPimenov}]{%
  \includegraphics[width=.39\linewidth]{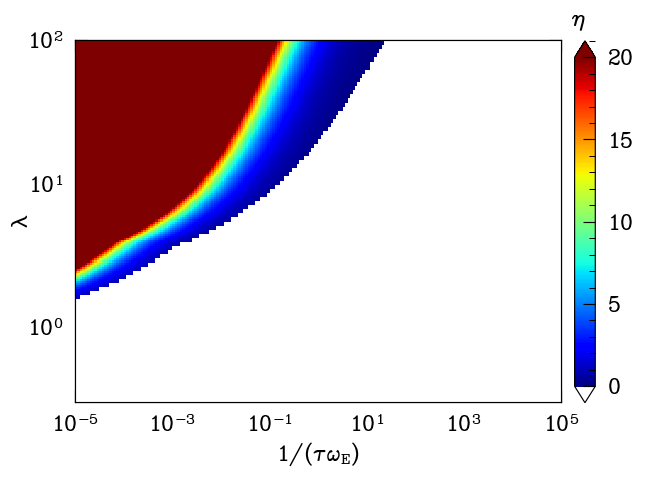}%
}\\
\hspace{3.75mm}\subfloat[Holstein phase diagram\label{FigHolstein}]{%
  \includegraphics[width=.39\linewidth]{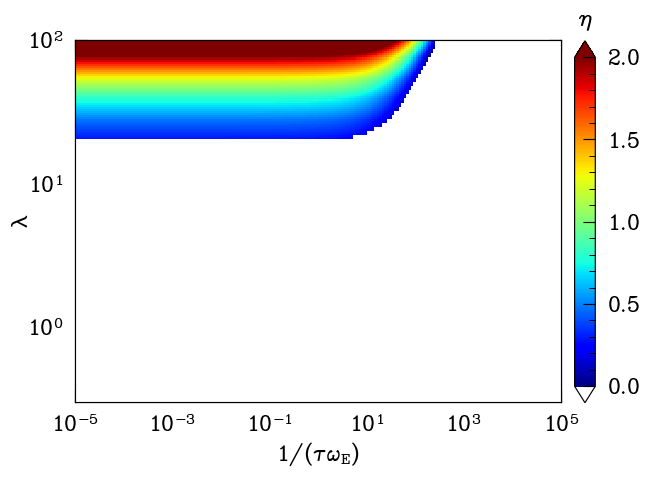}%
}\hspace{-4.5mm}
\subfloat[Planckian dissipation\label{FigPlanck}]{%
\vspace{1mm} \hspace{2mm} \includegraphics[width=.39\linewidth]{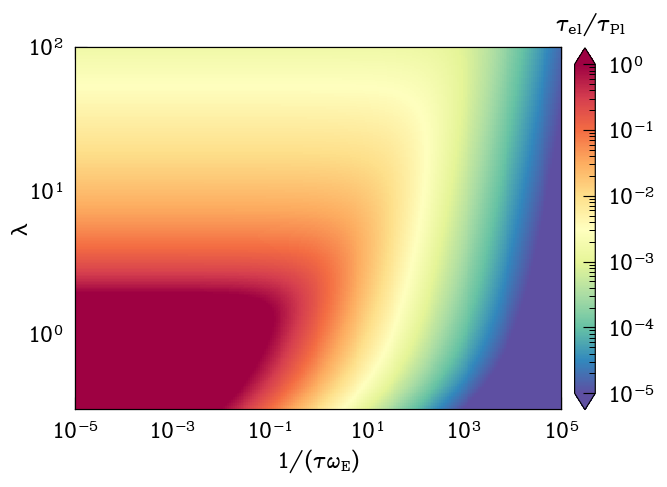}%
}
\caption{(a,b,c) Phase diagrams for scaling relations of the form Eq.~\eqref{eq1}. The color denotes the value of the Homes, Pimenov, and Holstein slopes{. The white region is where the respective linear scaling relation breaks down, as determined by our numerical algorithm (see Sec.~VI B of the Supplemental Material).}
In all instances, universal scaling exists in some regime of the strong-coupling limit. (d) The ratio of the normal-state scattering time $\tau_{\textrm{el}}$ to the Planckian lifetime $\tau_{\textrm{Pl}}\equiv \hbar/(k_B T_c)$ on the $\lambda$ versus $1/(\tau\omega_E)$ grid. We identify $\tau_{\textrm{el}}$ as $\zeta(\tau, \lambda)$ given in Eq.~\eqref{eqn3}.
}
\label{Fig4}
\end{figure*}

The previous result motivates us to consider scaling relations between $n_s(\tau, \lambda)/n$ and $\sigma(\tau, \lambda)$ for a wide range of $\tau$ and $\lambda$. We consider scaling relations of the form Eq.~\eqref{eq1} with $\psi=T_c$ (Homes), $\psi=\tau^{-1}$ (Pimenov) and $\psi=\omega_E$ (Holstein). Results for these scaling relations are shown in Figs. \ref{FigHomes}, \ref{FigPimenov}, and \ref{FigHolstein}, respectively. The slope of the superfluid density versus $\sigma(\tau,  \lambda)\cdot \psi/[\omega_p^2/(8\pi^2)]$ is plotted on a grid of $\lambda$ versus $1/(\tau\omega_E)$, with the lack of a color denoting a breakdown of scaling between $n_s(\tau, \lambda)/n$ and $\sigma(\tau, \lambda)$. In Fig. \ref{FigHomes}, we see that Homes scaling is obeyed in the dirty weak-coupling limit, as predicted by BCS theory \cite{Nam1967,Kogan2013}. In the clean strong-coupling limit, we see that Homes scaling is obeyed for $\lambda\gtrapprox 2\times 10^1$, with Holstein scaling also emerging in a similar regime of the "phase diagram". Pimenov scaling appears to remain valid in the strong-coupling regime, although strong $\lambda$ dependence emerges in the clean limit. 

Our numerical results indicate that universal scaling relations of the form given in Eq.~\eqref{eq1} can be explained within the framework of Eliashberg theory; namely, by virtue of strong interactions between electrons and Einstein phonons. Note that Homes scaling fails only in the clean weak-coupling limit and for certain intermediate values of $\lambda$ and $1/(\tau\omega_E)$. The former violation occurs due to the superfluid density "flattening" to unity as both interactions and the scattering rate are decreased. The latter violation of Homes scaling is more non-trivial, and is the result of non-linear "back-bending" phenomena~\cite{JHRB_SM}. Finally, the work of Zaanen~\cite{Zaanen2004Jul} argued that Homes scaling is a consequence of Planckian dissipation in the normal state. In Fig.~\ref{FigPlanck}, we plot the ratio of the total scattering time over the Planckian time, and find no correlation between scaling behavior and the onset of Planckian dissipation in the normal state. Violation of the Planckian bound (and onset of a "super-Planckian" timescale) in Fig.~\ref{FigPlanck} agrees with Ref.~\cite{Hartnoll2022Nov}.  
\\\\
\indent{\it The origin of Homes scaling.}--
To understand the physical origin of Homes scaling {observed in our numerical calculations,} we perform a semi-analytical calculation of the $T=0$ superfluid density on the imaginary frequency axis. Taking $\Delta(i\omega_n)\approx \lim\limits_{T\rightarrow 0}\Delta(i\omega_0)\equiv \Delta_0$ and $Z(i\omega_n)\approx \lim\limits_{T\rightarrow 0}Z(i\omega_0)\equiv Z_0$, we find \cite{JHRB_SM}
\begin{align}
&\dfrac{n_{s}(\tau, \lambda)}{n}=\dfrac{\pi}{2Z_0\gamma_0} \left[1+\dfrac{4}{\pi\sqrt{1-\gamma_0^2}} \arctan\left(\dfrac{\gamma_0-1}{\sqrt{1-\gamma_0^2}}\right)\right],\label{eqn4}
\end{align}

\noindent where $\gamma_0\equiv 1/(2\tau \Delta_0 Z_0)$. A similar result may be derived on the real frequency axis assuming a constant complex gap \cite{JHRB_SM, Fibich1965,Fibich1965b,Kresin1987Aug,Boyack2023Mar}. In the dirty limit of Eq.~\eqref{eqn4}, $\gamma_0>>1$, and thus the above expression simplifies to $\sim \pi\tau\Delta_0$, in analogy to Nam's result for the dirty BCS superfluid density \cite{Nam1967, Nam1967b}. We identify the Homes slope in this scenario with the ratio $\Delta_0/(2T_c)$. Taking the clean limit ($\gamma_0<<1$), Eq. \eqref{eqn4} reduces to $\sim 1/Z_0$. In the BCS limit, $Z_0=1$, leading to a breakdown of Homes scaling in the clean limit due to a vanishing Homes slope. However, for finite $\lambda$, the superfluid density is suppressed below unity, and scales as the inverse of $Z(i\omega_0)$. As such, our semi-analytical estimate for the clean superfluid density agrees with the results given in Fig.~\ref{Fig1}.

We emphasize that the renormalization $Z_0$ is a crucial ingredient for the realization of Homes scaling in the clean strong-coupling limit. This can be seen by recalling Eq.~\eqref{eqn3}, from which a rough prediction of the clean Homes proportionality factor may be calculated (up to a constant) to be $\eta_H(\lambda)\sim I^{-1}(\lambda)\cdot (\lambda/Z_0)$, where $I(\lambda)$ is the dimensionless integral introduced in Eq.~\eqref{eqn3} with $T=T_c$. We note that $Z_0\sim \sqrt{\lambda}$ and $I(\lambda)\sim 1$ as $\lambda\rightarrow \infty$ \cite{Combescot1995May,Mayrhofer2024Mar}. As such, the Homes proportionality factor $\eta_H(\tau, \lambda)$ is found to be a slowly varying function of $\lambda$ in this limit. Such weak dependence on $\lambda$ is a hallmark of Homes scaling induced by strong electron-phonon coupling, and sets it apart from Homes scaling in the weakly coupled dirty system, where the Homes slope is a constant set at $\boldsymbol{\varDelta}_{0}/(2T_c)\sim 0.8825$~\cite{Nam1967} and where $\boldsymbol{\varDelta}_{0}$ is the $T=0$ BCS gap. Weak $\lambda$ dependence in the Homes slope is similarly observed at large $\lambda$ and small $1/(\tau \omega_E)$ in Fig.~\ref{FigHomes}.

If $Z_0$ is set to unity, then $\eta_H(\tau, \lambda)\sim \lambda$ in the clean strong-coupling limit, and thus Homes scaling breaks down. Homes scaling also breaks down in the case of clean superconductors where superconductivity is mediated by bosons with a finite momentum-dependent dispersion~\cite{Raines2023}, in which electromagnetic vertex corrections in the superfluid density cancel any dependence on the mass renormalization. However, in the case of a dynamical gap mediated by dispersionless Einstein bosons, this cancellation does not occur~\cite{Raines2023}. This motivates us to propose that Homes scaling, while a poor signature of high $T_c$, normal-state Planckian dissipation, or a high impurity concentration, is instead a universal hallmark of {some general} Galilean non-invariance{ and momentum relaxation. The former ensures a non-unity superfluid density~\cite{Leggett1965}, while the latter ensures a finite DC conductivity~\cite{Maslov2016Dec}.} In the present paper, {both Galilean non-invariance and momentum relaxation are} achieved by virtue of elastic scattering of electrons by impurities or inelastic scattering of electrons by Einstein phonons. Note that the opposite is not universally true; i.e., Galilean non-invariance{/momentum relaxation} does not always result in a linear Homes slope, as evident from Fig.~\ref{FigHomes}~\footnote{We do not consider localization or Umklapp-scattering effects, which can also dramatically alter the behavior of a superconductor~\cite{SadovskiiBook}}. 

{We emphasize that our main focus is not to explain Homes scaling in high-$T_c$ superconductors. Rather, we demonstrate linear Homes scaling in a strongly interacting model by virtue of a mechanism that is not solely due to impurity scattering. Similarly, we note that the Coulomb interaction should not have an appreciable effect on the Homes slope in electron-phonon superconductors, as this interaction does not significantly affect observables in the weak-coupling limit~\cite{Marsiglio2020Jun, Allen1975Aug} and is negligible compared to the divergent electron-phonon coupling in the ASETh limit~\cite{Combescot1995May}. In this way, Homes scaling can be seen as a robust consequence of Cooper pairing and some general disorder/dissipation.}

{\it The asymptotically strong limit.}-- The consideration of an Einstein model is important for the strong-coupling analysis, as the $\lambda\rightarrow \infty$ limit is universally described by an Einstein spectrum satisfying $\lambda \omega_E^2=2$~\cite{Combescot1995May}. For $\lambda\rightarrow \infty$, the Homes slope in the dirty limit reduces to a universal constant given by $\sim 1/(3a)\not = \Delta_0/(2T_c)$~\cite{JHRB_SM}, where $a\approx 0.256$~\cite{Combescot1989}. However, in the clean limit, then $\sigma(\lambda)T_c/[\omega_p^2/(8\pi^2)]=\lambda^{-1}$ for all $\tau$, while the superfluid density scales as $\lambda^{-1/2}$. This results in a diverging Homes proportionality factor proportional to $\sqrt{\lambda}$, and the breakdown of Homes scaling as $\lambda\rightarrow \infty$. 
Similar analysis suggests that Pimenov and Holstein scaling break down in the dirty and clean limits as $\lambda\rightarrow \infty$.

\begin{figure}[t!]
\hspace{0mm} \includegraphics[width=1\columnwidth]{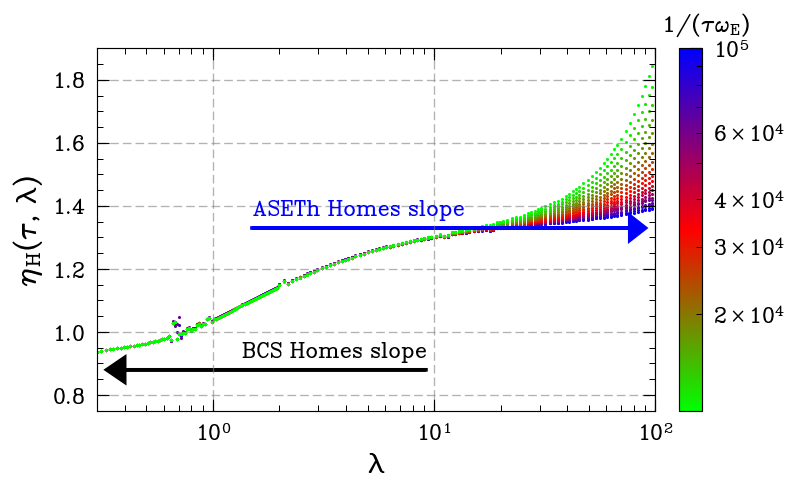}
\vspace{-5mm} \caption{ { Homes proportionality factor $\eta_H(\tau, \lambda)$ in the dirty limit plotted  versus the electron-phonon coupling $\lambda$. As the system becomes dirtier in the large-$\lambda$ limit, the Homes proportionality factor agrees with the asymptotic prediction.}}
\label{Main_HomesSlope}
\end{figure}

In Fig.~\ref{Main_HomesSlope}, we show the dirty limit of the Homes factor plotted versus $\lambda$, from $\lambda=0.3$ to $\lambda=100$. Extrapolation of the small-$\lambda$ data to $\lambda\rightarrow 0$ gives a Homes factor of $\sim 0.88$, in agreement with our theoretical BCS prediction. Extrapolation of the large-$\lambda$ data to $\tau \omega_E\rightarrow 0$ yields a dirty Home slope of $\sim 1.35$ for $\lambda=100$, which is in agreement with our prediction for $\lambda\rightarrow \infty$ via the asymptotic Eliashberg equations~\cite{JHRB_SM}. {Once again, we note that our main conclusions do not depend upon the precise value of some upper cutoff $\lambda_c$. Rather, our value of $\eta_H\sim 1/(3a)$ should be interpreted as a universal upper limit to the dirty Homes slope in Eliashberg theory, regardless of $\lambda_c$.}
\\\\
\indent {\it Conclusions.}--By combining numerical and analytical techniques for electron-phonon superconductors at weak and strong coupling and arbitrary scattering rates, we find that {Homes-like} scaling relations are not solely correlated with large $T_c$~\cite{Homes2004Jul}, normal-state Planckian dissipation~\cite{Zaanen2004Jul}, or some {large scattering rate}~\cite{Kogan2013}. Instead, we find that Homes scaling is closely connected to {Galilean non-invariance and momentum relaxation}, and thus remains valid in the clean limit for large electron-phonon coupling $\lambda$ assuming an Einstein phonon model~\footnote{We emphasize that our work is not solely focused on the large-$\lambda$ limit of Eliashberg theory. Rather, the large-$\lambda$ limit is used as a proof of concept for Homes-type linear scaling in the clean and asymptotically strong dirty limits.}. Pimenov and Holstein scaling are shown to emerge for strong enough electron-phonon coupling for certain values of the scattering rate, while the Homes slope approaches a universal constant for $\lambda\rightarrow \infty$ in the dirty limit. Our numerical values of the Homes slope for $\lambda\sim 0.3$ and $\lambda\sim 100$ are in agreement with the theoretical predictions of BCS theory and ASETh, respectively.

\vspace{2mm}
\noindent {\it Acknowledgments.}-- We thank Alexander Balatsky, Andrey Chubukov, Nikolay Gnezdilov, Christopher Homes, Frank Marsiglio, Pavel Volkov, and Emil Yuzbashyan for insightful conversations. This research was supported through funds provided by Dartmouth College. 

\bibliography{main}

\end{document}


\title{Supplemental Material: \\ Universal scaling relations in electron-phonon superconductors}

\author{Joshuah T. Heath}
\author{Rufus Boyack}
\affiliation{Department of Physics and Astronomy, Dartmouth College, Hanover, New Hampshire 03755, USA}

\date{\today}

\pacs{1}

\maketitle
\tableofcontents

\renewcommand{\theequation}{S.\arabic{equation}}

\section{Preliminaries}

\subsection{Overview of the Eliashberg equations}

In this subsection, we present a brief discussion of the Eliashberg equations. Since there is a wealth of literature with derivations~\cite{RickayzenBook, Parks1, AlexandrovBook, Bennemann, Marsiglio2020Jun, Protter2021} of these equations, our presentation will be minimal. In addition, the next subsection presents careful definitions of the "clean" and "dirty" limits in regard to electron-phonon superconductivity. 

In Eliashberg theory (ETh), $\lambda$ is a dimensionless parameter that characterizes the strength of the electron-phonon coupling. For simplicity, the phonon is assumed to be described by a Holstein model with electron-phonon coupling $\alpha^2 F(\omega)=A\delta(\omega-\omega_{E})$, where $A=\lambda\omega_{E}/2$, and the bosonic propagator is $\lambda(i\Omega_{m})=2A\omega_{E}/(\omega_{E}^2+\Omega_{m}^2)$. The fermionic and bosonic Matsubara frequencies are defined by $\omega_{n}=(2n+1)\pi T$ and $\Omega_{n}=2n\pi T$, respectively, where $n\in\mathbb{Z}$ and $T$ is the temperature. Throughout the Supplemental Material, we use Natural units $\hbar=k_{B}=c=1$. 

Under the assumptions of particle-hole symmetry and a constant electronic density of states, the single-band Eliashberg equations, formulated on the imaginary frequency axis, have the form~\cite{Bennemann, Marsiglio2020Jun}:
\begin{subequations}
\begin{align}
\Delta(i\omega_{n})Z(i\omega_{n})  &= \pi T\sum_{m=-\infty}^{\infty}
\lambda(i\omega_{n}-i\omega_m)\frac{\Delta({i\omega_m})}{\sqrt{\omega_m^2 + \Delta^2({i\omega_m})}}, 
  \label{eq:EthEqnsIm1}\\
Z(i\omega_{n})  &=  1 + \frac{\pi T}{\omega_{n}}\sum_{m=-\infty}^{\infty} 
\lambda(i\omega_{n}-i\omega_m) \frac{\omega_m}{\sqrt{\omega_m^2  + \Delta^2({i\omega_m})} }. 
\label{eq:EthEqnsIm2}
\end{align}
\end{subequations}
The gap function $\Delta(i\omega_n)$ and the renormalization function $Z(i\omega_n)$ are related to $\phi(i\omega_n)$ and $\widetilde{\omega}(i\omega_n)$ by 
$\phi(i\omega_n)\equiv\Delta(i\omega_n) Z(i\omega_n)$ and $\widetilde{\omega}(i\omega_n)=i\omega_{n} Z(i\omega_n)$. In terms of real frequencies, the Eliashberg equations are given by~\cite{Marsiglio1988, Marsiglio1991, Bennemann, Marsiglio2020Jun}
\begin{subequations}
\begin{align}
\Delta(\omega + i \delta)Z(\omega + i \delta)  &= \pi T\sum_{m=-\infty}^{\infty}
\lambda(\omega-i\omega_m)\frac{\Delta({i\omega_m})}{\sqrt{\omega_m^2 + \Delta^2({i\omega_m})}}\nonumber \\
 &\quad+  \frac{i\pi}{2} \lambda\omega_E  \Biggl\{\left[N(\omega_E)+f(\omega_E-\omega)\right] 
  \frac{\Delta(\omega-\omega_E + i \delta)}{\sqrt{(\omega - \omega_E + i\delta)^2 - \Delta^2(\omega-\omega_E + i \delta)}}\nonumber \\
 &\hspace{1.7cm} + \left[N(\omega_E)+f(\omega_E+\omega)\right]  
 \frac{\Delta(\omega+\omega_E + i \delta)}{\sqrt{(\omega + \omega_E + i\delta)^2  - \Delta^2(\omega+\omega_E + i \delta)}}\Biggr\},
 \label{eq:EthEqnsRe1}\\
Z(\omega + i\delta)  &=  1 + \frac{i\pi T}{\omega}\sum_{m=-\infty}^{\infty} 
\lambda(\omega-i\omega_m) \frac{\omega_m}{\sqrt{\omega_m^2  + \Delta^2({i\omega_m})} } \nonumber \\
 &\hspace{0.6cm}+  i\pi \lambda\frac{\omega_E}{2\omega} \Biggl\{\left[N(\omega_E)+f(\omega_E-\omega)\right]
  \frac{(\omega-\omega_E)}{\sqrt{(\omega-\omega_E + i\delta)^2 - \Delta^2(\omega-\omega_E+i\delta)}} \nonumber \\ 
&\hspace{2cm}  + \left[N(\omega_E)+f(\omega_E+\omega)\right]\frac{(\omega+\omega_E)}{ \sqrt{(\omega+\omega_E + i\delta)^2 - \Delta^2(\omega+\omega_E + i \delta)}}\Biggr\}.
\label{eq:EthEqnsRe2}
\end{align}
\end{subequations}
Analytical continuation of Eqs.~\eqref{eq:EthEqnsRe1}-\eqref{eq:EthEqnsRe2} from the real frequency axis to the imaginary frequency axis, via $\omega+i\delta\rightarrow i\omega_n$, leads to the imaginary axis equations given in Eqs.~\eqref{eq:EthEqnsIm1}-\eqref{eq:EthEqnsIm2}~\cite{Combescot1995May}. Note, however, one cannot obtain the real frequency axis equations by just replacing $i\omega_{n}$ by $\omega+i\delta$ in the imaginary frequency axis Eliashberg equations because the analytical continuation is not unique and the result would also have simple poles~\cite{Marsiglio1988, Combescot1995May}. 

In the presence of impurity scattering, where $1/\tau$ is the impurity scattering rate, the gap and renormalization functions can be written as~\cite{Marsiglio1992}:
\begin{align}
\phi(i\omega_{n}) &= \phi_{\textrm{cl}}(i\omega_{n})+\dfrac{1}{2\tau}\dfrac{\Delta(i\omega_{n})}{\sqrt{\omega_{n}^{2}+\Delta^{2}(i\omega_{n})}}, \\
\widetilde{\omega}(i\omega_{n}) &= \widetilde{\omega}_{\textrm{cl}}(i\omega_{n})+\dfrac{1}{2\tau}\dfrac{\omega_{n}}{\sqrt{\omega_{n}^{2}+\Delta^{2}(i\omega_{n})}}.
\end{align}
The subscript $\textrm{cl}$ denotes the corresponding clean limit of a given quantity. In the subsequent analysis, we shall drop the cl subscript (thus, all of the gap and renormalization functions shall refer to their expressions in the clean limit) and we shall incorporate the $\tau$ dependence explicitly.

\begin{figure}[!t]
\hspace{-10mm} \includegraphics[width=0.7\columnwidth]{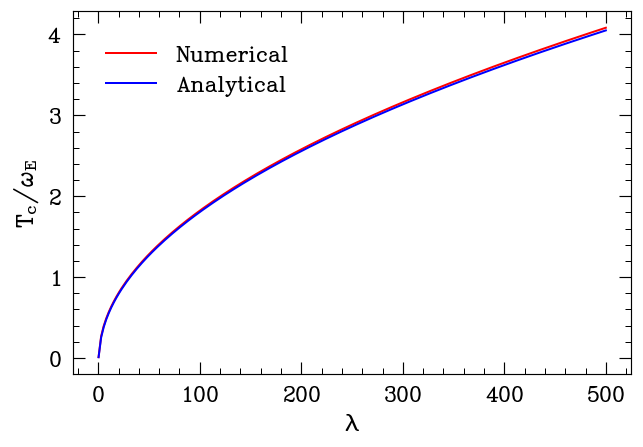}
\caption{Comparison of the numerical result for $T_c$ (red) and the analytical result proposed by Combescot (blue). Near-exact agreement is found for values of $\lambda$ from $\lambda=0.3$ to $\lambda=500$.}
\label{fig:Combescot}
\end{figure}

The transition temperature $T_c$~\cite{Allen1983Jan} can be obtained by numerically solving the linearized Eliashberg equations. The value of $T_c$ may also be determined analytically from Combescot's~\cite{Combescot1989} interpolation formula, which assumes an Einstein phonon model. This formula is given by
\begin{equation}
\dfrac{T_c}{\omega_E}=\dfrac{a}{\sqrt{\exp(2/\lambda)-1}},  \quad \textrm{where\ } a\approx 0.256.
\label{eq:Combescot}
\end{equation}
Interestingly, the above formula is an accurate prediction for $T_c$ for all $\lambda$. In the limit that $\lambda\rightarrow \infty$, the Allen-Dynes formula~\cite{Allen1975Aug} $T_c/\omega_E \sim 0.18 \sqrt{\lambda}$ is recovered. In addition, in the limit that $\lambda\rightarrow0$, the weak-coupling result~\cite{Karakozov1975, Combescot1989, Combescot1990, Marsiglio2018Jul} $T_c/\omega_E \sim 0.25\exp(-1/\lambda)$ is recovered. In Fig~\ref{fig:Combescot}, we show a comparison between the numerically determined result and Combescot's interpolation formula for $T_c$, and find near-perfect agreement for all $\lambda$ values considered.

\subsection{Formal definition of the clean and dirty limits}

Throughout the main text and the Supplemental Material, we make reference to the clean and dirty limits when discussing  the zero-temperature superfluid density and the normal-state conductivity at $T=T_c$. In this subsection, we formally define these limits for arbitrary coupling $\lambda$. For future reference, we let $\ell$ denote the electron mean free path and $\xi$ the electromagnetic coherence length in a superconductor. 

Formally, the coherence length is defined by~\cite{TinkhamBook}:
\begin{equation}
\xi = \dfrac{3\pi}{4}K(0,\mathbf{0})\cdot\left(\underset{q\rightarrow\infty}{\lim} 
q K(0, \mathbf{q})\right)^{-1},
\label{eq:CohLength}
\end{equation}
\noindent where $K(\Omega_{m},\bf{q})$ is the current-current correlation function (with both paramagnetic and diamagnetic contributions). For an isotropic system, the static correlation function depends on $q^2$. 
Following the derivation in Sec.~37 of Ref.~\cite{AGDBook}, while modifying the gap to be frequency dependent and including the renormalization function, the electromagnetic response kernel within Eliashberg theory is:
\begin{equation}
K(0,\mathbf{q}) = \dfrac{3\pi}{4}T\sum_{n=-\infty}^{\infty}
\int_{-1}^1\dfrac{\phi^2(i\omega_n)}
{\widetilde{\omega}_n^2+\phi^2(i\omega_n)}\cdot
\dfrac{1-x^2}{\sqrt{\widetilde{\omega}_n^2+\phi^2(i\omega_n)}+\dfrac{1}{4}q^2v_{F}^2x^2} dx.
\label{eq:Kernel}
\end{equation}
Using this result for the current-current correlation function, the numerator and denominator in Eq.~\eqref{eq:CohLength} are straightforward to compute and, as shown by Rickayzen \textit{et al.}~\cite{Lemberger1978Dec}, the coherence length can be written as
\begin{equation}
\xi = \frac{1}{2} v_F\sum_{n=-\infty}^\infty \dfrac{\Delta^2(i\omega_n)}{Z(i\omega_n)\left(\omega_n^2+\Delta^2(i\omega_n)\right)^{3/2}} \cdot \left( \sum_{n=-\infty}^\infty \dfrac{\Delta^2(i\omega_n)}{\omega_n^2 +\Delta^2(i\omega_n)}\right)^{-1}. 
\label{coherence}
\end{equation}

In the case of BCS theory, the gap is independent of frequency, thus in the limit that $T=0$ the coherence length reduces to its well-known~\cite{TinkhamBook} value~$\xi_{0, \mathrm{BCS}} = \hbar v_F/(\pi \DeltaOBCS)$ (where we have restored $\hbar$). In this case,
\begin{equation}
\dfrac{\ell}{\xi_{0, \mathrm{BCS}}}=\dfrac{\tau v_F}{v_F/(\pi \DeltaOBCS)}\sim \tau \DeltaOBCS,
\label{eq:CohBCS}
\end{equation}  
where the mean free path is $\ell=\tau v_F$ and $\tau$ is the elastic scattering time. In the case where $\tau\DeltaOBCS\rightarrow \infty$, the mean free path diverges due to low instances of scattering events. There are formally two such limits~\cite{Klein1994}: the Pippard limit (for clean type-I superconductors) and the London limit (for clean type-II superconductors). As we will be concerned with superconductors with a non-zero penetration depth, our definition of the clean limit is synonymous with the London limit. In the limit where $\tau \DeltaOBCS\rightarrow 0$, many elastic scattering events occur, and thus the mean free path is small compared to the coherence length. This case is defined to be the local or dirty limit~\cite{Klein1994}.

\begin{figure}[!t]
\hspace{-10mm} \includegraphics[width=0.7\columnwidth]{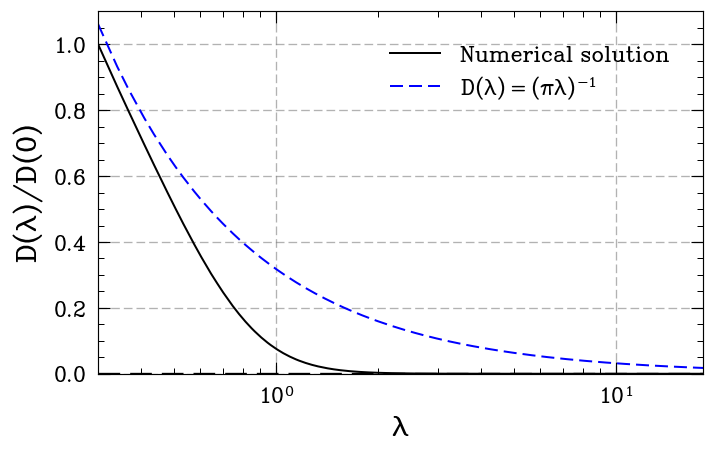}
\caption{The value of $D(\lambda)/D(0)$ versus $\lambda$, where $D(\lambda)$ is given by Eq.~\eqref{D_eq} and $D(0)$ is approximated by the value of $D(\lambda)$ for $\lambda=0.3$. The value of the inverse coupling constant is plotted for comparison, which follows from the strong-coupling analysis discussed later. Suppression of the ratio for $\lambda \gtrapprox 1$ results in an amplified value of $\ell/\xi$ compared to the BCS case, and thus the dirty/clean limit in strong-coupling Eliashberg theory is dependent upon $\lambda$.}
\label{D_func}
\end{figure}

For a frequency-dependent gap, we use Eq.~\eqref{coherence} to define 
\begin{equation}
D(\lambda) \equiv \frac{1}{2} \omega_E \sum_{n=-\infty}^\infty \dfrac{\Delta^2(i\omega_n)}{Z(i\omega_n) \left(\omega_n^2+\Delta^2(i\omega_n)\right)^{3/2}} \cdot \left(\sum_{n=-\infty}^\infty \dfrac{\Delta^2(i\omega_n)}{\omega_n^2 +\Delta^2(i\omega_n)}\right)^{-1}.
\label{D_eq}
\end{equation}
We include the Einstein frequency $\omega_E$ to ensure that $D(\lambda)$ is dimensionless. In the limit that $\lambda\rightarrow0$, a BCS-like result as found in Eq.~\eqref{eq:CohBCS} is obtained. For moderate electron-phonon coupling, however, the clean and dirty limits are modified from the BCS case. In Fig.~\ref{D_func}, we plot the Eliashberg value of $D(\lambda)$ divided by the weak-coupling value of $D(\lambda)$, defined as $D(0)\equiv \underset{\lambda\rightarrow0}{\lim}D(\lambda)$, in which case we find the ratio decreases upon increasing $\lambda$ above $\lambda\approx 0.3$, and hence in the strong-coupling regime we must define the dirty and clean limits as follows:
\begin{subequations}
\begin{equation}
\textrm{Dirty limit:}\qquad \tau \omega_E D^{-1}(\lambda)\rightarrow 0\label{dirtydef},
\end{equation}
\begin{equation}
\textrm{Clean limit:}\qquad \tau \omega_E D^{-1}(\lambda)\rightarrow \infty\label{cleandef}.
 \end{equation}
\end{subequations}
For values of $\lambda \lessapprox 1$, the dirty and clean limits are well-approximated by $\tau\omega_E\ll 1$ and $\tau \omega_E\gg 1$, respectively. However, for $\lambda\gtrapprox 1$, we find that $\tau\omega_E \ll D(\lambda)$ defines the dirty limit, which is a more stringent constraint upon the mean free path than in weak-coupling Eliashberg theory as $D(\lambda)$ is a rapidly decreasing function of $\lambda$. Similarly, the clean limit is defined by $\tau \omega_E \gg D(\lambda)$, which is a more lenient constraint upon the mean free path as compared to the weak-coupling Eliashberg case. The precise definitions of the dirty and clean limits as defined in Eqs.~\eqref{dirtydef} and \eqref{cleandef}, respectively, are important whenever we consider physical quantities in the presence of a finite electron-phonon coupling, especially for $\lambda\gg1$. In practice, we shall fix $\lambda$ between $0.3-100$ (excluding the asymptotic limit $\lambda\rightarrow \infty$), and let $\tau\omega_{E}\rightarrow0$ or $\tau\omega_{E}\rightarrow\infty$, which then reduce to the conventional dirty and clean limits, respectively.

\section{Superfluid density in Eliashberg theory}

\subsection{General formulation of the superfluid density on the imaginary and real frequency axes}

The superfluid density $n_{s}$ is defined as the long-wavelength limit of the static electromagnetic response~\cite{RickayzenBook, AGDBook}: 
\begin{equation}
\dfrac{n_s}{n} \equiv \lim_{{\bf q}\rightarrow 0} \Lambda K(\Omega_{m}=0, \mathbf{q}).
\label{eq:SupDens}
\end{equation}
$\Lambda=m/(ne^2)$ is the London parameter. Since we consider uniform systems, the superfluid-density tensor is diagonal and has diagonal entries given by $n_{s}/n$; the current-current correlation function is similarly diagonal. Extending the result in Eq.~\eqref{eq:Kernel} to incorporate impurities, the normalized superfluid density within Eliashberg theory is
\begin{align}
\dfrac{n_s(\tau, \lambda, T)}{n}&=\dfrac{3\pi}{4}T\sum_{n=-\infty}^{\infty}
\int_{-1}^1\dfrac{\phi^2(i\omega_n)}
{\widetilde{\omega}_n^2+\phi^2(i\omega_n)}\cdot
\dfrac{1-x^2}{\sqrt{\widetilde{\omega}_n^2+\phi^2(i\omega_n)}+\dfrac{1}{2\tau}} dx \notag\\
&=\pi T \sum_{n=-\infty}^{\infty}\dfrac{\Delta^{2}(i\omega_{n})}{\omega_{n}^{2}+\Delta^{2}(i\omega_{n})}\cdot\dfrac{1}{Z(i\omega_{n})\sqrt{\omega_{n}^{2}+\Delta^{2}(i\omega_{n})}+\dfrac{1}{2\tau}}. 
\label{eq:EThSupDens1}
\end{align}
It is convenient to express the superfluid density in terms of the gap and renormalization functions on the real frequency axis. To do this, we consider the following Matsubara frequency summation identity for a function $g(i\omega_n)$, where $\mathcal{C}$ is a closed contour which encloses all of the fermionic Matsubara frequencies~\cite{AGDBook}: 
\begin{align}
T\sum_{n=-\infty}^{\infty}g(i\omega_n)=\dfrac{1}{4\pi i}\oint_{\mathcal{C}} \tanh\left(\dfrac{z}{2T}\right)g(z)dz.\label{ContourEq}
\end{align}
There are no singularities in the complex plane (i.e., there are no complex solutions such that $\Delta(z)=z$), and thus the superfluid density can be represented as a closed contour integral along the branch cut~\cite{Eliashberg1960, AGDBook}: 
\begin{align}
\dfrac{n_{s}(\tau,\lambda, T)}{n} & =2\pi T\sum_{n=0}^{\infty}\dfrac{\Delta^{2}(i\omega_{n})}{\omega_{n}^{2} + \Delta^{2}(i\omega_{n})}\cdot\dfrac{1}{Z(i\omega_{n})\sqrt{\omega_{n}^{2}+\Delta^{2}(i\omega_{n})}+\dfrac{1}{2\tau}} \notag\\
&=\dfrac{1}{2i}\oint \dfrac{\Delta^{2}(z)}{\Delta^{2}(z) - z^{2}}\cdot\dfrac{\tanh\left(\dfrac{z}{2T}\right)}{Z(z)\sqrt{\Delta^{2}(z)-z^{2}}+\dfrac{1}{2\tau}}\ dz \notag\\
 &=-\dfrac{1}{2}\int_{0}^{\infty}\dfrac{\Delta^{2}(\omega+i\delta)}{\omega^{2}-\Delta^{2}(\omega+i\delta)}
\cdot\dfrac{\tanh\left(\dfrac{\omega}{2T}\right)}{Z(\omega+i\delta)\sqrt{\omega^{2}-\Delta^{2}(\omega+i\delta)}+\dfrac{i}{2\tau}}\ d\omega \notag\\
&\quad -\dfrac{1}{2}\int_{-\infty}^{0}\dfrac{\Delta^{2}(\omega+i\delta)}{\omega^{2}-\Delta^{2}(\omega+i\delta)}\cdot\dfrac{\tanh\left(\dfrac{\omega}{2T}\right)}{-Z(\omega+i\delta)\sqrt{\omega^{2}-\Delta^{2}(\omega+i\delta)}+\dfrac{i}{2\tau}} \ d\omega.
 \label{Appendix1}
\end{align}	
Note that in the last step of the above, we have used the general property~\cite{Marsiglio1992}:
\begin{equation}
\sqrt{\Delta^2-z^2}=-i\textrm{sgn}[\textrm{Re}(z)]\sqrt{z^2-\Delta^2}.
\end{equation}
In the last line of Eq.~\eqref{Appendix1}, we have already explicitly put in the minus sign due to the sgn function. 
\newpage
From the relations $\phi(z)=\phi(-z)$, $\phi^*(z)=\phi(z^*)$, $Z(z)=Z(-z)$, and $Z^*(z)=Z(z^*)$~\cite{Marsiglio1992}, Eq.~\eqref{Appendix1} then simplifies to the following expression:

\begin{align}
\dfrac{n_{s}(\tau, \lambda, T)}{n} &= -\dfrac{1}{2}\int_{0}^{\infty}\dfrac{\Delta^{2}(\omega+i\delta)}{\omega^{2}-\Delta^{2}(\omega+i\delta)}\cdot\dfrac{\tanh\left(\dfrac{\omega}{2T}\right)}{Z(\omega+i\delta)\sqrt{\omega^{2}-\Delta^{2}(\omega+i\delta)}+\dfrac{i}{2\tau}}\ d\omega\notag\\
&\phantom{=}	-\dfrac{1}{2}\int_{0}^{\infty}\dfrac{\Delta^{2}(-\omega+i\delta)}{\omega^{2}-\Delta^{2}(-\omega+i\delta)}\cdot\dfrac{\tanh\left(\dfrac{\omega}{2T}\right)}{Z(-\omega+i\delta)\sqrt{\omega^{2}-\Delta^{2}(-\omega+i\delta)}-\dfrac{i}{2\tau}}\ d\omega\notag\\
&= -\mathrm{Re}\left[\int_{0}^{\infty}\dfrac{\Delta^{2}(\omega+i\delta)}{\omega^{2}-\Delta^{2}(\omega+i\delta)}\cdot\dfrac{\tanh\left(\dfrac{\omega}{2T}\right)}{Z(\omega+i\delta)\sqrt{\omega^{2}-\Delta^{2}(\omega+i\delta)}+\dfrac{i}{2\tau}}\right] d\omega \notag\\
&\equiv-\mathrm{Re}\left[\int_{0}^{\infty}\dfrac{\Delta^{2}(\omega)}{\omega^{2}-\Delta^{2}(\omega)}\cdot\dfrac{\tanh\left(\dfrac{\omega}{2T}\right)}{Z(\omega)\sqrt{\omega^{2}-\Delta^{2}(\omega)}+\dfrac{i}{2\tau}}\right]d\omega.
\label{eq:EThSupDens2}
\end{align}

\noindent Note that, $Z(\omega)$ and $\Delta(\omega)$ are shorthand notation for $Z(\omega+i\delta)$ and $\Delta(\omega+i\delta)$, respectively. Equation~\eqref{eq:EThSupDens2} defines the general form of the superfluid density for a frequency-dependent gap and for general impurity scattering time $\tau$. If a prefactor of $1/\left[i\sqrt{\Delta^2(\omega)-\omega^2}\right]$ is factored out of the expression, then the subsequent result agrees with Ref.~\cite{Combescot1996}. 

It is helpful to rewrite the integrand in Eq.~\eqref{eq:EThSupDens2} as follows:

\begin{equation}
\dfrac{\Delta^{2}(\omega)}{\omega^{2}-\Delta^{2}(\omega)}\cdot\dfrac{1}{Z(\omega)\sqrt{\omega^{2}-\Delta^{2}(\omega)}+\dfrac{i}{2\tau}} = -2i\tau \dfrac{\Delta^2(\omega)}{\omega^2-\Delta^2(\omega)} + \dfrac{\Delta^{2}(\omega)}{Z(\omega)\sqrt{\omega^{2}-\Delta^{2}(\omega)}}\cdot\bigg(\dfrac{1}{\theta}\dfrac{1}{\theta - i\sqrt{\omega^{2}-\Delta^{2}(\omega)}}\bigg),
 \end{equation}
 
\noindent where we have defined $\theta\equiv 1/\left[2\tau Z(\omega)\right]$. Using this identity, the zero-temperature superfluid density then becomes

\begin{equation}
\dfrac{n_{s}(\tau,\lambda)}{n}=2\tau\mathrm{Re}\bigg[i\int_{0}^{\infty}\dfrac{\Delta^{2}(\omega)}{\omega^{2}-\Delta^{2}(\omega)}d\omega\bigg]-\mathrm{Re}\bigg[\int_{0}^{\infty}\dfrac{\Delta^{2}(\omega)}{Z(\omega)\sqrt{\omega^{2}-\Delta^{2}(\omega)}}
\cdot\bigg(\dfrac{1}{\theta}\dfrac{1}{\theta - i\sqrt{\omega^{2}-\Delta^{2}(\omega)}}\bigg)d\omega\bigg]. 
\label{eq:EThSupDens3}
\end{equation}

\noindent Equation~\eqref{eq:EThSupDens3} can be used to easily determine the zero-temperature superfluid density in the dirty $(\tau\omega_{E} \ll 1)$ and the clean $(\tau\omega_{E} \gg 1)$ limits:

\begin{align}
\dfrac{n_s(\tau,\lambda)}{n}=
\begin{cases}
& \displaystyle2\tau\mathrm{Re}\bigg[i\int_{0}^{\infty}\dfrac{\Delta^{2}(\omega)}{\omega^{2}-\Delta^{2}(\omega)}d\omega\bigg],\quad \tau\omega_{E} \ll 1\\ &\\
& -\displaystyle\mathrm{Re}\bigg[\int_{0}^{\infty}\dfrac{\Delta^{2}(\omega)}{Z(\omega)\left(\omega^{2}-\Delta^{2}(\omega)\right)^{3/2}}d\omega\bigg],\quad \tau\omega_{E} \gg 1. 
\end{cases}
\end{align}

In the above, we have implicitly assumed that $\Delta(\omega)$ is a complex quantity, that is, $\mathrm{Im}[\Delta(\omega)]\neq0$; in the next section, we shall consider the superfluid density in the BCS limit, where the gap is a real constant. In the dirty limit, $n_s(\tau,\,\lambda)/n$ scales as $\tau\omega_{E}$ and is increasingly suppressed as the impurity scattering rate increases. The expression for $n_s(\tau,\,\lambda)/n$ in the clean limit agrees with Ref.~\cite{Karakozov1991}. In the clean limit, the integrand appearing in the superfluid density is inversely proportional to the renormalization function $Z(\omega)$. As we shall show, the appearance of $1/Z(\omega)$ in $n_s(\tau,\,\lambda)/n$ has significant implications for superfluid response in clean Eliashberg systems. In particular, the presence of the $1/Z(\omega)$ factor will lead to Homes scaling in the clean regime for certain electron-phonon coupling strengths.  
\newpage

\subsection{Superfluid density in the BCS limit}
\label{sec:SupDensBCS}

In the previous section, Eq.~\eqref{eq:EThSupDens1} and Eq.~\eqref{eq:EThSupDens2} express the general form of the superfluid density on the imaginary and real frequency axes, respectively. In the case of BCS theory, $Z(\omega)=1$ and $\Delta(\omega)=\DeltaBCS$ is independent of frequency and thus the superfluid density on the imaginary frequency axis reduces to the following expression~\cite{Abrikosov1959}:
\begin{equation}
\left.\dfrac{n_s(\tau, T)}{n}\right|_{\textrm{BCS}}=\pi T \sum_{n=-\infty}^{\infty} \dfrac{\DeltaBCS^2}
{\omega_n^2+\DeltaBCS^2}\cdot \dfrac{1}{\sqrt{{\omega}_n^2+\DeltaBCS^2}+\dfrac{1}{2\tau}}.
\end{equation}
The Matsubara frequency summation can be converted to a contour integral and, using Eq.~\eqref{eq:EThSupDens3}, the result is~\cite{Abrikosov1959, Berlisnky1993, Dutta2022}:
\begin{align}
\dfrac{n_{s}(\tau,\,T)}{n} &= 2\tau\mathrm{Re}\left[i\int_{0}^{\infty}\dfrac{\DeltaBCS^{2}}{\left(\omega+i\delta\right)^{2}-\DeltaBCS^{2}}\tanh\left(\dfrac{\omega}{2T}\right)\ d\omega\right] - 2\tau\mathrm{Re}\left[\int_{0}^{\infty}\dfrac{\DeltaBCS^{2}}{\sqrt{\omega^{2}-\DeltaBCS^{2}}}\cdot\dfrac{\tanh\left(\dfrac{\omega}{2T}\right)}{\dfrac{1}{2\tau} - i\sqrt{\omega^{2}-\DeltaBCS^{2}}}\ d\omega\right] \notag \\
& = \pi\DeltaBCS\tau\tanh\left(\dfrac{\DeltaBCS}{2T}\right) -\int_{\DeltaBCS}^\infty\dfrac{\DeltaBCS^{2}}{\sqrt{\omega^2-\DeltaBCS^2}}\cdot \dfrac{\tanh\left(\dfrac{\omega}{2T}\right)}{\omega^2-\DeltaBCS^2 + \left(\dfrac{1}{2\tau}\right)^2}\ d\omega.
\label{eq:BCSSupDens1}
\end{align}
Note that, the presence of the $i\delta$ term $(\delta\rightarrow0)$ is not required in the case of Eliashberg theory because in that instance $\Delta$ has a non-zero imaginary component and thus there are no poles on the real frequency axis. For BCS theory, the analytical continuation to real frequencies requires care because $\DeltaBCS$ is a purely real quantity. In terms of the zero-temperature BCS gap $\DeltaOBCS$, the zero-temperature superfluid density in the BCS limit is~\cite{Kogan2013, Tao2017, Dutta2022}:
\begin{align}
\dfrac{n_s(\tau)}{n} = \pi \DeltaOBCS\tau -\int_{\DeltaOBCS}^\infty\dfrac{\DeltaOBCS^{2}}{\sqrt{\omega^2-\DeltaOBCS^2}}\cdot \dfrac{1}{\omega^2-\DeltaOBCS^2+\left(\dfrac{1}{2\tau}\right)^2}\ d\omega. 
\label{eq:BCSSupDens2}
\end{align}
The closed-form expression for the above integral can be easily obtained by making a hyperbolic trigonometric substitution and using the identity in Eq.~(4) of Sec.~2.451 of Ref.~\cite{GRBook}. We omit the explicit evaluation of this integral for the BCS case; for details see Refs.~\cite{Kogan2013, Tao2017, Dutta2022}. In the next section, however, we shall evaluate an analogous integral for the case of Eliashberg theory.

At a general temperature $T$, we can further simplify Eq.~\eqref{eq:BCSSupDens1} by taking the dirty and clean limits:
\begin{align}
\dfrac{n_s(\tau, T)}{n}=
\begin{cases}
\pi \DeltaBCS \tau\tanh \left(\dfrac{\DeltaBCS}{2T}\right),\qquad &\DeltaBCS\tau\rightarrow 0\\ &\\ 
  1+2\bigintss_{\DeltaBCS}^\infty\,\dfrac{\omega}{\sqrt{\omega^2-\DeltaBCS}}f'(\omega, T)\ d\omega,\qquad &\DeltaBCS\tau\rightarrow\infty,
\end{cases} 
\end{align}
where $f(\omega, T)=\left(e^{\omega/T}+1\right)^{-1}$ is the Fermi-Dirac distribution function and $f^{\prime}(\omega, T)$ denotes its frequency derivative. In the dirty limit, the superfluid density becomes increasingly small and scales as $\DeltaBCS\tau$. In the clean limit, the superfluid density goes to unity as $T\rightarrow0$.

\subsection{Superfluid density on the imaginary frequency axis in the \texorpdfstring{$T\rightarrow 0$\ }\ limit}

From Eq.~\eqref{eq:EThSupDens1}, the zero-temperature superfluid density is defined as
\begin{equation}
\dfrac{n_s(\tau, \lambda)}{n}
=\pi \lim_{T\rightarrow 0} T \sum_{n=-\infty}^{\infty}\dfrac{\Delta^{2}(i\omega_{n})}{\omega_{n}^{2}+\Delta^{2}(i\omega_{n})}\cdot\dfrac{1}{Z(i\omega_{n})\sqrt{\omega_{n}^{2}+\Delta^{2}(i\omega_{n})}+\dfrac{1}{2\tau}}.
\end{equation}
Performing the Matsubara frequency summation for the frequency-dependent gap is theoretically challenging, thus, as a simple approximation, we shall set 
$\Delta(i\omega_{n})\approx\underset{T\rightarrow 0}{\lim}\Delta(i\omega_0)\equiv\Delta_{0}$ and $Z(i\omega_{n})\approx\underset{T\rightarrow 0}{\lim}Z(i\omega_0)\equiv Z_{0}$; that is, we approximate the gap to be the zero-temperature limit of the gap with zero Matsubara index. The justification for this approximation shall ultimately be based upon comparison with the exact numerical evaluation of the superfluid density. The summation can be evaluated in the zero-temperature limit by converting the Matsubara frequency sum to an appropriate integral, as discussed in Ref.~\cite{AGDBook}. Therefore, we now obtain
\begin{equation}
\dfrac{n_s(\tau,\lambda)}{n} =\int_0^\infty\dfrac{\Delta_0^{2}}{\omega^{2}+\Delta_0^{2}}\cdot\dfrac{1}{Z_0\sqrt{\omega^{2}+\Delta_0^{2}}+\dfrac{1}{2\tau}}d\omega.\label{ImagSFD}
\end{equation}
As with the frequency-dependent gap, we can split up the integrand in a similar fashion:
\begin{equation}
\dfrac{\Delta_0^{2}}{\omega^{2}+\Delta_0^{2}}\cdot\dfrac{1}{Z_0\sqrt{\omega^{2}+\Delta_0^{2}}+\dfrac{1}{2\tau}} = 2\tau \dfrac{\Delta_0^2}{\omega^2+\Delta_0^2} - \dfrac{\Delta_0^{2}}{Z_0\sqrt{\omega^{2}+\Delta_0^{2}}}\cdot\bigg(\dfrac{1}{\theta_0}\dfrac{1}{\theta_0 + \sqrt{\omega^{2}+\Delta_0^{2}}}\bigg),
\end{equation}
where we have defined $\theta_0\equiv 1/(2\tau Z_0)$. The integral of the first term can be evaluated with the residue theorem, while for the integral of the second term we evaluate it by using a substitution given below:
\begin{align}
\dfrac{n_s(\tau, \lambda)}{n}
&=\int_0^\infty\dfrac{\Delta_0^{2}}{\omega^{2}+\Delta_0^{2}}\cdot\dfrac{1}{Z_0\sqrt{\omega^{2}+\Delta_0^{2}}+\dfrac{1}{2\tau}}d\omega\notag\\
&\notag\\
&=2\tau \int_0^\infty \dfrac{\Delta_0^2}{\omega^2+\Delta_0^2} d\omega -\dfrac{\Delta_0^2}{Z_0\theta_0}\int_0^\infty \dfrac{1}{\sqrt{\omega^2+\Delta_0^2}}\cdot \left(
\dfrac{1}{\theta_0+\sqrt{\omega^2+\Delta_0^2}}
\right)d\omega\notag\\
&\notag\\
&=\pi \Delta_0 \tau +\dfrac{2\Delta_0^2}{Z_0\theta_0\sqrt{\theta_0^2-\Delta_0^2}}\int_0^\infty \dfrac{d}{d\omega}\left(\dfrac{\omega-\sqrt{\omega^2+\Delta_0^2}}{\sqrt{\theta_0^2-\Delta_0^2}}\right)\cdot \left[1-\left(\dfrac{\theta_0-\omega+\sqrt{\omega^2+\Delta_0^2}}{\sqrt{\theta_0^2-\Delta_0^2}}\right)^2\right]^{-1}d\omega\notag\\
&\notag\\
&=\pi \Delta_0 \tau -\dfrac{2\Delta_0^2}{Z_0\theta_0\sqrt{\theta_0^2-\Delta_0^2}}\int_{(\theta_0+\Delta_0)/\left(\sqrt{\theta_0^2-\Delta_0^2}\right)}^{\theta_0/\sqrt{\theta_0^2-\Delta_0^2}}\dfrac{1}{1-u^2}\,du\notag\\
&\notag\\
&=\pi \Delta_0 \tau -\dfrac{2\Delta_0^2}{Z_0\theta_0 \sqrt{\theta_0^2-\Delta_0^2}}\cdot \left[\textrm{arctanh}\left(\dfrac{\theta_0}{\sqrt{\theta_0^2-\Delta_0^2}}\right)-\textrm{arctanh}\left(\dfrac{\theta_0+\Delta_0}{\sqrt{\theta_0^2-\Delta_0^2}}\right) \right]\notag\\
&\notag\\
&=\pi \Delta_0\tau -\dfrac{2\Delta_0^2}{Z_0\theta_0\sqrt{\theta_0^2-\Delta_0^2}}\cdot \textrm{arctanh}\left(\dfrac{\theta_0-\Delta_0}{\sqrt{\theta_0^2-\Delta_0^2}}\right)\notag\\
&\notag\\
&=\dfrac{\pi}{2Z_0\gamma_0}\cdot \left[1+\dfrac{4}{\pi}\dfrac{1}{\sqrt{1-\gamma_0^2}}\cdot \arctan\left(\dfrac{\gamma_0-1}{\sqrt{1-\gamma_0^2}}\right)\right]. 
\label{eq:EThSupDens5}
\end{align}

\noindent In the last step we have defined $\gamma_0\equiv 1/(2\tau \Delta_0Z_0)$.

This expression reproduces Eq.~(3) in the main text. If we set $Z=1$ and $\Delta_0=\DeltaOBCS$, then Eq.~\eqref{eq:EThSupDens5} reproduces the BCS result~\cite{Kogan2013, Tao2017}. However, our result has two differences, both of which have fundamental significance to explaining scaling laws in superconductors. First, $\Delta_{0}$ depends on the electron-phonon coupling strength. In the weak-coupling limit, $\Delta_0$ is modified from the BCS result only by a constant prefactor~\cite{Boyack2023Mar}; however, in the strong-coupling limit, $\Delta_0\sim\sqrt{\lambda}$ and thus the superfluid density will be very different from the BCS counterpart. Second, Eq.~\eqref{eq:EThSupDens5} incorporates the renormalization function, which is a function of electron-phonon coupling $\lambda$. In BCS theory, as $T\rightarrow 0$, $n_s/n$ is unity in the clean limit and thus no Homes scaling occurs~\cite{Kogan2013}. However, in ETh, since the superfluid density depends on $\lambda$ through the renormalization parameter $Z_0$, Homes scaling is possible because both the conductivity and the superfluid density can scale proportionately (with some residual $\lambda$ dependence); this is a non-trivial result that goes beyond BCS theory and has not been previously explored. 

As argued in Refs.~\cite{Leggett1965, Raines2023}, for a Galilean-invariant system the normalized zero-temperature superfluid density is equal to unity. Since an electron-boson theory with an Einstein-phonon propagator breaks Galilean invariance, the presence of $1/Z_{0}$ in the zero-temperature superfluid density is not surprising. Indeed, the electromagnetic vertex corrections which restore the zero-temperature superfluid density to be unity are in fact zero for the Einstein-phonon theory~\cite{Raines2023}. The result $n_s(\lambda)/n\approx1/Z_0$ in the clean limit is in agreement with our numerical calculations using Eq.~\eqref{eq:EThSupDens1}.

\subsection{Superfluid density on the real frequency axis in the constant-gap approximation}

The superfluid density is a thermodynamic (static) quantity, meaning that it is evaluated at zero external Matsubara frequency, as indicated in Eq.~\eqref{eq:EThSupDens1}. It can be expressed in terms of quantities on the imaginary frequency axis, without the need for analytical continuation. We have also presented the real-axis formulation of the superfluid density, as this is occasionally implemented in the context of Eliashberg theory~\cite{Karakozov1991, Combescot1996}. In the previous section, we formulated a simple constant-gap approximation using the imaginary frequency axis gap. For completeness, here we perform a similar analysis for the real-axis formulation. As a simple approximation, suppose that $\Delta(\omega) \approx \Delta \equiv \Delta_1 + i\Delta_2$ and $Z(\omega) \approx Z \equiv Z_1 + iZ_2$, where these quantities are defined by~\cite{Fibich1965,Fibich1965b,Boyack2023Mar}: 
\begin{align}
\Delta_{1}(T)&=\mathrm{Re}\left[\Delta(\omega=\Delta_1,T)\right], \quad 
\Delta_{2}(T)=\mathrm{Im}\left[\Delta(\omega=\Delta_1,T)\right]; \\
Z_{1}(T)&=\mathrm{Re}\left[Z(\omega=\Delta_1,T)\right], \quad Z_{2}(T)=\mathrm{Im}\left[Z(\omega=\Delta_1,T)\right].
\end{align}
There is a large contribution to the integral in Eq.~\eqref{eq:EThSupDens2} in the vicinity of frequencies where $\omega=\Delta_1$, since the denominator is small in this region. Substituting the constant-gap approximation into Eq.~\eqref{eq:EThSupDens3}, we have
\begin{align}
\dfrac{n_{s}(\tau, \lambda)}{n}
&=2\tau\mathrm{Re}\bigg[i\int_{0}^{\infty}\dfrac{\Delta^{2}(\omega)}{\omega^{2}-\Delta^{2}(\omega)}\ d\omega\bigg]-\mathrm{Re}\bigg[\int_{0}^{\infty}\dfrac{\Delta^{2}(\omega)}{Z(\omega)\sqrt{\omega^{2}-\Delta^{2}(\omega)}}\cdot\bigg(\dfrac{1}{\vartheta}\cdot\dfrac{1}{\vartheta - i\sqrt{\omega^{2}-\Delta^{2}(\omega)}}\bigg)\ d\omega\bigg]\notag\\
&\notag\\
&\approx2\tau\mathrm{Re}\bigg[i\int_{0}^{\infty}\dfrac{\Delta^{2}}{\omega^{2}-\Delta^{2}}\ d\omega\bigg]-\mathrm{Re}\bigg[\int_{0}^{\infty}\dfrac{\Delta^{2}}{Z\sqrt{\omega^{2}-\Delta^{2}}}\cdot\bigg(\dfrac{1}{\vartheta}\cdot\dfrac{1}{\vartheta - i\sqrt{\omega^{2}-\Delta^{2}}}\bigg)\ d\omega\bigg] \notag\\
&\notag\\
&=\pi\Delta_{1}\tau-\mathrm{Re}\bigg[\dfrac{1}{\vartheta Z}\int_{0}^{\infty}\dfrac{\Delta^{2}}{\sqrt{\omega^{2}-\Delta^{2}}}\cdot\bigg(\dfrac{1}{\vartheta - i\sqrt{\omega^{2}-\Delta^{2}}}\bigg)\ d\omega\bigg],
\label{eq:EThSupDens4}
\end{align}
where in the above we define $\vartheta \equiv 1/(2\tau Z)$. The first term in Eq.~\eqref{eq:EThSupDens4} is almost identical to Eq.~\eqref{eq:BCSSupDens2}, with $\Delta_0$ replaced by $\Delta_1$. The second term in Eq.~\eqref{eq:EThSupDens4} differs from Eq.~\eqref{eq:BCSSupDens2} in that the complex renormalization function is present, and the gap is complex. The above integral can be evaluated analytically as follows:
\begin{align}
\int_{0}^{\infty}\dfrac{\Delta^{2}}{\sqrt{\omega^{2}-\Delta^{2}}}\cdot\dfrac{1}{\vartheta-i\sqrt{\omega^{2}-\Delta^{2}}}\ d\omega  
 &=-\dfrac{2i\Delta^2}{\sqrt{\vartheta^2-\Delta^2}}\int_{{i(\vartheta+\Delta)}/{\sqrt{\vartheta^2-\Delta^2}}}^{{i\vartheta}/{\sqrt{\vartheta^2-\Delta^2}}} \dfrac{du}{1+u^2} \notag\\
&=\dfrac{2\Delta^{2}}{\sqrt{\vartheta^{2}-
\Delta^{2}}}\cdot\textrm{arctanh}\bigg(\dfrac{\sqrt{\vartheta^{2}-\Delta^{2}}}{\vartheta+\Delta}\bigg).
\end{align}

\noindent The approximate zero-temperature superfluid density in Eliashberg theory, for a constant complex gap, is given by

\begin{align}
\dfrac{n_{s}(\tau, \lambda)}{n} 
&=\mathrm{Re}\bigg[\dfrac{\pi}{2Z\gamma}\bigg\{1+\dfrac{4}{\pi}\dfrac{1}{\sqrt{1-\gamma^{2}}}\textrm{arctan\ensuremath{\bigg(}\ensuremath{\ensuremath{\dfrac{\gamma-1}{\sqrt{1-\gamma^{2}}}}}\bigg)}\bigg\}\bigg],
\label{eq:EThSupDens5Real}
\end{align}

\noindent where we have defined $\gamma\equiv 1/\left(2\tau\Delta Z\right)$. The above has close similarities to  Eq.~\eqref{eq:EThSupDens5}, except with $\gamma_0$ and $Z_0$ (which are both quantities that are defined on the imaginary frequency axis) replaced with $\gamma$ and $Z$ (which are both quantities that are defined on the real frequency axis).

\section{The effect of electron-phonon coupling on the normal-state electrical conductivity}

\subsection{Normal-state electrical conductivity}

The electrical conductivity for general electron-phonon coupling $\lambda$, scattering time $\tau$, and external frequency $\nu$ is given by~\cite{Klein1994, Marsiglio1994, Marsiglio1995, Marsiglio1996}:
\begin{align}
\sigma(\nu,\tau,\lambda,T)&=
\dfrac{\omega_p^2}{8\pi \nu}\cdot \bigg\{
\int_0^\infty \tanh\left(\dfrac{\omega}{2T}\right)\cdot \left[
\dfrac{1-N(\omega)N(\omega+\nu)-M(\omega)M(\omega+\nu)}{-i\epsilon(\omega)-i\epsilon(\omega+\nu)+1/\tau}
\right]d\omega\notag\\
&\quad + \int_0^\infty \tanh\left(\dfrac{\omega+\nu}{2T}\right)\cdot \left[
\dfrac{1-N^*(\omega)N^*(\omega+\nu)-M^*(\omega)M^*(\omega+\nu)}{-i\epsilon^*(\omega)-i\epsilon^*(\omega+\nu)-1/\tau}
\right]d\omega \notag\\
&\quad + \int_0^\infty \left[
\tanh\left(\dfrac{\omega+\nu)}{2T}\right)-\tanh\left(\dfrac{\omega}{2T}\right)
\right]\cdot \left[
\dfrac{1+N^*(\omega)N(\omega+\nu)+M^*(\omega)M(\omega+\nu)}{i\epsilon^*(\omega)-i\epsilon(\omega+\nu)+1/\tau}
\right]d\omega\notag\\
&\quad + \int_{-\nu}^0 \tanh\left(\dfrac{\omega+\nu}{2T}\right)\cdot \bigg[
\dfrac{1-N^*(\omega)N^*(\omega+\nu)-M^*(\omega)M^*(\omega+\nu)}{-i\epsilon^*(\omega)-i\epsilon^*(\omega+\nu)-1/\tau} \notag\\
&\quad + \dfrac{1+N^*(\omega)N(\omega+\nu)+M^*(\omega)M(\omega+\nu)}{i\epsilon^*(\omega)-i\epsilon(\omega+\nu)+1/\tau}\bigg]d\omega \bigg\},
\label{eq:SigmaCond}
\end{align}
where we define~\cite{Marsiglio1995}:
\begin{align}
&\epsilon(\omega)\equiv\sqrt{\widetilde{\omega}^{2}(\omega+i\delta)-\phi^{2}(\omega+i\delta)},\quad N(\omega)\equiv\dfrac{\widetilde{\omega}(\omega+i\delta)}{\epsilon(\omega)},\quad M(\omega)\equiv\dfrac{\phi(\omega+i\delta)}{\epsilon(\omega)}, \\
&\widetilde{\omega}(\omega+i\delta)\equiv\omega Z(\omega+i\delta),\qquad  \phi(\omega+i\delta)\equiv\Delta(\omega+i\delta)Z(\omega+i\delta). 
\end{align}
The plasma frequency $\omega_{p}$ is defined by $\omega_{p}=\sqrt{4\pi ne^{2}/m}$. 
In the above equations, all quantities implicitly depend upon $\lambda$ and $T$. From now on, we will also use the shorthand $\Delta(\omega)\equiv \Delta(\omega+i\delta)$ and $Z(\omega)\equiv Z(\omega+i\delta)$.

As discussed in the main text, Homes and Pimenov scaling are relationships between the superfluid density and the real part of the zero frequency (DC), normal-state ($T=T_c+0^{+}$) electrical conductivity. To take the normal-state limit of Eq.~\eqref{eq:SigmaCond}, the gap is set to zero, and to take the DC limit the frequency is set to zero. The final result is 
\begin{equation}
\sigma_1(\tau,\lambda,T) = -\dfrac{\omega_{p}^{2}}{4\pi}\bigintsss_{0}^{\infty}\dfrac{f'(\omega, T)}{\dfrac{1}{2\widetilde{\tau}(\omega, \lambda, T)}+\dfrac{1}{2\tau}}\ d\omega;\qquad \dfrac{1}{2\widetilde{\tau}(\omega, \lambda, T)}\equiv\omega \textrm{Im}\left[Z(\omega+i\delta)\right] . 
\label{eq:SigmaNormalState}
\end{equation}
As in Sec.~\ref{sec:SupDensBCS}, $f'(\omega, T)$ is the $\omega$-derivative of the Fermi-Dirac distribution function.  To clarify our notation, we will use the conventional notation that $\sigma_1(\tau, \lambda, T)$ denotes the real part of the conductivity and $\sigma_2(\tau, \lambda, T)$ denotes the imaginary part. In the main text and in the discussion on scaling in Sec.~\ref{sec:ScalingLaws}, we use $\sigma(\tau, \lambda, T)$ to denote the real part of the conductivity for simplicity. Similarly, we drop the explicit $T$ dependence for $\sigma_1$ and $\sigma_2$ if $T=T_c$, and for the superfluid density $n_s/n$ if $T=0$. Omitting the $\tau$-dependence in either the conductivity or the superfluid density will imply that the clean limit has been taken. Additionally, omitting the $\lambda$-dependence will imply that the BCS limit is being considered. Omitting any frequency dependence will imply that the DC limit is being considered.

To obtain the complete $\lambda$ dependence of the electrical conductivity, it remains to determine the imaginary part of the renormalization function. For an Einstein model, the normal-state expression for $Z(\omega+i\delta)$ is given by~\cite{Marsiglio1991, Marsiglio1995} 
\begin{align}
Z(\omega+i\delta) &=	1+\frac{i\pi T}{\omega}\sum_{m=-\infty}^{\infty} \lambda\left(\omega-i\omega_{m}\right)\textrm{sgn}\left(\omega_{m}\right) \notag\\
&\quad 	+\frac{i\pi\lambda}{2}\frac{\omega_{E}}{\omega} \left[2N\left(\omega_{E}\right)+f\left(\omega_{E}-\omega\right)+f\left(\omega_{E}+\omega\right)\right] \notag\\
&= 1+\frac{i\pi T}{\omega}\lambda\omega_{E}^{2}\sum_{m=0}^{\infty}\left[\frac{1}{\omega_{E}^{2}-\left(\omega-i\omega_{m}\right)^{2}}-\frac{1}{\omega_{E}^{2}-\left(\omega+i\omega_{m}\right)^{2}}\right] \notag \\ &\quad +\frac{i\pi\lambda}{2}\frac{\omega_{E}}{\omega}
\left[2N\left(\omega_{E}\right)+f\left(\omega_{E}-\omega\right)+f\left(\omega_{E}+\omega\right)\right] \notag\\ 
&= 	1+\frac{1}{2}\lambda\frac{\omega_{E}}{\omega}
\textrm{Re}\left[\psi\left(\frac{1}{2}-i\frac{\omega_{E}+\omega}{2\pi T}\right)-\psi\left(\frac{1}{2}+i\frac{\omega_{E}-\omega}{2\pi T}\right)\right] \notag \\ &\quad +\frac{i\pi\lambda}{2}\frac{\omega_{E}}{\omega}
\left[2N\left(\omega_{E}\right)+f\left(\omega_{E}-\omega\right)+f\left(\omega_{E}+\omega\right)\right].
\end{align}
Here, $\psi$ denotes the digamma function. Taking the imaginary part of this expression and simplifying, we obtain 
\begin{align}
2\omega\mathrm{Im}\left[Z(\omega+i\delta)\right] &= \pi\lambda\omega_{E} \left[2N\left(\omega_{E}\right)+f\left(\omega_{E}-\omega\right)+f\left(\omega_{E}+\omega\right)\right] \notag\\
&= \pi\lambda\omega_{E}\left[\coth\left(\frac{\omega_{E}}{2T}\right)+\frac{1}{2}\tanh\left(\frac{\omega-\omega_{E}}{2T}\right)-\frac{1}{2}\tanh\left(\frac{\omega_{E}+\omega}{2T}\right)\right].
\end{align}
The denominator in the integrand of the conductivity represents an enhanced scattering rate~\cite{Millis1988, Marsiglio1991, Bennemann}, where
\begin{align}
\dfrac{1}{\widetilde{\tau}(\omega, \lambda, T)} = \pi\lambda\omega_{E}\left[\coth\left(\frac{\omega_{E}}{2T}\right)-\frac{1}{2}\tanh\left(\frac{\omega_{E}-\omega}{2T}\right)-\frac{1}{2}\tanh\left(\frac{\omega_{E}+\omega}{2T}\right)\right].
\end{align}
In summary, the DC normal-state conductivity reduces to~\cite{Millis1988}:
\begin{equation}
\sigma_{1}(\tau, \lambda, T) =\frac{\omega_{p}^{2}}{4\pi} \cdot \frac{1}{2\pi T\lambda} \bigintss_{0}^{\infty} \frac{\textrm{sech}^{2}\left(x\dfrac{\omega_{E}}{2T}\right)}{\coth\left(\dfrac{\omega_{E}}{2T}\right) - \dfrac{1}{2} 
\bigg\{\textrm{tanh}\left[\dfrac{\omega_E}{2T} \left(1-x\right)\right]+\textrm{tanh}\left[\dfrac{\omega_E}{2T}\left(1+x\right)\right]\bigg\} 
+ \dfrac{1}{\pi \lambda \tau\omega_E}}dx.
\label{eq:SigmaNormalState2}
\end{equation}
As in Eq.~(3) of the main text, we can write 
$\sigma(\tau, \lambda, T)\equiv (\omega_p^2/(4\pi))\cdot \zeta(\tau, \lambda, T)$ with $\zeta(\tau, \lambda,\,T)$ given by
\begin{align}
\zeta(\tau, \lambda, T)&\equiv\dfrac{1}{2\pi \lambda T}
\bigintss_{0}^{\infty}\,\frac{\textrm{sech}^2\left(
x\dfrac{\omega_E}{2T}\right)}{
\textrm{coth} \left( \dfrac{\omega_E}{2T} \right) -\dfrac{1}{2} 
\bigg\{\textrm{tanh}\left[\dfrac{\omega_E}{2T} \left(1-x\right)\right]+\textrm{tanh}\left[\dfrac{\omega_E}{2T}\left(1+x\right)\right]\bigg\} 
+ \dfrac{1}{\pi \lambda \tau\omega_E}}dx.
\label{eq:ZetaEq}
\end{align}

\begin{figure}[!t]
\hspace{-10mm} \includegraphics[width=0.7\columnwidth]{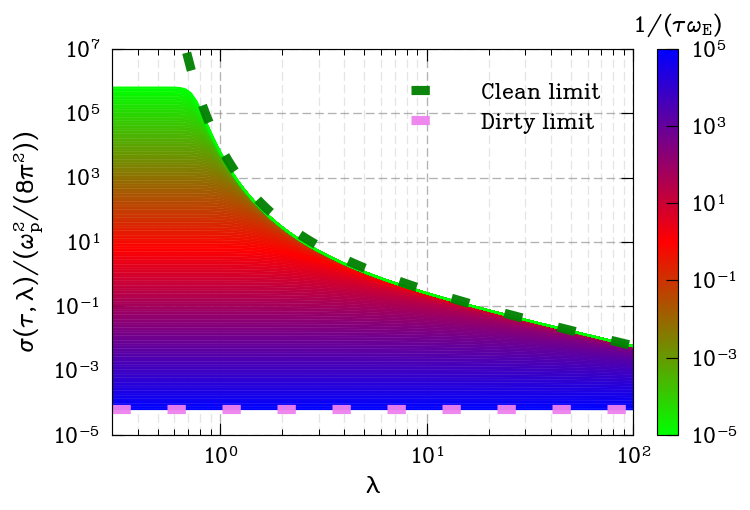}
\caption{Normal-state conductivity in units of $\omega_E$ plotted versus electron-phonon coupling strength $\lambda$ for different values of $1/(\tau \omega_E)$. The estimate $2\pi \tau \omega_E$ (green dashed line) in the dirty limit and the estimate $(\omega_E/\lambda)\sinh(\omega_E/T_c)$ (violet dashed line) in the clean limit are in good agreement with the numerical results.}
\label{SMFig_Cond}
\end{figure}

\noindent As mentioned in the main text, the value of $\sigma_1(\tau, \lambda)/(\omega_p^2/(8\pi^2))$ reduces to $2\pi \tau$ in the dirty limit and $\sim (1/\lambda) \sinh(\omega_E/T_c)$ in the clean limit. In Fig.~\ref{SMFig_Cond}, we plot the normal-state conductivity at $T=T_c$ versus $\lambda$, and find good agreement with the dirty and clean estimates.

\subsection{Superfluid density, Ferrell-Glover-Tinkham sum rule, and the \texorpdfstring{$\mu$}\ SR relaxation rate}

For a system to be a superconductor, the electromagnetic response must satisfy two criteria~\cite{RickayzenBook}: (i) $\underset{\mathbf{q}\rightarrow0}{\lim}K_{ij}(0,\mathbf{q})\neq0$ and (ii) $\underset{\Omega\rightarrow0}{\lim}K_{ij}(\Omega,\mathbf{0})\neq0$. Nam~\cite{Nam1967} showed that, in a general Eliashberg theory of superconductivity, these two limits commute with each other. The first limit corresponds to the presence of a non-zero superfluid density, as indicated in Eq.~\eqref{eq:SupDens}. In addition, Nam~\cite{Nam1967, Nam1967b} showed that the superfluid density is equivalent to the $\mu$SR relaxation rate defined by $\underset{\nu\rightarrow0}{\lim}\nu\sigma_2(\nu)$, where $\nu$ is the (real) frequency of the external vector potential. In this section, we shall use the Ferrell-Glover-Tinkham (FGT) sum rules~\cite{Glover1957, Ferrell1958} to explicitly prove that the superfluid density given in Eq.~\eqref{eq:EThSupDens5} is in agreement with the aforementioned conductivity constraint, for general $\lambda$, $\tau$, and $T$. 

Let $\sigma_{1}(\nu)$ and $\sigma_{2}(\nu)$ denote the real and imaginary parts of the complex conductivity, respectively. Note that here we consider a general superconducting system, and thus only the frequency dependence is written explicitly in the conductivity. The FGT sum rules are given by~\cite{Glover1957, Ferrell1958}:
\begin{align}
\sigma_{1}(\nu) &= \dfrac{1}{\pi}\mathcal{P}\int_{-\infty}^{\infty}\dfrac{\omega\sigma_{2}(\omega)}{\omega^{2}-\nu^{2}}\ d\omega, \\
\sigma_{2}(\nu) &= -\dfrac{\nu}{\pi}\mathcal{P}\int_{-\infty}^{\infty}\dfrac{\sigma_{1}(\omega)}{\omega^{2}-\nu^{2}}\ d\omega.
\end{align}
Here, $\mathcal{P}$ denotes the Cauchy principal value. These relations follow directly from the Kramers-Kronig relations for the electrical conductivity, under appropriate assumptions~\cite{Glover1957, Ferrell1958} such as $\sigma(\nu)\rightarrow0$ as $\nu\rightarrow0$ etc. Using the FGT sum rules, we can simplify the imaginary contribution by multiplying $\sigma_{2}(\nu)$ by $\nu$ and simplifying: 
\begin{align}
\nu\sigma_{2}(\nu)&=-\dfrac{1}{\pi}\mathcal{P}\int_{-\infty}^{\infty}\dfrac{\nu^{2}\sigma_{1}(\omega)}{\omega^{2}-\nu^{2}}\ d\omega	\notag\\
&=\dfrac{1}{\pi}\mathcal{P}\int_{-\infty}^{\infty}\left(1-\dfrac{\omega^{2}}{\omega^{2}-\nu^{2}}\right)\sigma_{1}(\omega)\ d\omega \notag\\
&=\dfrac{1}{\pi}\mathcal{P}\int_{-\infty}^{\infty}\left(1-\dfrac{\omega^{2}}{\omega^{2}-\nu^{2}}\bigg)\cdot\bigg(\dfrac{n_{s}}{n}\pi\Lambda^{-1}\delta(\omega)+\sigma_{1}'(\omega)\right)\ d\omega	\notag\\
&=\dfrac{n_{s}}{n}\Lambda^{-1}+\dfrac{1}{\pi}\int_{-\infty}^{\infty}\sigma_{1}'(\omega)\left(1-\dfrac{\omega^{2}}{\omega^{2}-\nu^{2}}\right)\ d\omega. 
\end{align}
In the above, we have split the real component of the conductivity into two components: a Dirac-delta  function component proportional to both $n_s/n$ and the inverse London parameter, and a non-divergent part denoted by $\sigma_1'(\omega)$. Taking the limit that $\nu\rightarrow0$, the result is $\underset{\nu\rightarrow0}{\lim}\nu\sigma_{2}(\nu)=\Lambda^{-1}\dfrac{n_{s}}{n}$~\cite{Nam1967, Nam1967b}.  

Let us now explicitly verify the sum rule relation $\underset{\nu\rightarrow0}{\lim}\nu\sigma_{2}(\nu)=\Lambda^{-1}\dfrac{n_{s}}{n}$ in the context of Eliashberg theory. 
Consider the electrical conductivity in Eq.~\eqref{eq:SigmaCond}; multiply this expression by $\nu$ and then take the limit that $\nu\rightarrow0$ to obtain
\begin{align}
\underset{\nu\rightarrow0}{\lim}\nu\sigma\left(\nu, \tau, \lambda, T\right) &= \frac{\omega_{p}^{2}}{8\pi}\biggl\{ \int_{0}^{\infty}\tanh\left(\frac{\omega}{2T}\right)\left[\frac{1-N^{2}\left(\omega\right)-M^{2}\left(\omega\right)}{-2i\epsilon\left(\omega\right)+1/\tau}\right]\ d\omega \notag\\ &\qquad \qquad+\int_{0}^{\infty}\tanh\left(\frac{\omega}{2T}\right)\left[\frac{1-\left(N^{*}\left(\omega\right)\right)^{2}-\left(M^{*}\left(\omega\right)\right)^{2}}{-2i\epsilon^{*}\left(\omega\right)-1/\tau}\right]\ d\omega\biggr\} \nonumber \\
&= \frac{i\omega_{p}^{2}}{4\pi}\mathrm{Re}\int_{0}^{\infty}\left[\frac{1-N^{2}\left(\omega\right)-M^{2}\left(\omega\right)}{2\epsilon\left(\omega\right)+i/\tau}\right]\tanh\left(\frac{\omega}{2T}\right)d\omega\nonumber \\
&= \frac{i\omega_{p}^{2}}{4\pi}\mathrm{Re}\int_{0}^{\infty}\frac{1}{2\epsilon\left(\omega\right)+i/\tau}\left[1-\frac{\omega^{2}+\Delta^{2}\left(\omega\right)}{\omega^{2}-\Delta^{2}\left(\omega\right)}\right]\tanh\left(\frac{\omega}{2T}\right) d\omega\nonumber \\
&= -\frac{i\omega_{p}^{2}}{4\pi}\mathrm{Re}\int_{0}^{\infty}\dfrac{\Delta^{2}\left(\omega\right)}{\omega^{2}-\Delta^{2}\left(\omega\right)}\dfrac{\tanh\left(\dfrac{\omega}{2T}\right)}{\epsilon\left(\omega\right)+\dfrac{i}{2\tau}}\ d\omega.
\end{align}

Inserting the expressions for the plasma frequency $\omega_{p}^{2}=4\pi ne^{2}/m$ and the inverse London parameter $\Lambda^{-1}=ne^{2}/m$, and then taking the imaginary part, we find
\begin{align}
\underset{\nu\rightarrow0}{\lim}\nu\sigma_{2}\left(\nu, \tau, \lambda, T\right) &= -\frac{ne^{2}}{m}\mathrm{Re}\left[\int_{0}^{\infty}\dfrac{\Delta^{2}(\omega)}{\omega^{2}-\Delta^{2}(\omega)}\cdot\dfrac{\tanh\left(\dfrac{\omega}{2T}\right)}{Z(\omega)\sqrt{\omega^{2}-\Delta^{2}(\omega)}+\dfrac{i}{2\tau}}\right]d\omega \notag\\
&= \Lambda^{-1}\frac{n_{s}}{n}.
\end{align}
Thus, within the context of Eliashberg theory, we have explicitly verified the constraint relating the imaginary part of the electrical conductivity and the superfluid density, for all $\lambda$, $\tau$, and $T$. This confirms our previous result for the Eliashberg superfluid density.

\section{Scaling relations between the superfluid density and electrical conductivity}
\label{sec:ScalingLaws}

\subsection{Homes, Pimenov, and Holstein scaling in the BCS limit}

\begin{figure}[!t]
\hspace{-10mm} \includegraphics[width=0.7\columnwidth]{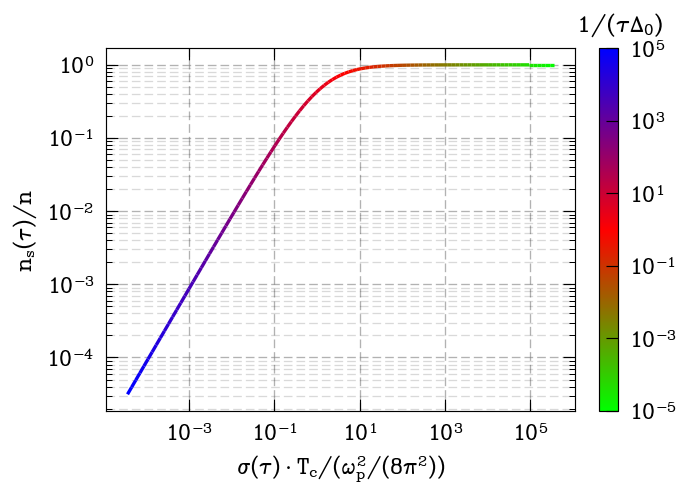}
\caption{Homes diagram for a BCS superconductor, with $\DeltaOBCS/T_c \approx 1.76$. In the dirty limit (i.e., $\tau \DeltaOBCS \ll 1$), Homes scaling is obeyed. In the clean limit (i.e., $\tau \DeltaOBCS \gg 1$), the superfluid density saturates to unity, and Homes scaling breaks down.}
\label{SMFig_BCS_Homes}
\end{figure}

Before considering Homes scaling within the context of Eliashberg theory, we review scaling laws within the context of BCS theory. We define the Homes proportionality factor $\eta_{H}(\tau)$ as the ratio of the superfluid density $n_s(\tau)/n$ to the quantity $\sigma(\tau)T_c/(\omega_p^2/(8\pi^2))$. In the scenario where $\eta_{H}(\tau)$ is a constant upon varying $n_s(\tau)/n$ and varying $\sigma(\tau)T_c$, the proportionality factor will be considered a linear slope. The DC normal-state conductivity in BCS theory reduces to the Drude result~\cite{Zimmerman1991}, thus from Eq.~\eqref{eq:EThSupDens5} we can write down the BCS Homes factor for general $\tau$ as follows:
\begin{equation}
\eta_H(\tau)\equiv \dfrac{n_s(\tau)}{n}\cdot\left[\dfrac{\sigma(\tau)T_c}{\omega_p^2/\left(8\pi^2\right)}\right]^{-1} = \dfrac{\DeltaOBCS}{2T_c}\cdot \left\{1+\dfrac{4}{\pi}\dfrac{1}{\sqrt{1-\gamma_{\textrm{BCS}}^2}}\arctan
\left(\dfrac{\gamma_{\textrm{BCS}}-1}{\sqrt{1-\gamma_{\textrm{BCS}}^2}}\right) \right\},
\end{equation}
where we define $\gamma_{\mathrm{BCS}}\equiv 1/(2\tau \DeltaOBCS)$. A plot of the normalized superfluid density versus $\sigma(\tau)T_c$ is given in Fig.~\ref{SMFig_BCS_Homes}. Note that, to generate this diagram, we continuously tune the parameter $\tau$, and thereby change $n_s(\tau)/n$ and $\sigma(\tau)$ to generate the Homes plot. In this way, the Homes scaling plot may be thought of as a parametric plot of a curve with $x$ and $y$ coordinates given by $\sigma(\tau)T_c$ and $n_s(\tau)/n$, respectively, which are functions of $\tau$. In the clean limit (i.e., $\DeltaOBCS\tau\rightarrow \infty$), the Homes factor reduces to $1/(2\pi T_c\tau)$, and hence approaches zero in the clean BCS limit. Intuitively, this result arises because $n_s(\tau)/n\rightarrow 1$ as $\DeltaOBCS\tau\rightarrow \infty$. In the dirty limit (i.e., $\DeltaOBCS\tau\rightarrow 0$), the Homes factor reduces to a constant slope of $\eta_H\equiv\underset{\DeltaOBCS\tau\rightarrow 0}{\lim}\eta(\tau)=\DeltaOBCS/(2T_c)$. Note that there are often subtleties in regard to taking $\lambda\rightarrow 0$ and identifying this limit with BCS theory, as the gap and critical temperature for these two scenarios differ by a factor of $\sqrt{e}$~\cite{Marsiglio2018Jul,PhysRevB.101.064506}. However, as the dirty Homes slope in the above scales as the ratio $\DeltaOBCS/(2T_c)$, this discrepancy does not come into play when considering this regime of Homes scaling.

In BCS theory, Homes scaling is present only in the dirty limit, while linear Homes scaling breaks down in the clean limit due to the proportionality constant plateauing at unity. Within the context of BCS theory, Homes scaling in the dirty limit directly follows from the early works of de Gennes~\cite{DeGennes1966} and Nam~\cite{Nam1967}. In de Gennes' book, he explains that the connection between the superfluid density and the electrical conductivity is because they are gauge-invariant quantities that arise from matrix elements of the same current operator, and in the local limit they are proportional to each other. 
In this limit, Nam showed that $\underset{T\rightarrow0}{\lim}\ n_s(\tau,T)/n\equiv n_{s}(\tau)/n=\pi \tau \DeltaOBCS$. Combining this formula with the Drude result for the conductivity~\cite{Zimmerman1991} leads to the BCS Homes scaling previously defined.

Let us now consider BCS theory in a slightly more general context, where we allow the interaction strength to be arbitrary. The variability of the Homes slope in the BCS limit is determined by the variability of $\DeltaOBCS/T_c$. We may solve for the gap ratio by recalling the form of the BCS gap equation for an arbitrary temperature $T$~\cite{AGDBook}:
\begin{equation}
\dfrac{1}{g}=T\sum_{n=-\infty}^{\infty}\bigintsss\dfrac{1}{\omega_n^2+\xi_k^2+\DeltaBCS^2}\dfrac{d^3k}{(2\pi)^3}
 = \dfrac{mk_F}{2\pi^2}\bigintsss_{0}^{\omega_{E}} \dfrac{\tanh\left(\dfrac{1}{2T}\sqrt{\xi^2+\DeltaBCS^2}\right)}{\sqrt{\xi^2+\DeltaBCS^2}} d\xi.
\label{eq:BCSGap}
\end{equation}
Here, $g$ is the attractive electron-electron potential, $\DeltaBCS$ is the temperature-dependent BCS gap, $\xi$ is the single-particle dispersion relative to the Fermi energy, $m$ is the mass of the electron, and $k_F$ is the Fermi wavevector. The cutoff is taken to be the Einstein frequency $\omega_E$~\cite{AGDBook}.

To determine the Homes slope in the BCS limit, we must find the ratio $\DeltaOBCS/T_c$ for an arbitrary coupling strength. This will provide us with the full range of possible Homes slopes permitted within BCS theory in the dirty limit. Let $T=T_c$ and define $\lambda\equiv gN(\epsilon_F)$, where $N(\epsilon_F)$ is the density of states at the Fermi energy $\epsilon_F$. Linearizing Eq.~\eqref{eq:BCSGap} leads to the following equation for $T_c$~\cite{AGDBook}:
\begin{equation}
\dfrac{1}{\lambda}=\int_0^{1}\dfrac{1}{x}\tanh\left(x \frac{\omega_{E}}{2T_c}\right) dx.
\end{equation}
In realistic materials described by BCS theory, $\lambda$ is typically small; i.e., $\lambda\lessapprox 0.3$~\cite{DeGennes1966}. However, we will proceed assuming that some "asymptotically strong BCS theory" remains applicable for $\lambda\gg 1$. In the limit that $\lambda\rightarrow\infty$, the tanh function can be expanded for small arguments, and after evaluating the above integral the result is $T_c=\lambda \omega_{E}/2$~\cite{Bennemann}. This is the form of the critical temperature in the strong-coupling limit of BCS theory. 

\begin{figure*}
\hspace{-5mm}\subfloat[\label{BCS_Pimenov}]{%
\includegraphics[width=.48\linewidth]{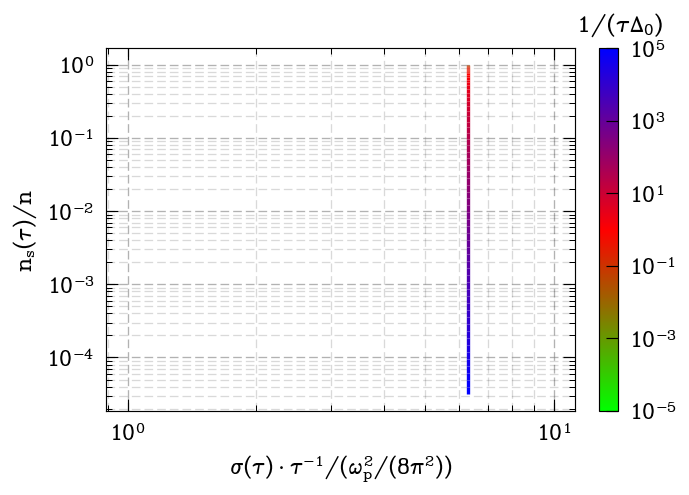}}\hspace{-1mm}
\subfloat[\label{BCS_Holstein}]{\includegraphics[width=.48\linewidth]{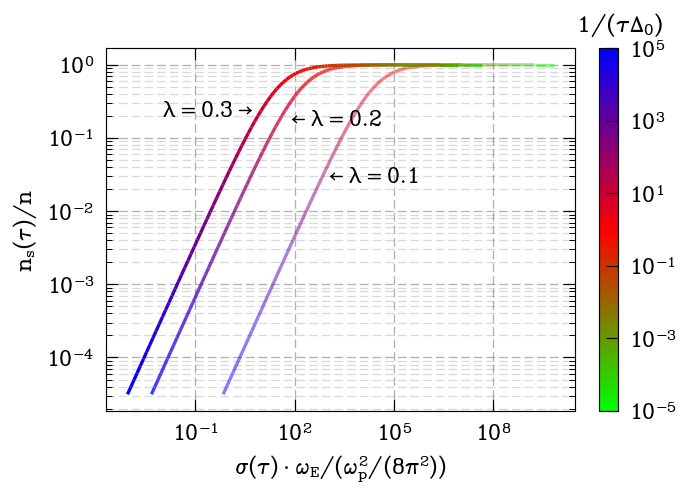}} \\ 
\subfloat[\label{BCS_Slopes}]{\includegraphics[width=.48\linewidth]{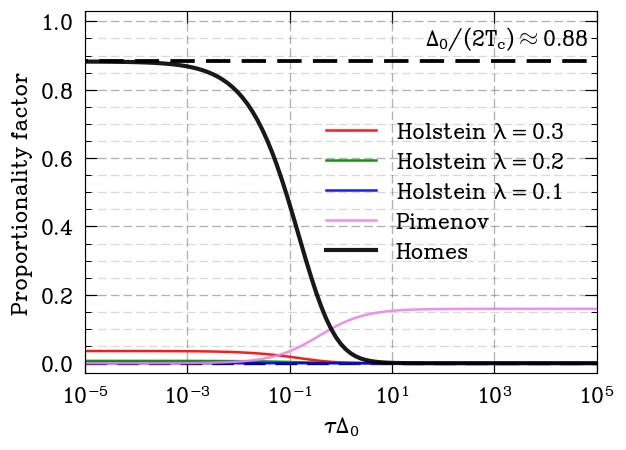}}%
\caption{(a) Pimenov diagram in the BCS limit. As $\sigma(\tau)\cdot \tau^{-1}$ is a constant, the BCS Pimenov plot traces out a vertical line that terminates at unity as we tune $\tau\DeltaOBCS$. (b) Holstein plot in the BCS limit for three different values of $\lambda\equiv g N(\epsilon_F)$. As $\lambda$ becomes smaller, the Holstein slope in the dirty limit becomes smaller. (c) The Homes, Pimenov, and Holstein proportionality factors versus $\tau \DeltaOBCS$. The Homes and Holstein factors are found by taking the numerical gradient of the instantaneous Homes and Holstein slopes, while the Pimenov factor is found by dividing the superfluid density by $\sigma(\tau)\cdot \tau^{-1}$. For small $\tau \DeltaOBCS$, the value of the Homes factor approaches the theoretical value of $\sim 0.88$. The Holstein factor remains small for physically-relevant values of $\lambda$, regardless of whether or not the system is in the dirty or clean limits. The Pimenov factor remains small in the dirty limit, and then increases to plateau at $1/(2\pi)$ in the clean limit.}
\label{fig:BCS_slopes}
\end{figure*}

To find the value of $\DeltaOBCS$, we return to the gap equation~\eqref{eq:BCSGap} and take the zero temperature limit:

\begin{equation}
\dfrac{1}{\lambda}=\int_0^{\omega_{E}} \dfrac{1}{\sqrt{\xi^2+\DeltaOBCS^2}} d\xi = -\log\left(\frac{\DeltaOBCS}{\omega_{E}+\sqrt{\omega_{E}^{2}+\DeltaOBCS^{2}}}\right). 
\end{equation}
As such, we find that $\DeltaOBCS=2\omega_{E} \exp(-1/\lambda)/[1-\exp(-2/\lambda)]$. As $\lambda\rightarrow \infty$, this reduces to $\DeltaOBCS=\omega_{E}\lambda$, which gives the $T=0$ BCS gap in the asymptotically strong limit. The BCS gap ratio then goes as $\DeltaOBCS/T_c=2$ in an asymptotically strong system, which results in a Homes proportionality of $\eta_H=1$ in the dirty limit. This result shows that, in BCS theory, the Homes slope as we have defined it is bounded above by unity.

Finally, we turn to the concept of Pimenov~\cite{Pimenov1999Oct} and Holstein scaling in the BCS limit. We define the Pimenov and Holstein factors as $\eta_P(\tau)$ and $\eta_{\textrm{Hol}}(\tau)$, respectively, where

\begin{subequations}
\begin{equation}
\eta_P(\tau)\equiv \dfrac{n_s(\tau)}{n}\cdot\left[\dfrac{\sigma(\tau)\tau^{-1}}{\omega_p^2/\left(8\pi^2\right)}\right]^{-1}=\eta_H(\tau)\cdot T_c\tau,
\label{eq:Pimenov}
\end{equation}
\begin{equation}
\eta_{\textrm{Hol}}(\tau)\equiv \dfrac{n_s(\tau)}{n}\cdot\left[\dfrac{\sigma(\tau)\omega_E}{\omega_p^2/\left(8\pi^2\right)}\right]^{-1}=\eta_{H}(\tau)\cdot \dfrac{T_c}{\omega_E}.
\label{eq:Holstein}
\end{equation}
\end{subequations}

\noindent As with the Homes diagram, the Pimenov and Holstein diagrams may be generated by considering a parametric plot of $n_s(\tau)/n$ versus $\sigma(\tau)\tau^{-1}$ and $n_s(\tau)/n$ versus $\sigma(\tau)\omega_E$ (respectively) via continuously tuning $\tau$ (see Fig.~\ref{fig:BCS_slopes}). As previously stated, $\sigma(\tau)=ne^2\tau/m$ in the normal-state and thus there is no notion of a Pimenov slope or linear Pimenov scaling in the BCS limit, as a plot of $n_s(\tau)/n$ versus $\sigma(\tau)\tau^{-1}=\omega_p^2/4\pi$ results in a vertical line running from $n_s(\tau)/n=0$ to $n_s(\tau)/n=1$ as $\DeltaOBCS\tau$ increases. The Pimenov factor may still be defined as the above ratio, in which case $\underset{ \DeltaOBCS\tau\rightarrow \infty}{\lim}\eta_P(\tau)= 1/(2\pi)$ in the clean limit whereas $\underset{\DeltaOBCS\tau\rightarrow 0}{\lim}\eta_{P}= \DeltaOBCS\tau/2\ll1$ in the dirty limit. For the Holstein factor, $\underset{\DeltaOBCS\tau\rightarrow \infty}{\lim}\eta_{\textrm{Hol}}(\tau)=1/(2\pi \tau \omega_E)\ll1$ in the clean limit whereas $\underset{\DeltaOBCS\tau\rightarrow 0}{\lim}\eta_{\textrm{Hol}}(\tau)\sim \DeltaOBCS/(2\omega_E)$ in the dirty limit. As we find in Fig.~\ref{fig:BCS_slopes}, the Holstein slope remains very small in the dirty BCS limit for physically relevant values of $\lambda$.

\subsection{Homes, Pimenov, and Holstein scaling at arbitrary \texorpdfstring{$\tau$\ }\ and \texorpdfstring{$\lambda$}\ }
\label{sec:EThScaling}

We now consider scaling relations in Eliashberg theory. Beyond BCS theory, Homes scaling is typically considered in high-$T_c$ superconductors such as the pnictides~\cite{Dordevic2022Jun} and, most notably, the cuprates~\cite{Homes2004Jul}. In our study, we emphasize that we do not aim to explain Homes scaling in these high-$T_c$ materials. Instead, our main endeavor is to understand Homes scaling itself in a strongly-correlated electron-phonon system where scaling behavior emerges via a mechanism distinct from impurity scattering. Note that the electron-phonon superconductors we consider in our work differ from the cuprates for three important reasons. First, there is strong evidence that the force-mediating bosons in the cuprates are not phonons, but are instead antiferromagnetic spin fluctuations~\cite{Hwang2021Jun}. Second, note that Coulomb repulsion is typically strong in cuprate superconductors, with the magnitude of Coulomb interactions comparable to the conduction band width~\cite{Pickett1989Apr}. In our work, we ignore the effects of Coulomb interactions in the Eliashberg equations given in Eqs.~\ref{eq:EthEqnsIm1} and \ref{eq:EthEqnsIm2}. Finally, it is well established now that the cuprate superconductors exhibit a $d$-wave order parameter~\cite{RevModPhys.72.969}, and as a consequence both elastic and inelastic scattering events may result in strong pair breaking~\cite{PhysRev.131.563,Millis1988}. In our work, we assume that the gap is isotropic (i.e., $s$-wave), and that there are no magnetic impurities~\cite{Kogan2013b}. As a consequence, Anderson's theorem~\cite{Anderson1959} holds, and electron-impurity scattering does not result in pair breaking. 

\begin{figure*}
\hspace{0mm}\subfloat[\label{SM_Homes_1}]{%
\includegraphics[width=.45\linewidth]{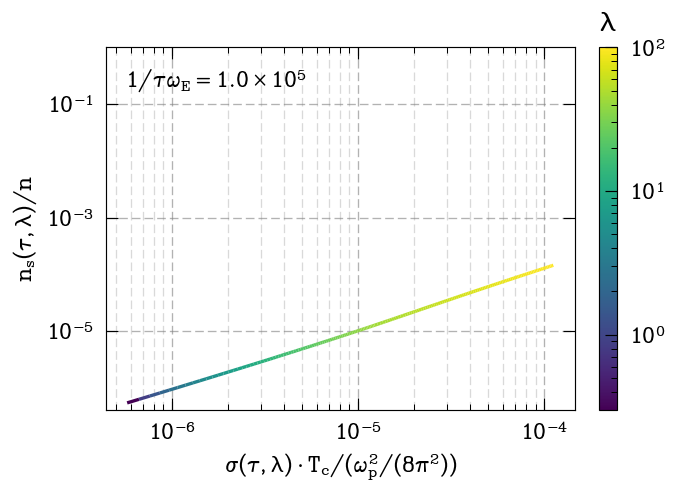}}\hspace{-1mm}
\subfloat[\label{SM_Homes_2}]{\includegraphics[width=.45\linewidth]{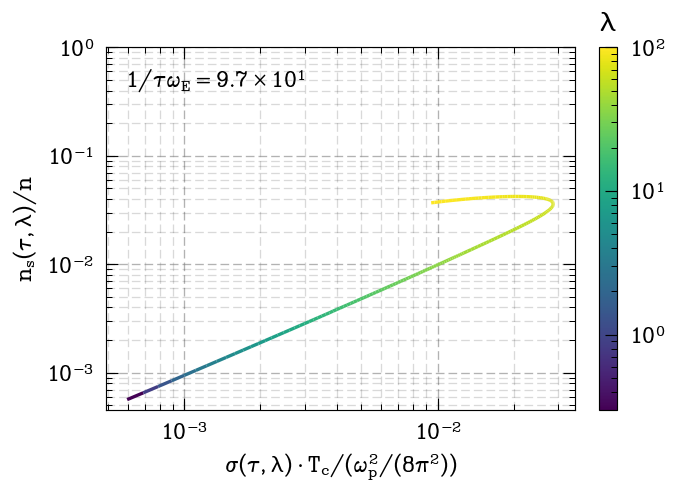}}\\
\hspace{0mm}\subfloat[\label{SM_Homes_3}]{\includegraphics[width=.45\linewidth]{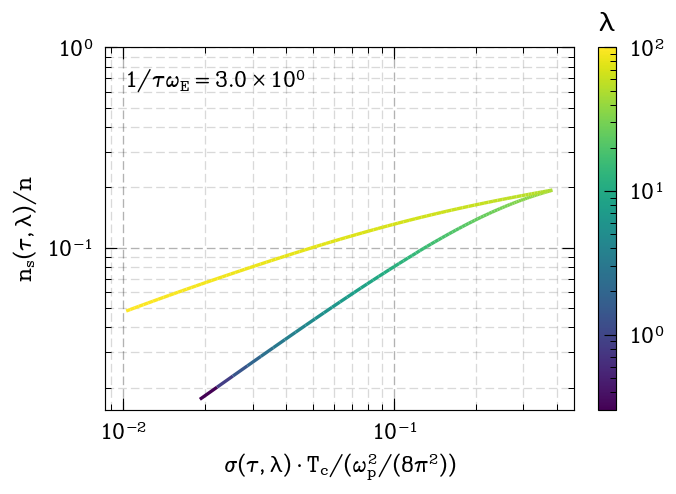}}\hspace{-4.5mm}
\subfloat[\label{SM_Homes_4}]{\vspace{1mm} \hspace{2mm} \includegraphics[width=.45\linewidth]{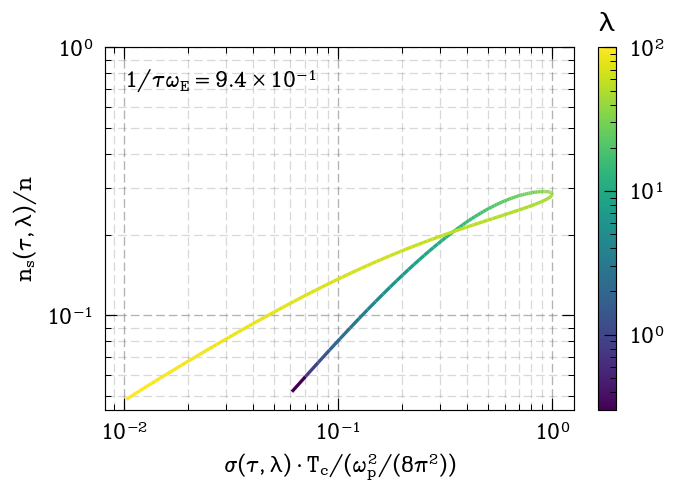}}\\
\hspace{0mm}\subfloat[\label{SM_Homes_5}]{\includegraphics[width=.45\linewidth]{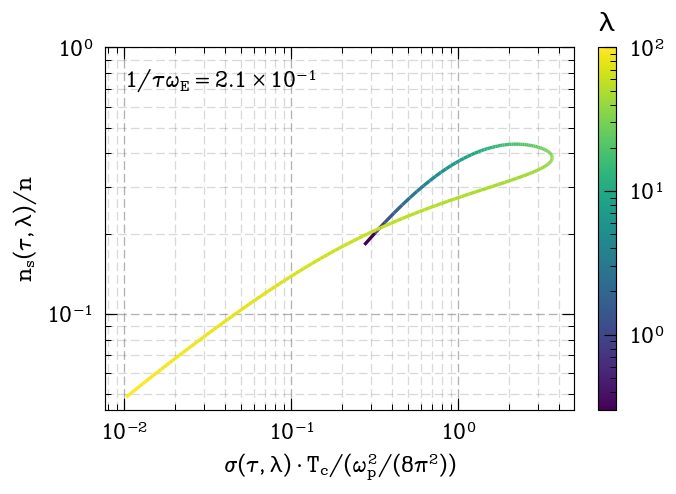}}\hspace{-4.5mm}
\subfloat[\label{SM_Homes_6}]{\vspace{1mm} \hspace{2mm} \includegraphics[width=.45\linewidth]{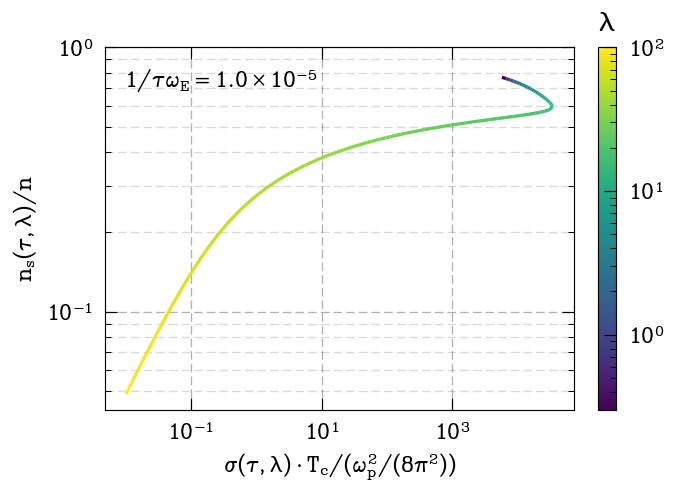}}
\caption{Examples of the Homes plot with different values of the scattering rate. (a) For a large value of $1/(\tau \omega_E)$, linear Homes scaling remains applicable for all $\lambda$ considered. (b) Upon decreasing $1/(\tau \omega_E)$, linear Homes scaling begins to break down at some large coupling strength. Homes scaling remains at low-coupling. (c-d) Further decreasing $1/(\tau \omega_E)$ results in linear Homes scaling, which is only applicable at weak and strong coupling due to the "back-bending" phenomenon. (e-f) For very small values of $1/(\tau \omega_E)$, linear Homes scaling is only present in the strong-coupling regime. In the weak-coupling regime, the Homes proportionality factor approaches a constant.}
\label{HomesPlots}
\end{figure*}

Within the context of Eliashberg theory, the DC normal-state conductivity is given by $\sigma(\tau, \lambda, T)\equiv (\omega_p^2/(4\pi))\cdot \zeta(\tau, \lambda, T)$ with $\zeta(\tau, \lambda, T)$ defined in Eq.~\eqref{eq:ZetaEq}. To calculate the zero-temperature superfluid density $n_s(\tau, \lambda)/n$, the first step is to determine the gap function $\Delta(i\omega_n)$ and the renormalization function $Z(i\omega_n)$ on the imaginary frequency axis. Once these functions are obtained, they can be inserted into Eq.~\eqref{eq:EThSupDens1} to obtain the superfluid density. Examples of the Homes plot of $n_s(\tau, \lambda)/n$ versus $\sigma(\tau, \lambda) T_c$ for various values of $1/(\tau \omega_E)$ are shown in Fig.~\ref{HomesPlots}. These figures were obtained by varying the electron-phonon coupling strength $\lambda$ for fixed $1/(\tau\omega_E)$. In Eliashberg theory, the electron-phonon coupling provides a second parameter that can be tuned (besides the scattering rate) by which we may numerically produce such Homes plots. In this way, the Homes slope is found by plotting a parametric curve with $x$ and $y$ coordinates given by $\sigma(\tau, \lambda)T_c$ and $n_s(\tau, \lambda)/n$ as a function of changing $\lambda$ for fixed $\tau$.

In Fig.~\ref{HomesPlots}, we observe that, as $1/(\tau \omega_E)$ increases, there is a "back-bending" of the Homes proportionality as a function of $\lambda$. That is, as we tune $\lambda$ from weak to strong coupling for intermediate values of $1/(\tau \omega_E)$, the Homes proportionality factor is an increasing function of $\lambda$ before "bending back" at some intermediate coupling strength and returning to the origin.

To better understand the back-bending phenomena and the linear Homes slope itself, let us recall the form of the normal-state conductivity at $T=T_c$. Using Eq.~\eqref{eq:SigmaNormalState2}, this expression may be written as $\sigma(\tau, \lambda)=(\omega_p^2/(4\pi))\cdot \zeta(\tau, \lambda)$, where we define $\zeta(\tau, \lambda) \equiv [{1}/{(2\pi \lambda T_c)}]\cdot I(\tau,\lambda)$. The integral $I(\tau, \lambda)$ may be evaluated by assuming that the dominant weight of the integrand is at $x=0$:

\begin{align}
I(\tau,\lambda)&\equiv \bigintsss_{0}^{\infty}\,\frac{\textrm{sech}^2
\left(x\dfrac{\omega_E}{2T} \right)}{\textrm{coth} \left( \dfrac{\omega_E}{2T} \right)
-\dfrac{1}{2} \left\{\textrm{tanh}\left[\dfrac{\omega_E}{2T} \left(1-x\right)\right]+\textrm{tanh}\left[\dfrac{\omega_E}{2T}\left(1+x\right)\right]\right\} + \dfrac{1}{\pi \lambda \tau\omega_E}} dx\notag\\
&\notag\\
&\approx\frac{T_c/\omega_E}{\textrm{csch} \left( \dfrac{\omega_E}{T_c} \right) + 
\dfrac{1}{2\pi \lambda \tau\omega_E}}. 
\label{eq:IInt}
\end{align}
Using this result, the product of the normalized electrical conductivity and the transition temperature then becomes 
\begin{align}
\dfrac{\sigma(\tau,\,\lambda)T_c}{\omega_p^2/(8\pi^2)}
&\approx\dfrac{T_c}{\omega_E\lambda }\cdot\frac{1}{
\textrm{csch} \left( \dfrac{\omega_E}{T_c} \right)
+\dfrac{1}{2\pi \lambda \tau\omega_E}} \notag\\ 
&=\begin{cases}
2\pi T_c\tau,\qquad & \tau \omega_E\rightarrow 0 \\    &\\
\dfrac{T_c}{\omega_E \lambda}\sinh\bigg(\dfrac{\omega_E}{T_c}\bigg),\qquad & \tau \omega_E\rightarrow \infty.
\end{cases}
\end{align}
Note that in the limits above, we take the limit of $\tau\omega_E\rightarrow 0$ and $\tau \omega_E\rightarrow \infty$ at fixed $\lambda$. In the analysis below, whenever we consider similar limits for the normal-state conductivity, we will also assume that we hold $\lambda$ fixed.

The superfluid density on the imaginary frequency axis is given in Eq.~\eqref{eq:EThSupDens5}, which simplifies as follows based upon the limiting values of $\gamma_0\equiv 1/(2\tau \Delta_0Z_0)$:

\begin{align}
\dfrac{n_s(\tau,\,\lambda)}{n}=\dfrac{\pi}{2Z_0\gamma_0}\cdot \left[1+\dfrac{4}{\pi}\dfrac{1}{\sqrt{1-\gamma_0^2}}\cdot \arctan\left(\dfrac{\gamma_0-1}{\sqrt{1-\gamma_0^2}}\right)\right]\approx\begin{cases}
\pi \tau\Delta_0 ,\qquad & \tau \Delta_0 Z_0\rightarrow 0 \\
&\\
\dfrac{1}{Z_0},\qquad &\tau \Delta_0 Z_0\rightarrow \infty. 
\end{cases}
\end{align}
where $\Delta_0$ is defined in the previous section: $\Delta_0\equiv\underset{T\rightarrow 0}{\lim}\Delta(i\omega_{0})$. As discussed in the main text, in the clean limit the superfluid density does not approach unity, but instead is approximately given by the inverse of the real part of the renormalization function.

\begin{figure*}
\hspace{-5mm}\subfloat[\label{Gap_sqrt_lamb}]{%
\includegraphics[width=.48\linewidth]{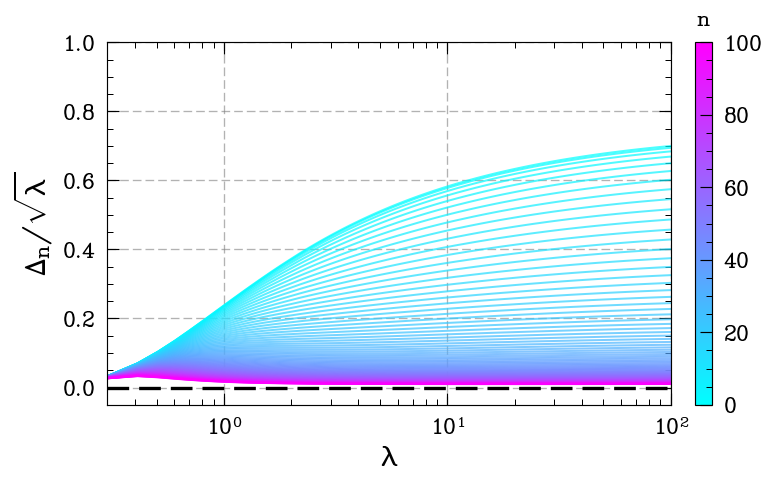}}\hspace{-1mm}
\subfloat[\label{Z_sqrt_lamb}]{\includegraphics[width=.48\linewidth]{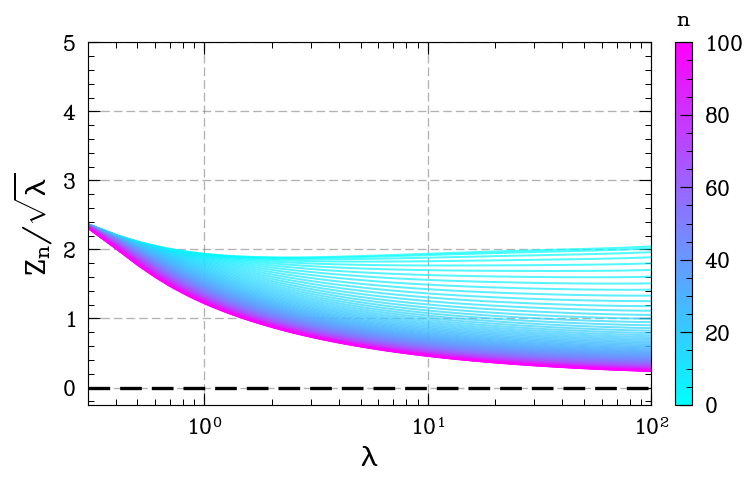}} \\ 
\subfloat[\label{GapZProd_sqrt_lamb}]{\includegraphics[width=.48\linewidth]{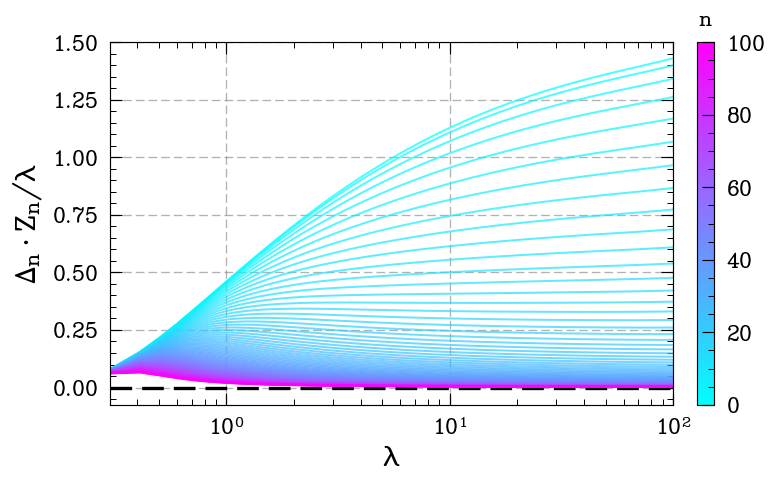}}%
\caption{ Plot of (a) the gap function and (b) the renormalization function over $\sqrt{\lambda}$ versus $\lambda$ in the zero-temperature limit for different Matsubara frequencies $\omega_{n}=(2n+1)\pi T$. Both quantities at lower Matsubara frequencies plateau to a constant for $\lambda\gg 1$. (c) The product of the gap function and the renormalization function, renormalized by $\lambda$. As $\lambda$ grows, the value of the product for lower Matsubara terms approaches a constant near $\pi/2$. Note that in the above figure (as well as in Figs.~\ref{fig:ASEth_quantities} and \ref{Fig5}), the gap function and the Matsubara frequencies are normalized by the Einstein frequency $\omega_E$.}
\label{fig:Sqrt_lamb}
\end{figure*}

Before discussing the Homes slope, note that the above analysis gives a physical explanation of the back-bending phenomenon. First, consider the limit $\tau \omega_E\rightarrow 0$. In this case, $\sigma(\tau, \lambda)T_c\sim T_c\tau$ and $n_s(\tau, \lambda)/n\sim \pi \tau \Delta_0$, and thus the Homes proportionality factor is an increasing function of $\lambda$, and originates from the origin. This is observed in Fig.~\ref{HomesPlots}(a). In the limit that $\tau\omega_E\rightarrow \infty$, however, note that $n_s(\tau, \lambda)/n\sim 1/Z_0$. As $Z_0$ increases with increasing $\lambda$, the superfluid density continuously approaches zero as we increase the electron-phonon coupling strength. Likewise, for large values of $\lambda$, $T_c/\omega_E\sim \sqrt{\lambda}$, and thus $\sigma T_c\sim (1/\sqrt{\lambda})\cdot \sinh(1/\sqrt{\lambda})$, which approaches zero as $\lambda\rightarrow \infty$. As a consequence, we see that, as $\tau \omega_E\rightarrow \infty$ for fixed $\lambda$, the Homes proportionality factor is a decreasing function of $\lambda$ which approaches the origin as $\lambda\rightarrow \infty$, as seen in Fig.~\ref{HomesPlots}(c) and Fig.~\ref{HomesPlots}(d). From our numerical data, we see that intermediate values of the scattering rate results in the back-bending phenomenon, as the Homes proportionality factor approaches the origin as $\tau \omega_E\rightarrow \infty $ and $\lambda\gg 1$. Thus, our analytical result for the superfluid density agrees with our numerical finding of the so-called back-bending. 

Now, we are in a position to discuss the Homes proportionality factor $\eta_H(\tau, \lambda)$ for some finite electron-phonon coupling $\lambda$. From the above, we find the Homes proportionality factor to simplify to the following:

\begin{align}
\eta_H(\tau, \lambda)\equiv \dfrac{n_s(\tau, \lambda)}{n}\cdot\left[\dfrac{\sigma(\tau, \lambda)T_c}{\omega_p^2/\left(8\pi^2\right)}\right]^{-1} =
\begin{cases}
\dfrac{\Delta_0}{2T_c},\qquad &\tau \Delta_0 Z_0\rightarrow  0\quad \&\quad \tau \omega_E\rightarrow 0 \\    &\\
\dfrac{\lambda}{Z_0}\cdot \dfrac{\omega_E}{T_c}\textrm{csch}\left(\dfrac{\omega_E}{T_c}\right),\qquad &\tau \Delta_0 Z_0\rightarrow \infty \quad \& \quad \tau \omega_E\rightarrow \infty.
\end{cases}
\label{Eth_Homes}
\end{align}

\noindent In the first limit in Eq.~\eqref{Eth_Homes}, the Homes proportionality factor is nearly identical to the dirty BCS result, with $\DeltaOBCS$ being replaced by $\Delta_0$. The proportionality factor in the second limit, however, is more complicated than the BCS result due to the fact that the $\lambda$ dependence in $\sigma(\tau, \lambda)T_c$ dominates over the $\tau$-dependence. We can simplify $\eta_H(\tau, \lambda)$ by once again noting that $T_c/\omega_E\sim \sqrt{\lambda}$ in the strong-coupling limit, in  which case the Homes proportionality factor for $\tau \Delta_0 Z_0\rightarrow \infty$ and $\tau \omega_E\rightarrow \infty$ in the limit of large $\lambda$ reduces to
\begin{equation}
\eta_H(\tau, \lambda)\approx \dfrac{\lambda}{Z_0}.
\end{equation}

To simplify further, we will now study the explicit $\lambda$ dependence of $Z_0$ in the weak and strong-coupling limits. In the weak-coupling limit, it is known that $Z_0\sim 1+\lambda$~\cite{Marsiglio2018Jul}. This results in a "flattening" of the Homes slope already seen in the clean BCS system. In the strongly interacting limit, the renormalization function instead goes as $Z_0\sim \sqrt{\lambda}$. This can be seen by first noting that, from the form of the Eliashberg equations on the imaginary axis, the product of $\Delta(i\omega_m)Z(i\omega_m)$ scales as $\lambda$ in the limit of $T\rightarrow 0$. Thus, $Z_0\Delta_0\sim \lambda$. As mentioned in Ref.~\cite{Kresin1987Aug}, the behavior of the gap edge for $\lambda\gg 1$ scales as $\Delta_0\sim \sqrt{\lambda}$, in the same manner as $T_c$. As such, it stands to reason that $Z_0\sim \sqrt{\lambda}$ as well for large electron-phonon coupling. Similar scaling of the renormalization function is noted in Ref.~\cite{Mayrhofer2024Mar} for a $T=0$ 2D Fermi liquid close to a $q=0$ charge quantum critical point. In the context of our work, we can check such scaling of the renormalization function numerically by plotting the gap and renormalization functions versus $\lambda$. As found in Fig.~\ref{fig:Sqrt_lamb}, both $\Delta(i\omega_0)/(\sqrt{\lambda}\omega_E)$ and $Z(i\omega_0)/\sqrt{\lambda}$ flatten to constant values in the strong-coupling limit. 
\begin{figure*}
\hspace{-5mm}\subfloat[\label{ETh_Pimenov}]{%
\includegraphics[width=.48\linewidth]{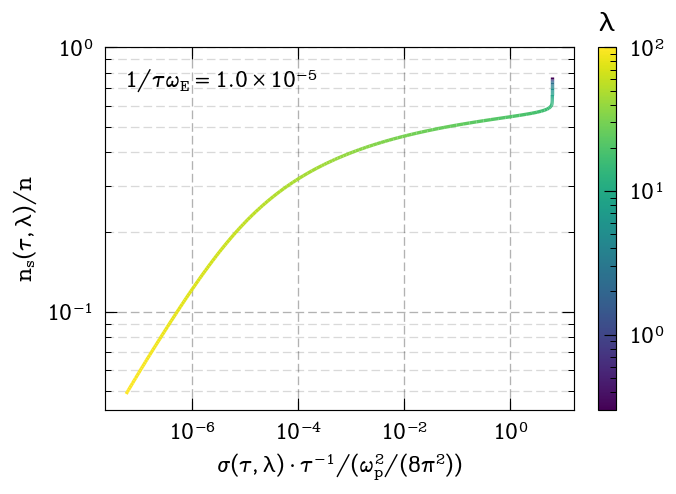}}\hspace{-1mm}
\subfloat[\label{ETh_Holstein}]{\includegraphics[width=.48\linewidth]{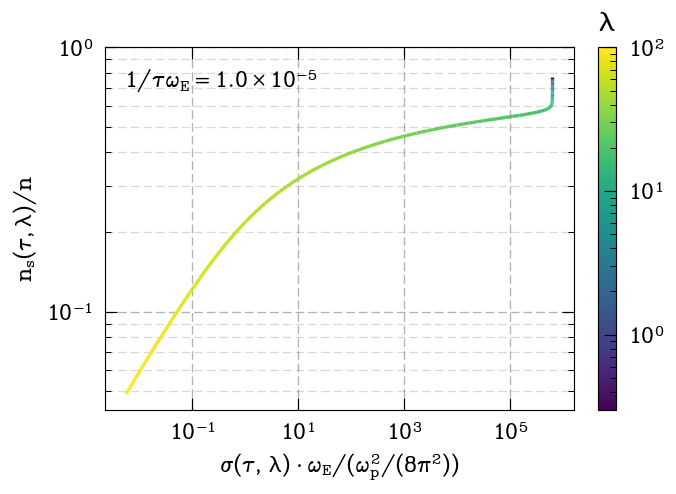}}%
\caption{Pimenov and Holstein plots in the clean limit. As we tune $\lambda$ from weak to strong coupling, linear Pimenov and Holstein scaling begins to emerge.}
\label{fig:ETh_slopes}
\end{figure*}
As discussed in the main text, this suggests that the Homes proportionality factor is a slowly varying function of $\lambda$ away from the weak-coupling limit. The weak $\lambda$-dependence discussed above therefore sets the Homes scaling in the Eliashberg system apart from Homes scaling in the dirty BCS limit. In the next section we investigate the asymptotic Eliashberg equations to provide further understanding on the strong-coupling behaviour of $\eta_{H}(\tau,\,\lambda)$. 

We now discuss Pimenov and Holstein scaling within the framework of Eliashberg theory. From the previous section, we may write the Pimenov and Holstein factors as follows:

\begin{subequations}
\begin{equation}
\eta_P(\tau, \lambda)=
\begin{cases}
\dfrac{\Delta_0\tau}{2},&\tau \Delta_0 Z_0\rightarrow  0\quad \&\quad \tau \omega_E\rightarrow 0 \\ &\\
\dfrac{\lambda}{Z_0}\cdot {\tau\omega_E}\textrm{csch}\left(\dfrac{\omega_E}{T_c}\right), 
&\tau \Delta_0 Z_0\rightarrow  \infty\quad \&\quad \tau \omega_E\rightarrow \infty 
\end{cases}
\end{equation}
\begin{equation}
\eta_{\textrm{Hol}}(\tau, \lambda)=
\begin{cases}
\dfrac{\Delta_0}{2\omega_E},&\tau \Delta_0 Z_0\rightarrow  0\quad \&\quad \tau \omega_E\rightarrow 0 \\   &\\
\dfrac{\lambda}{Z_0}\cdot 
\textrm{csch}\left(\dfrac{\omega_E}{T_c}\right), 
& \tau \Delta_0 Z_0\rightarrow  \infty \quad \&\quad \tau \omega_E\rightarrow \infty.
\end{cases}
\end{equation}
\end{subequations}

\noindent Both Pimenov and Holstein proportionality factors reduce to the dirty BCS result for $\tau \Delta_0 Z_0\rightarrow 0$ and $\tau \omega_E\rightarrow 0$, except with $\DeltaOBCS$ replaced by $\Delta_0$. In the limit of $\tau\Delta_0Z_0\rightarrow 0$ and $\tau \omega_E\rightarrow \infty$, however, both Pimenov and Holstein factors display stronger $\lambda$-dependence which leads to the breakdown of these linear scaling relations for $\lambda \rightarrow \infty$, $\tau \Delta_0 Z_0\rightarrow \infty$, and $ \tau \omega_E\rightarrow \infty$. In Fig.~\ref{fig:ETh_slopes}, we observe that the clean limit of the Pimenov and Holstein plots both exhibit scaling as the electron-phonon coupling strength is increased.

\subsection{Planckian dissipation at arbitrary $\tau$ and $\lambda$}

The Planckian timescale $\tau_{\textrm{Pl}}=\hbar/(k_B T)$ is often considered to be a fundamental bound on dissipation in many-electron systems~\cite{Hartnoll2022Nov}. This timescale was first called "Planckian" in the work of Zaanen~\cite{Zaanen2004Jul}, who proposed such a lower bound on the scattering time as a possible explanation for Homes scaling in the cuprates~\cite{Homes2004Jul}. As discussed by Hartnoll and Mackenzie~\cite{Hartnoll2022Nov}, strong electron-phonon coupling may lead to a violation of the so-called "Planckian bound", as long as the temperature $T$ is not below $T_{\textrm{ph}}=\hbar\omega_E/k_B$. This provides motivation to consider the Planckian bound in the context of the present work, and explore any possible connection to Homes scaling.

Recall that $\zeta(\tau, \lambda)=[1/(2\pi \lambda T_c)]\cdot I(\tau, \lambda)$, where the integral $I(\tau,\lambda)$ is defined in Eq.~\eqref{eq:IInt}. Note that, as mentioned in the main text, $\zeta(\tau, \lambda)$ reduces to $\tau$ in the dirty limit and $\zeta(\tau, \lambda)=[1/(2\pi\lambda \omega_E)]\cdot \sinh(\omega_E/T_c)$ in the clean limit. As $\tau_{\textrm{Pl}}=1/T_c$ in Natural units for $T=T_c$, the ratio of $\tau_{\textrm{el}}\equiv \zeta(\tau,\,\lambda)$ to $\tau_{\textrm{Pl}}$ becomes
\begin{align}
\dfrac{\tau_{\textrm{el}}}{\tau_{\textrm{Pl}}}=
\begin{cases}
T_c\tau,&\tau \omega_E\rightarrow 0\\      &\notag\\
\dfrac{T_c}{2\pi \lambda \omega_E}\sinh\left(\dfrac{\omega_E}{T_c}\right),&\tau \omega_E\rightarrow \infty. 
\end{cases}
\end{align}

\begin{figure}[!t]
\hspace{-10mm} \includegraphics[width=0.7\columnwidth]{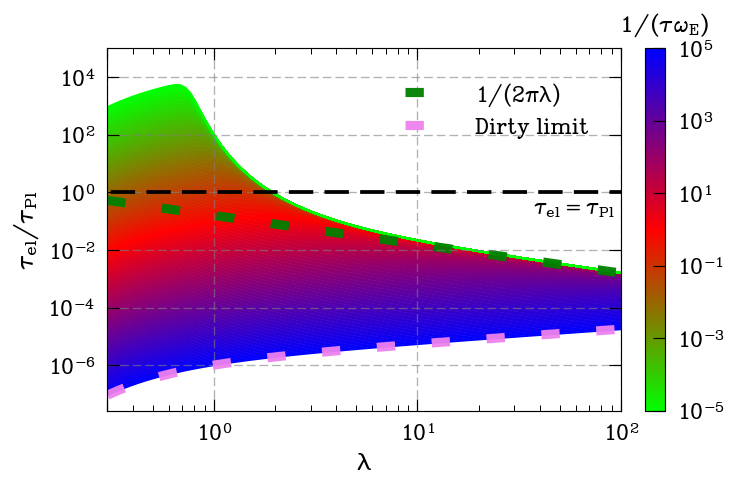}
\caption{The ratio $\tau_{\textrm{el}}/\tau_{\textrm{Pl}}$ versus the electron-phonon coupling strength $\lambda$. In the clean strong coupling limit, $\tau_{\textrm{el}}/\tau_{\textrm{Pl}}$ approaches $1/(2\pi\lambda)$ (green dashed line). In the dirty limit, the ratio approaches $T_c \tau$ (violet dashed line).}
\label{SMPlanckian}
\end{figure}

\noindent In Fig.~\ref{SMPlanckian}, we plot the above ratio versus $\lambda$ for various values of $\tau \omega_E$. For $\lambda\lessapprox 2$, $\tau_{\textrm{el}}> \tau_{\textrm{Pl}}$ in the clean limit, while for $\lambda \gtrapprox 2$, the clean value of $\tau_{\textrm{el}}$ decreases to a value much less than $\tau_{\textrm{Pl}}$. As a consequence, from our previous discussion in this Supplemental Material, we find that there is no correlation between Homes, Pimenov, and Holstein scaling and the presence of Planckian dissipation in Eliashberg theory.

\section{Asymptotically strong electron-phonon coupling}

\subsection{Universal form of the asymptotic Eliashberg equations}

We consider the Eliashberg equations in the asymptotic limit (i.e., $\lambda\rightarrow \infty$) by building upon the work of Marsiglio and Carbotte~\cite{Marsiglio1991, Marsiglio1988b} and Combescot~\cite{Combescot1995May,Combescot1989}. Eliashberg superconductivity in the limit that $\lambda\rightarrow\infty$, which we shall call asymptotically strong Eliashberg theory (ASETh), has been described as a "self-trapping" effect~\cite{Combescot1995May, Chubukov2020Jun}. There is only one relevant scale, $\omega_{E}$, and the theory can be cast in the form of a universal parameterless set of equations by defining $\overline{Q}=Q/(\omega_E\sqrt{\lambda/2})$. The Eliashberg equations then become
\begin{subequations}
\begin{align}
\overline{\Delta}(i\overline{\omega}_{n})Z(i\overline{\omega}_{n})&=\pi \overline{T}\sum_{m=-\infty}^{\infty}\lambda(i\overline{\omega}_{n}-i\overline{\omega}_{m})\cdot\dfrac{\overline{\Delta}(i\omega_{m})}{\sqrt{\overline{\omega}_{m}^{2}+\overline{\Delta}^{2}(i\overline{\omega}_{m})}}, \\
Z(i\overline{\omega}_{n})&=1+\dfrac{\pi \overline{T}}{\overline{\omega}_{n}}\sum_{m=-\infty}^{\infty}\lambda(i\overline{\omega}_{n}-i\overline{\omega}_{m})\cdot\dfrac{\overline{\omega}_{m}}{\sqrt{\overline{\omega}_{m}^{2}+\overline{\Delta}^{2}(i\overline{\omega}_{m})}}.
\end{align}
\end{subequations}
In terms of the redefined variables, the bosonic propagator is given by
\begin{equation}
\lambda(i\overline{\Omega}_{m}) = \dfrac{2}{\overline{\omega}_E^2 + \overline{\Omega}_{m}^2},
\end{equation}
Formally, the asymptotic limit can be implemented by taking $\overline{\omega}_{E}\rightarrow0$~\cite{Marsiglio1991}. As noted by Combescot, any Eliashberg superconductor is described by the Einstein model in the asymptotic limit. In the limit that $\overline{\omega}_E\rightarrow 0$, the asymptotic Eliashberg equations can be simplified to:
\begin{subequations}
\begin{align}
\overline{\Delta}(i\overline{\omega}_{n})Z(i\overline{\omega}_{n})&=\pi \overline{T}\underset{\overline{\omega}_E\rightarrow 0}{\lim}\sum_{m=-\infty}^{\infty} \dfrac{2}{\overline{\omega}_E^2+(\overline{\omega}_m-\overline{\omega}_n)^2} \cdot\dfrac{\overline{\Delta}(i\omega_{m})}{\sqrt{\overline{\omega}_{m}^{2}+\overline{\Delta}^{2}(i\omega_{m})}}, \\
 Z(i\overline{\omega}_n)&=1+\dfrac{\pi \overline{T}}{\overline{\omega}_n} \underset{\overline{\omega}_E\rightarrow 0}{\lim}\sum_{m=-\infty}^{\infty} \dfrac{2}{\overline{\omega}_E^2+(\overline{\omega}_m-\overline{\omega}_n)^2}\cdot \dfrac{\overline{\omega}_m}{\sqrt{\overline{\omega}_m^2 +\overline{\Delta}^2(i\overline{\omega}_m})}.
 \label{eq:ZAsym}
\end{align}
\end{subequations}
These are a new set of Eliashberg equations. By eliminating the renormalization function from the gap equation, one obtains the following asymptotic gap equation~\cite{Marsiglio1991, Combescot1995May, Chubukov2020Jun}:
\begin{equation}
\overline{\Delta}(i\overline{\omega}_{n}) = 2\pi \overline{T}\sum_{m\neq n}\dfrac{1}{[2\pi\overline{T}(n-m)]^2}\cdot\dfrac{\overline{\Delta}(i\omega_{m})-\dfrac{\overline{\omega}_{m}}{\overline{\omega}_{n}}\overline{\Delta}(i\overline{\omega}_{n})}{\sqrt{\overline{\omega}_{m}^{2}+\overline{\Delta}^{2}(i\omega_{m})}}.
\label{eq:GapEqnAsym}
\end{equation}
In the limit that $\overline{\omega}_{E}\rightarrow0$, there is a singularity in Eq.~\eqref{eq:ZAsym} arising from the term with $m=n$. To numerically avoid this singularity, one must retain the small $\omega_{E}$ dependence in the denominator, then perform the Matsubara frequency summation, and only then let $\overline{\omega}_{E}\rightarrow0$. In the analysis below, we determine the explicit form of $Z$ in the asymptotic limit. However, the asymptotic gap equation has a zero in the numerator and thus the term $n=m$ drops out from the summation in the limit that $\overline{\omega}_{E}\rightarrow0$. Thus, there is no singularity in the asymptotic gap equation~\cite{Chubukov2020Jun}. 

To determine the explicit form of $Z(i\overline{\omega}_n)$ in the asymptotic limit, we note that the term with $m=n$ term provides the dominant contribution to the series, thus we obtain
\begin{align}
Z(i\overline{\omega}_n)&=1+\dfrac{\pi \overline{T}}{\overline{\omega}_n}\underset{\overline{\omega}_E\rightarrow 0}{\lim}\sum_{m=-\infty}^{\infty} \dfrac{2}{\overline{\omega}_E^2+(\overline{\omega}_m-\overline{\omega}_n)^2}\cdot \dfrac{\overline{\omega}_m}{\sqrt{\overline{\omega}_m^2+\overline{\Delta}^2(i\overline{\omega}_m})} \notag\\
&\approx \dfrac{\pi \overline{T}}{\overline{\omega}_n} \cdot 
 \dfrac{\overline{\omega}_n}{ \sqrt{\overline{\omega}_n^2 +\overline{\Delta}^2(i\overline{\omega}_n})}\sum_{m=-\infty}^{\infty} \dfrac{2}{\overline{\omega}_E^2+(\overline{\omega}_m-\overline{\omega}_n)^2} \notag\\
&= \dfrac{\pi}{\overline{\omega}_E}\dfrac{1}{\sqrt{\overline{\omega}_n^2 +\overline{\Delta}^2(i\overline{\omega}_n)}} \coth\left(\dfrac{\overline{\omega}_E}{2\overline{T}}\right) \notag\\
&\approx \dfrac{\pi}{\overline{\omega}_E}\dfrac{1}{\sqrt{\overline{\omega}_n^2 +\overline{\Delta}^2(i\overline{\omega}_n)}}. 
\label{eq:ZAsym2}
\end{align}
In the last step we have taken $T\rightarrow 0$, and we have also used the fact that $\overline{\omega}_E^2=2/\lambda$. As mentioned earlier, the renormalization function is singular in the asymptotic limit. The above result, on the imaginary frequency axis, has a similar form to the result on the real frequency axis~\cite{Marsiglio1991, Combescot1995May}. For the $n=0$ Matsubara index, Eq.~\eqref{eq:ZAsym2} simplifies to 
\begin{equation}
\lim_{T\rightarrow 0}Z(i\overline{\omega}_0) = \dfrac{\pi}{\overline{\omega}_E}\dfrac{1}{\sqrt{\overline{\Delta}^2(i\overline{\omega}_0)}} 
\quad \Longrightarrow \quad \lim_{T\rightarrow 0}Z(i\overline{\omega}_0)\Delta(i\overline{\omega}_0) = \dfrac{\pi}{2}\lambda. 
\label{eq:ZAsym3}
\end{equation}
In the last step we have reintroduced $\lambda$ for convenience. In Fig.~\ref{fig:Sqrt_lamb}, a plot of the product of the gap function and the renormalization function, normalized by $\lambda$, is shown as a function  of $\lambda$. Fitting the large-$\lambda$ data (for values of $\lambda$ in the range $\lambda\approx75-100$) to a function of the form $a+b/\lambda+c/\lambda^2$, we find the value of the constant term to be $a\approx 1.546$, which is fairly close to the asymptotic value obtained theoretically in Eq.~\eqref{eq:ZAsym3} given by  $\pi/2$.

Finally, we close this section by noting the form of the dirty and clean limits in ASETh. Recall that the formal definitions of dirty and clean are described in Eqs.~\eqref{dirtydef} and \eqref{cleandef}, with the value of the function $D(\lambda)$ playing a fundamental role in both scenarios. Below, we write $D(\lambda)$ in the asymptotically strong limit for $T\rightarrow 0$, using the result from Eq.~\eqref{eq:ZAsym2}:

\begin{align}
\lim_{\lambda\rightarrow \infty}D(\lambda) &=\frac{1}{2} \omega_E \lim_{\lambda\rightarrow \infty} \sum_{n=-\infty}^\infty \dfrac{\Delta^2(i\omega_n)}{Z(i\omega_n) \left(\omega_n^2+\Delta^2(i\omega_n)\right)^{3/2}} \cdot \left(\sum_{n=-\infty}^\infty \dfrac{\Delta^2(i\omega_n)}{\omega_n^2 +\Delta^2(i\omega_n)}\right)^{-1}\notag\\
&= \frac{1}{2\pi} \omega_E \overline{\omega}_E\cdot \dfrac{1}{\omega_E}\sqrt{\dfrac{2}{\lambda}}\notag\\
&=\dfrac{1}{\pi \lambda}.
\end{align}

\noindent From Eqs.~\eqref{dirtydef} and \eqref{cleandef}, the formal definition of the dirty and clean limits for the asymptotically strong limit now reduce to $\pi \lambda\tau \omega_E \rightarrow 0$ and $\pi \lambda\tau \omega_E\rightarrow \infty$, respectively. For simplicity, we will drop the $\pi$. 

\subsection{Superfluid density and normal-state conductivity in the asymptotically strong limit}

Recall the form of the superfluid density on the imaginary frequency axis given in Eq.~\eqref{eq:EThSupDens1}. Using the definition $\overline{Q}=Q/(\omega_{E}\sqrt{\lambda/2})$, the superfluid density can be re-expressed as 
\begin{equation}
\dfrac{n_s(\tau, \lambda, T)}{n}
=\pi \overline{T} \sum_{m=-\infty}^\infty\dfrac{\overline{\Delta}^2(i\overline{\omega}_m)}{\overline{\omega}_m^2+\overline{\Delta}^2(i\overline{\omega}_m)}\cdot \dfrac{1}{Z(i\overline{\omega}_m)\sqrt{\overline{\omega}_m^2+\overline{\Delta}^2(i\overline{\omega}_m)}+\dfrac{1}{\tau \omega_E \sqrt{2\lambda}}}.
\end{equation}
At zero temperature, the asymptotic form of the renormalization function is given in Eq.~\eqref{eq:ZAsym2}. Inserting this result into the above equation and simplifying, we obtain 
\begin{equation}
\lim_{\lambda\rightarrow \infty }\dfrac{n_s(\tau, \lambda)}{n} = \dfrac{\overline{\omega}_{E}\overline{T}}{1+\dfrac{1}{\pi\lambda\tau\omega_{E}}}\cdot \sum_{m=-\infty}^{\infty}\dfrac{\overline{\Delta}^2(i\overline{\omega}_m)}{\overline{\omega}_m^2+\overline{\Delta}^2(i\overline{\omega}_m)}.\label{SFASETH}
\end{equation}

\begin{figure*}
\hspace{-0mm}\subfloat[\label{ASEth_Gap}]{  \includegraphics[width=.48\linewidth]{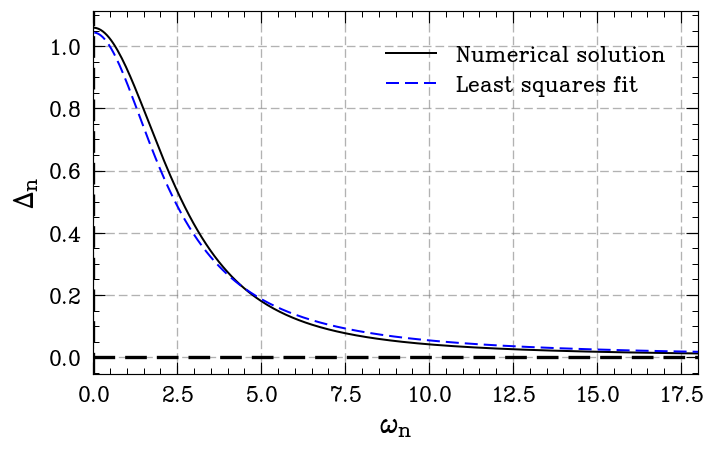}}\hspace{-1mm}
\subfloat[\label{ASEth_gap}]{\includegraphics[width=.55\linewidth]{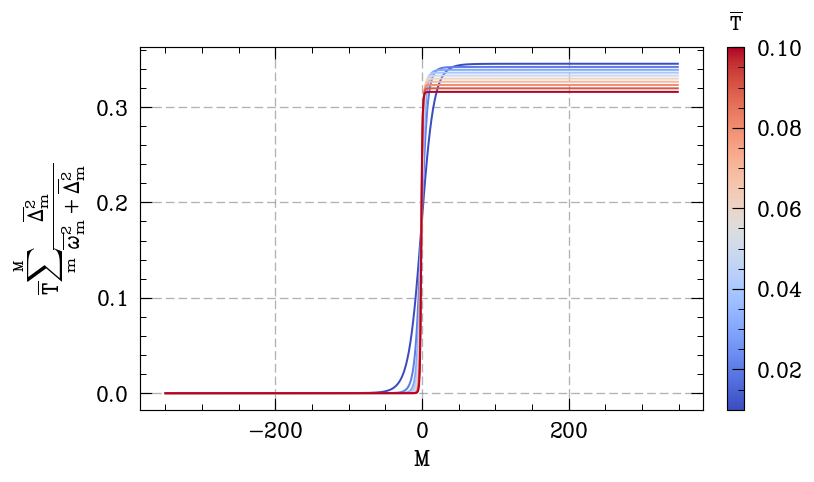}}
\caption{(a) Plot of the gap function versus the Matsubara frequency, with a least-squares fit given by Eq.~\eqref{eq:Lorentzian} shown in blue. (b) The value of the summation given in Eq.~\eqref{eq:Alpha}, from $M=-350$ up to $M=350$.}
\label{fig:ASEth_quantities}
\end{figure*}

\noindent We can further simplify the asymptotically strong superfluid density by rewriting Eq.~\eqref{SFASETH} as the following:

\begin{align}
\lim_{\lambda\rightarrow \infty }\dfrac{n_s(\tau, \lambda)}{n} &= \alpha \cdot 
\begin{cases}
\pi \tau \omega_E\sqrt{2\lambda} ,\qquad & \lambda \tau \omega_E\rightarrow 0\\
&\\
\sqrt{\dfrac{2}{\lambda}} ,\qquad & \lambda \tau \omega_E\rightarrow \infty
\end{cases}
\label{SFD}
\end{align}

\noindent where we have used the fact that $\overline{\omega}_E=\sqrt{2/\lambda}$ and where we have defined $\alpha$ to be the following $T\rightarrow 0$ summation over the Matsubara frequencies, which may be converted to an integral over a real variable $x$~\cite{AGDBook}:
\begin{equation}
  \alpha\equiv\lim_{\overline{T}\rightarrow 0} \alpha(\overline{T})=\lim_{\overline{T}\rightarrow 0}\overline{T}\sum_{m=-\infty}^\infty \dfrac{\overline{\Delta}^2(i\omega_m)}{\overline{\omega}_m^2+\overline{\Delta}^2(i\omega_m)} = \dfrac{1}{2\pi}\lim_{\overline{T}\rightarrow 0}\int^{\infty}_{-\infty}\dfrac{\overline{\Delta}^2(ix)}{\overline{\Delta}^2(ix)+x^2}\ dx. 
  \label{eq:Alpha}
\end{equation}
The explicit value of $\alpha$ defined in Eq.~\eqref{eq:Alpha} must still be determined. The simplest approximation is to assume a constant gap. If we let $\overline{\Delta}(ix)\approx \overline{\Delta}(i\omega_{0})$ and evaluate the integral in Eq.~\eqref{eq:Alpha}, then, using the numerically determined value $\overline{\Delta}(i\omega_0)\approx 1.05$ (where this result is for the $T=0$ case), we obtain $\alpha=\frac{1}{2}\overline{\Delta}(i\omega_{0})\approx0.525$. However, by numerically solving the asymptotic gap equation~\eqref{eq:GapEqnAsym}, we find that the form of the gap function $\overline{\Delta}(i\overline{\omega}_n)$ is a Lorentzian. Thus, we can refine the previous estimate by considering a gap function of the form: 
\begin{equation}
\overline{\Delta}(i\overline{\omega}_n)\approx \dfrac{A}{1+B\cdot \overline{\omega}_n^2}.
\label{eq:Lorentzian}
\end{equation}
The fit parameters $A$ and $B$ must be found numerically. As shown in Fig.~\ref{fig:ASEth_quantities}(a), the Lorentzian fit to the numerical solution of the asymptotic gap equation~\eqref{eq:GapEqnAsym} agrees very well. The approximate value for $\alpha$ is then given by

\begin{align}
\alpha &= \dfrac{1}{2\pi}\lim_{\overline{T}\rightarrow 0}\int_{-\infty}^{\infty}\dfrac{\overline{\Delta}^2(ix)}{\overline{\Delta}^2(ix)+x^2}\ dx \notag\\
&\approx \dfrac{1}{2\pi}\int_{-\infty}^{\infty}\dfrac{\dfrac{A^2}{(1+Bx^2)^2}}{\dfrac{A^2}{(1+Bx^2)^2}+x^2}\ dx\notag\\
&=\dfrac{1}{2\pi}\int_{-\infty}^{\infty}\dfrac{A^2}{A^2+x^2(1+Bx^2)^2}\ dx.
\end{align}

\noindent Using a least-squares minimization, the values of $A$ and $B$ are found to be $A\sim 1.044$ and $B\sim 0.1832$. Applying the residue theorem then gives $\alpha\approx 0.3323$. In Fig.~\ref{fig:ASEth_quantities}(b) we confirm this value numerically by plotting the value of $\alpha(\overline{T})$ in \eqref{eq:Alpha}  from $M=-350$ up to $M=350$ for several values of $\overline{T}$. As $\overline{T}\rightarrow 0$, the value of $\alpha\overline{T}$ approaches $\sim 0.33$ as the upper limit of summation limits to $M=350$.  This value of $\alpha$ is close to one third, and thus, for analytical convenience, we will set $\alpha \approx 1/3$. 

Next we consider the normal-state conductivity in the asymptotic limit. Using Combescot's interpolation formula in Eq.~\eqref{eq:Combescot}, in the asymptotic limit the critical temperature becomes 
${T_c}/{\omega_E} \sim a \sqrt{\lambda/{2}}$. Substituting this expression for $T_{c}$ into Eq.~\eqref{eq:IInt}, the integral $I(\tau,\,\lambda)$ then has the following form in the asymptotic limit:

\begin{equation}
\lim_{\lambda\rightarrow \infty}I(\tau,\,\lambda) \approx \lim_{\lambda\rightarrow \infty}\frac{
T_c/\omega_E}{\textrm{csch} \left( \dfrac{\omega_E}{T_c} \right)
+\dfrac{1}{2\pi \lambda \omega_E\tau}}
\approx\lim_{\lambda\rightarrow \infty}\frac{1} {1+\dfrac{1}{2\pi \lambda \tau T_c}}.
\end{equation}

\noindent Recall that $\sigma(\tau, \lambda)=(\omega_p^2/(4\pi))\cdot \zeta(\tau, \lambda)$ where $\zeta(\tau, \lambda) \equiv [1/{(2\pi \lambda T_c)}]\cdot I(\tau,\,\lambda)$. From the previous result, we then obtain the following value for the product of the normal-state conductivity and $T_c$ in the asymptotic limit:

\begin{align}
\lim_{\lambda\rightarrow \infty}\dfrac{\sigma (\tau, \lambda) T_c}{\omega_p^2/(8\pi^2)}=\begin{cases}
2\pi \tau \omega_E a \sqrt{\dfrac{\lambda}{2}},\qquad & \lambda \tau \omega_E \rightarrow 0\\
& \\
\dfrac{1}{\lambda},\qquad &\lambda \tau \omega_E \rightarrow \infty.
\end{cases}
\label{eq:ASEThCD}
\end{align}

\noindent Note that, in the above, $T_c$ scales as $\sqrt{\lambda}$ in the strong-coupling limit, and thus we have assumed that $\lambda^{3/2}\tau \omega_E\rightarrow 0$ and $\lambda^{3/2}\tau \omega_E\rightarrow \infty$ in the dirty and clean limits, respectively.

\subsection{Universal scaling relations in the asymptotically strong limit}

\begin{figure*}
\subfloat[\label{ASEth_Dirty1b}]{%
\includegraphics[width=.6\linewidth]{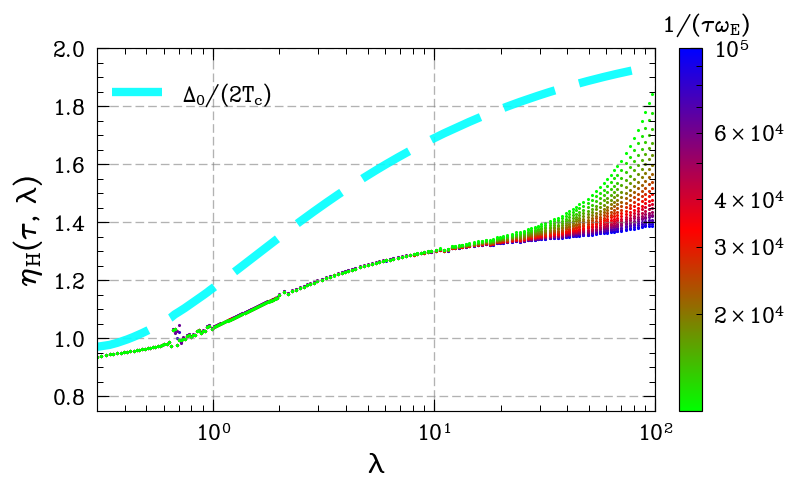}}\\
\vspace{-5mm}
\subfloat[\label{ASEth_Dirty2}]{\includegraphics[width=.6\linewidth]{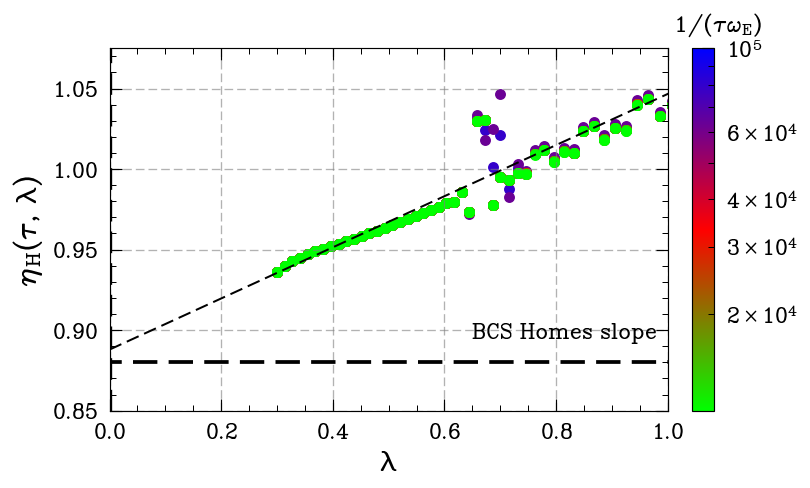}} \\
\vspace{1mm}
\hspace{-13mm}\subfloat[\label{ASEth_Dirty3}]{\includegraphics[width=.51\linewidth]{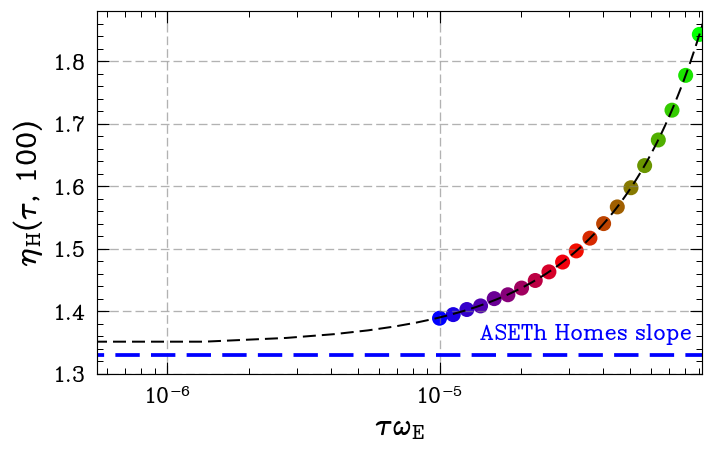}}\vspace{0mm}%
\caption{ (a) The value of the Homes proportionality factor $\eta_H(\tau,\,\lambda)$ versus $\lambda$ for various values of $1/(\tau\omega_E)$ in the dirty limit. For smaller values of $\lambda$, the factor approaches the BCS Homes slope of $\sim0.88$. For larger values of $\lambda$, the factor approaches the ASETh Homes slope of $\sim1.33$ for large values of $1/(\tau\omega_E)$. The dashed cyan line denotes the constant-gap approximation to the dirty Homes slope. The cyan line approaches the numerical value only in the weak-coupling Eliashberg regime. (b) Extrapolation of the BCS Homes slope data to the limit of $\lambda\rightarrow 0$. The extrapolation yields a value of $0.887$, which is very close to the predicted value in the dirty BCS limit. (c) Extrapolation of the ASETh Homes slope for $\lambda=100$ to the limit of $\tau\omega_E\rightarrow 0$. The extrapolation yields a value of $1.345$, which is very close to the predicted value in the dirty ASETh limit found by taking a Lorentzian fit to the asymptotic gap function.}
\label{ASEth_Dirty}
\end{figure*}

The superfluid density and the normal-state conductivity, in the asymptotically strong limit, are given by Eq.~\eqref{SFASETH} and Eq.~\eqref{eq:ASEThCD}, respectively. Using these two quantities, the asymptotically strong Homes proportionality factor is given by

\begin{align}
\lim_{\lambda\rightarrow \infty}\eta_H(\tau,\,\lambda)=\alpha\cdot
\begin{cases}
\dfrac{1}{a},&\lambda \tau \omega_E\rightarrow 0\\ &\\
\sqrt{2\lambda},&\lambda \tau \omega_E\rightarrow \infty.
\end{cases}
\label{ASETh_Homes_Cases}
\end{align}

\noindent In the clean scenario, the Homes proportionality factor exhibits some residual $\lambda$-dependence, and thus linear Homes scaling breaks down in this limit. In the dirty limit, the Homes proportionality factor is a universal constant equal to $\alpha/a$. Taking the constant-gap approximation, the Homes slope in the asymptotic limit is found to be $\alpha/a=\overline{\Delta}_0/(2\overline{T}_c)\sim 2.051$, which can be obtained from both the finite-$\lambda$ imaginary-axis expression Eq.~\eqref{Eth_Homes} in the $\lambda\rightarrow \infty$ limit and the asymptotic result given in Eq.~\eqref{ASETh_Homes_Cases}. Using the Lorentzian estimate for the gap function introduced in the previous section, we find a better estimate of the asymptotic Homes slope to be $\sim 1.33$. In Fig.~\ref{ASEth_Dirty3}, we find that an extrapolation of the Homes slope as $\tau\omega_E\rightarrow 0$ yields $1.345$ in the asymptotic limit, which is in near-perfect agreement with our estimate.

\begin{table}
\begin{center}\begin{tabular}{||c | c | c | c| c ||} 
 \hline
Defining limits & Name & Homes factor  & \,\,Pimenov factor\,\, & \,\,Holstein factor\,\,\rule[-2ex]{0pt}{6ex} \\ [2ex] 
 \hline\hline 
  $\DeltaOBCS \tau\rightarrow 0$; $\lambda\rightarrow 0$ & Dirty BCS & $\dfrac{\DeltaOBCS}{2T_c}$ & $\dfrac{\DeltaOBCS \tau}{2}$ & $\dfrac{\DeltaOBCS}{2\omega_E}$ \rule[-2ex]{0pt}{6ex} \\ [2ex]
 \hline 
 $\tau\Delta_0 Z_0 \rightarrow 0$; $\tau \omega_E\rightarrow 0$; $\lambda\lessapprox 1$ & Dirty WETh & $\dfrac{\Delta_0}{2T_c}$ & $\dfrac{\Delta_0 \tau}{2}$ & $\dfrac{\Delta_0}{2\omega_E}$ \rule[-2ex]{0pt}{6ex} \\ [2ex]
  \hline 
 $\lambda \tau\omega_E\rightarrow 0$; $\lambda\rightarrow \infty$ & \,\, Dirty ASETh \,\, & $\dfrac{\alpha}{a}$ & $\alpha \cdot \tau \omega_E \sqrt{\dfrac{\lambda}{2}}$ & $\alpha\cdot  \sqrt{\dfrac{\lambda}{2}}$ \rule[-2ex]{0pt}{6ex} \\ [2ex]
 \hline\hline
 $\DeltaOBCS\tau \rightarrow \infty$; $\lambda \rightarrow 0$ & Clean BCS & $\dfrac{1}{2\pi T_c \tau}$ & $\dfrac{1}{2\pi}$ & $\dfrac{1}{2\pi \tau \omega_E}$ \rule[-2ex]{0pt}{6ex} \\ [2ex]
 \hline 
 $\tau \Delta_0 Z_0 \rightarrow \infty$; $\tau\omega_E\rightarrow \infty$; $\lambda\lessapprox 1$ & Clean WETh & \,\,$\dfrac{\lambda}{Z_0}\cdot \dfrac{\omega_E}{T_c}\textrm{csch}\left(\dfrac{\omega_E}{T_c}\right)$\,\, & \,\,$\dfrac{\lambda}{Z_0}\cdot {\tau\omega_E}\textrm{csch}\left(\dfrac{\omega_E}{T_c}\right)$\,\, & \,\,$\dfrac{\lambda}{Z_0}\cdot 
    \textrm{csch}\left(\dfrac{\omega_E}{T_c}\right)$\,\,\rule[-3.2ex]{0pt}{8ex} \\ [2ex]
 \hline
$\lambda \tau\omega_E\rightarrow \infty$; $\lambda\rightarrow \infty$ & Clean ASETh & $\alpha\cdot \sqrt{2\lambda}$ & $\alpha \cdot a\lambda \tau \omega_E$ & $\alpha \cdot a\lambda$  \rule[-3.2ex]{0pt}{8ex} \\ [1ex]  \hline
\end{tabular}
\end{center}
\caption{Main results for the Homes, Pimenov, and Holstein proportionality factors in the clean and dirty limits, in addition to the ASETh limit and the weak-Eliashberg (WETh) limit. Recall that $\DeltaOBCS$ denotes the $T\rightarrow 0$ BCS gap, while $\Delta_0$ denotes the $T\rightarrow 0$ limit of the Eliashberg gap function at zero Matsubara index. In the dirty limit, Homes scaling is obeyed for all $\lambda$. In the clean limit, Homes scaling is violated in the weak-coupling limit, and has weak $\lambda$ dependence in the strong-coupling limit. Pimenov amd Holstein scaling are violated in the dirty limit for all entries considered above, while in the clean limit these scaling laws have strong $\lambda$ dependence for $\lambda \gg 1$. Note that there is no notion of linear scaling for the case of clean Homes, clean Holstein, and dirty or clean Pimenov scaling.
}
\label{Tab:MyTable}
\end{table}

As shown in Eq.~\eqref{eq:Pimenov} and Eq.~\eqref{eq:Holstein}, the Homes factor can be directly related to the Pimenov and Holstein factors, respectively.  Thus, using the previous result, we obtain:

\begin{align}
\lim_{\lambda\rightarrow \infty}\eta_P(\tau,\,\lambda)=\lim_{\lambda\rightarrow \infty}\eta_H(\tau,\,\lambda)\cdot T_c\tau=\alpha \cdot \begin{cases}
\tau \omega_E\sqrt{\dfrac{\lambda}{2}}, &\lambda\tau\omega_E\rightarrow 0 \\
& \\
a\lambda \tau \omega_E,&\lambda \tau\omega_E\rightarrow \infty
\end{cases}
\end{align}

\begin{align}
\lim_{\lambda\rightarrow \infty}\eta_{\textrm{Hol}}(\tau,\,\lambda)=\lim_{\lambda\rightarrow \infty}\eta_H(\tau,\,\lambda)\cdot \dfrac{T_c}{\omega_E}=
\alpha \cdot \begin{cases}
\sqrt{\dfrac{\lambda}{2}},&\lambda \tau \omega_E\rightarrow 0 \notag\\
&\notag\\
a\lambda,&\lambda \tau \omega_E\rightarrow \infty
\end{cases}
\end{align}

\noindent The Pimenov proportionality factor tends to zero in the limit that $\lambda\tau\omega_E\rightarrow 0$, whereas for all other cases the Pimenov and Holstein proportionality factors diverge in the asymptotic limit. The only scaling relation which exists as $\lambda\rightarrow \infty$ is the Homes relation in the dirty limit. In Table~\ref{Tab:MyTable}, we give a summary of the scaling relations considered in this work.

\section{Numerical details}

\subsection{Iterative solutions of the $T=0$ Eliashberg equations on the imaginary frequency axis}

Recall the form of the Eliashberg equations on the imaginary frequency axis, given in Eqs.~\eqref{eq:EthEqnsIm1} and \eqref{eq:EthEqnsIm2}. We rewrite these equations as a sum over a finite number of Matsubara frequencies, given by the following:

\begin{subequations}
\begin{align}
\Delta(i\omega_{n})Z(i\omega_{n}) &= \pi T\sum_{m=-M}^{M-1}\lambda(i\omega_{m}-i\omega_{n})\cdot\dfrac{\Delta(i\omega_{m})}{\sqrt{\omega_{m}^{2}+\Delta^{2}(i\omega_{m})}}, \\
Z(i\omega_{n}) &= 1+\dfrac{\pi T}{\omega_{n}}\sum_{m=-M}^{M-1}\lambda(i\omega_{m}-i\omega_{n})\cdot\dfrac{\omega_{m}}{\sqrt{\omega_{m}^{2}+\Delta^{2}(i\omega_{m})}}.
\end{align}
\end{subequations}

Note that we take the upper limit of the summation to be $m=M-1$ in order to ensure symmetry of the summation about the positive and negative Matsubara frequencies~\cite{Szczesniak2006}. 
The value of $M$ is truncated to a finite value that is large enough such that the gap and renormalization functions do not appreciably change between the $M$ and $M+1$ values. For the purposes of this work, we want to consider a large number of different electron-phonon coupling strengths $\lambda$, and we also want to ensure that the criteria for convergence are consistent for all values of $\lambda$. For this reason, we consider a feedback (or "Ouroboros") algorithm, which ensures both timely convergence of the iterative solution and a consistent criterion for termination of the iterative procedure. 

A schematic of the Ouroboros protocol is given in Fig.~\ref{fig:Worm}. Iteration begins with two basic input parameters: a list of $\lambda$ values to consider and the initial value for the summation truncation $m_{\mathrm{max}}=M_{\mathrm{init}}$. We will take the list of $\lambda$ values to be of length $N_\lambda$. We take a temperature $T/T_c$ that is close enough to zero  such that our results are a good approximation to the zero-temperature gap and renormalization functions. The iteration begins with an initial "guess" for the gap and renormalization functions, which is typically a list of ones of length $2M$. With these initial input parameters, the base algorithm proceeds as follows: 

\begin{figure*}[b]
\includegraphics[width=.55\linewidth]{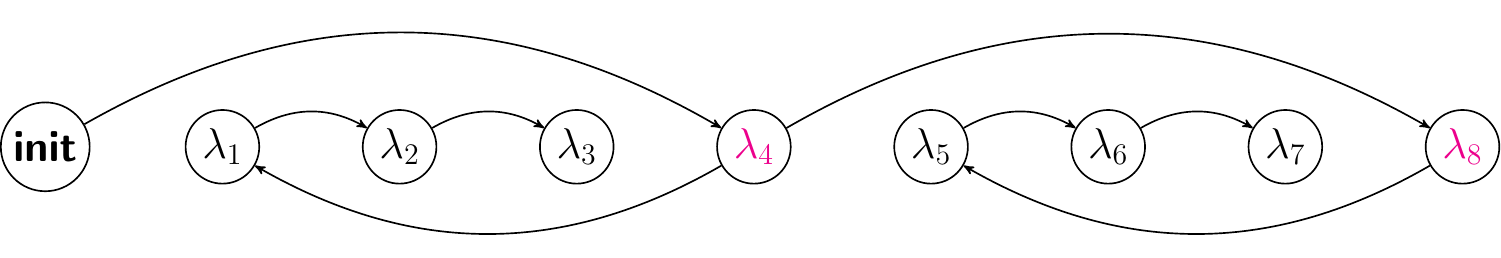}
\caption{Schematic of the "Ouroboros" feedback algorithm implemented to numerically solve the Eliashberg equations.}
\label{fig:Worm}
\end{figure*}

\begin{figure*}
\hspace{0mm}\subfloat[Comparison of tail parameters.\label{SM_tail}]{%
  \includegraphics[width=.45\linewidth]{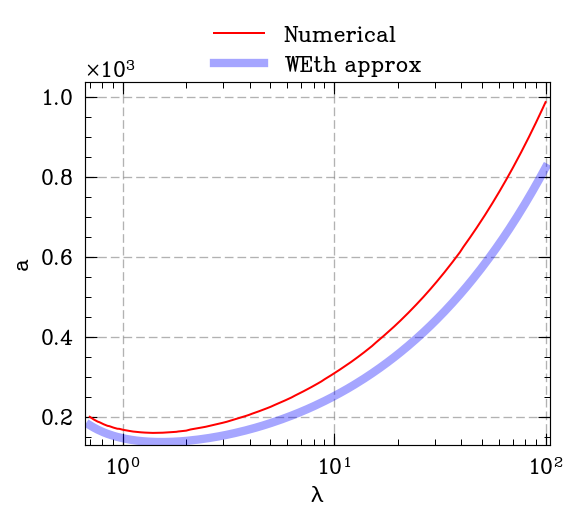}}\\ \vspace{4mm}
\hspace{0mm}\subfloat[Truncation values of the Matsubara sum.\label{SM_Matsubara}]{%
  \includegraphics[width=.45\linewidth]{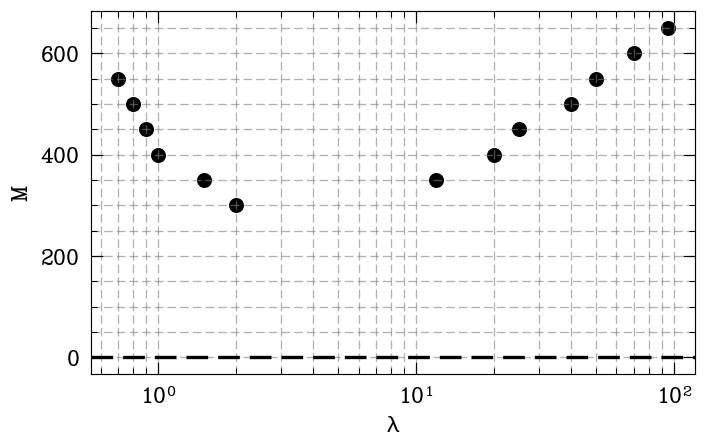}}\hspace{6mm}
\subfloat[Distance metric between tails.\label{SM_Distance}]{
  \includegraphics[width=.45\linewidth]{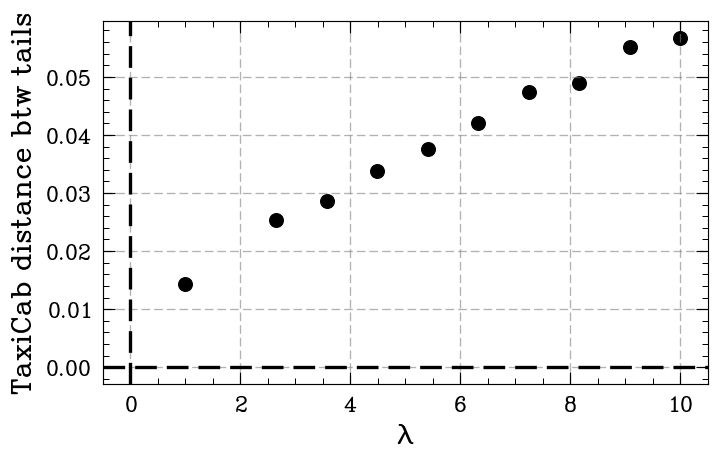}}
\caption{(a) Comparison of the tail parameters between those found iteratively (red) and via the weak Eliashberg (WETh) approximation. The same basic trend for the tail parameter vs. $\lambda$ is found up to $\lambda\sim 10^2$. (b) Truncation values of the Matsubara sum versus $\lambda$, with a black dot denoting whenver $M$ is updated by the Ouroboros algorithm. For intermediate values of $\lambda$, the algorithm converges for fewer Matsubara terms. (c) Distance metric between the gap function's "tail" and the estimate of the tail computed in the text. For the values of $M$ considered, the taxicab distance remains small well beyond the regime of weak Eliashberg theory.}
\label{Fig4}
\end{figure*}

\begin{enumerate}
    \item Iteratively solve the Eliashberg equations for $m_{\mathrm{max}}=M$ and $\lambda=\lambda_i$ until convergence is reached to the desired accuracy, and subsequently obtain $\Delta(i\omega_n)$ and $Z(i\omega_n)$ for $n\in [-M,\,M-1]$. Here, $\lambda_i$ is the $i$th term in the total list of $\lambda$, where $i\le N_{\lambda}$ (in Fig.~\ref{fig:Worm}, $i=4$ for example).
    \item Repeat step 1) for $m_{\mathrm{max}}=M+1$, and obtain $\Delta(i\omega_{n})$ and  $Z(i\omega_{n})$ for $n\in[-M-1,\,M]$ through iteration to the desired accuracy.
    \item If both $|\Delta(i\omega_{M})-\Delta(i\omega_{M+1})|$ and $|Z(i\omega_{M})-Z(i\omega_{M+1})|$ are below a desired threshold, accept the lists of $\Delta(i\omega_{n})$ and $Z(i\omega_{n})$ as the gap and renormalization functions for $\lambda_i$. If, however, $|\Delta(i\omega_{M})-\Delta(i\omega_{M+1})|$ and $|Z(i\omega_{M})-Z(i\omega_{M+1})|$ are not below the threshold, increase $M$ to $M+50$, and repeat steps 1) through 3).
    \item Upon successful completion of step 3), use the value of $M$ and the values of $\Delta(i\omega_{n})$ and $Z(i\omega_{n})$ for $n\in [-M,\,M-1]$ as the initial guess for the iterative procedure for $\lambda_{i-r}$, where $r$ is some integer (in Fig. \ref{fig:Worm}, $r=3$ as an example). Repeat step 1) using these new input parameters to find $\Delta(i\omega_{n})$ and $Z(i\omega_{n})$ for $\lambda_{i-r}$ and $n\in [-M,\,M-1]$.
    \item Use the solutions of the gap and renormalization functions found in step 4) as the initial guess for the iteration procedure for $\lambda_{i-r+1}$. Repeatedly feed in the previous result for the gap and renormalization functions for $\lambda_{i-r+2}$, $\lambda_{i-r+3}$, ... $\lambda_{i-1}$ to find the gap and renormalization functions for each coupling parameter.   
    \item Feed in the result for the gap and renormalization functions for $\lambda_i$ as the initial guess for the gap and renormalization functions for $\lambda_{i+r}$. Next, return to step 1) and repeat until the gap and renormalization functions for all $\lambda$ under consideration are found to the desired accuracy.
\end{enumerate}

We may refine the algorithm to determine the gap and renormalization functions for larger Matsubara terms, using the so-called "tail extension". Within Eliashberg theory, it has been noted previously that we may write the gap and renormalization functions as $\Delta(i\omega_n)\approx a_n/n^2$ and $Z(i\omega_n)\approx 1+{b_n}/{|n|}$~\cite{Marsiglio2018Jul} (note that this $a$ is taken to be a fit parameter, and has nothing to do with the asymptotic Homes slope in the previous discussion). In the limit of $n\gg 1$ and $n\gg \omega_E/{(2\pi T)}$, the dependence on $n$ effectively drops out of these coefficients, and $a_n$ and $b_n$ become constants independent of $n$, which we denote by $a$ and $b$, respectively.

\begin{figure*}
\hspace{3mm}\subfloat[Gap function versus Matsubara frequency\label{SM_Eth_Gap}]{%
\includegraphics[width=.55\linewidth]{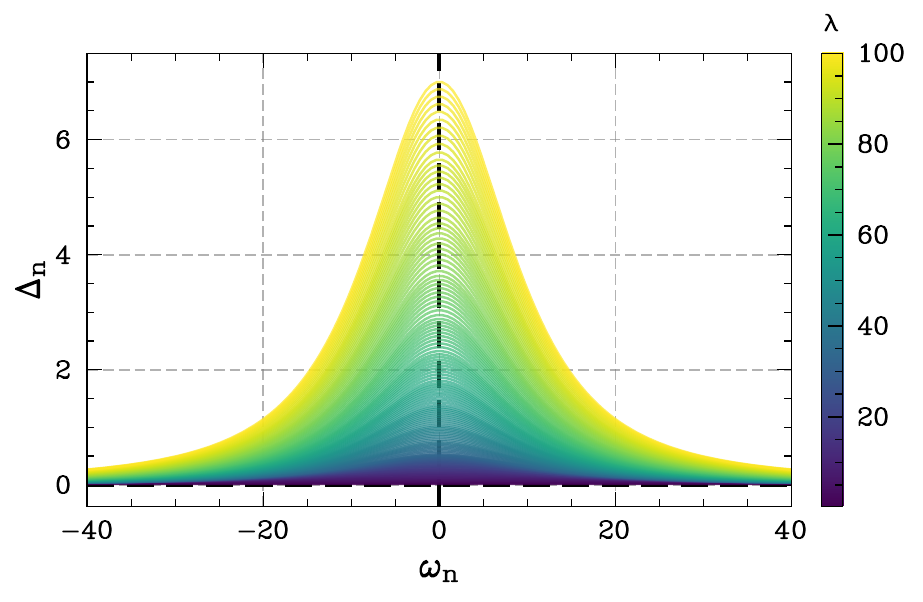}}
\hspace{-1mm}
\subfloat[Renormalization function versus Matsubara frequency \label{FigPimenov}]{%
\hspace{-2mm} \includegraphics[width=.55\linewidth]{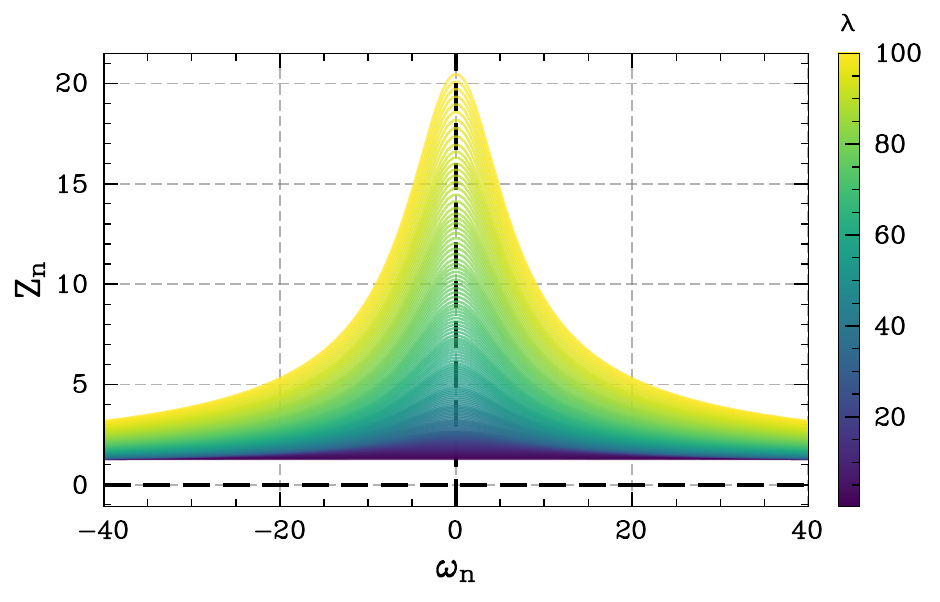}}
\caption{Numerical results for (a) the gap and (b) the renormalization on the imaginary frequency axis for $T/T_c=0.1$. In both of the above plots, the "Ouroboros" feed-back algorithm allows us to go up to $\lambda\sim100$ while still ensuring consistent convergence of the self-consistent Eliashberg equations.}
\label{Fig5}
\end{figure*}

Let us define the portion of the gap and renormalization functions where $\Delta(i\omega_n)\approx a/n^2$ and $Z(i\omega_n)\approx b/|n|$ as the "tail". We may then write an extension to the Ouroboros algorithm discussed above which allows us to trivially extend the gap and renormalization functions to higher values of $n$ without the need for iteration. We write this algorithm below as an extension of the above:

\begin{enumerate}
    \item Upon successfully finding the summation truncation $M$ based upon $|\Delta(i\omega_{M})-\Delta(i\omega_{M+1})|$ and $|Z(i\omega_{M})-Z(i\omega_{M+1})|$ (i.e., upon completion of step 3 given in the algorithm above), consider an interval of the gap function and renormalization function from some $m=m'<M-200$ to $M$. Obtain an estimate for the coefficients $a$ and $b$ through a least-squares fit, assuming the gap function goes as $m^{-2}$ and the renormalization function goes as $|m|^{-1}$.
    \item Given the value of $a/m^2$ and $b/|m|$ and the actual values of the gap and renormalization functions found iteratively, compute three quantities: i) the coefficient of determination between $a/m^2$ and $b/|m|$ and the iteratively-found solutions for $\Delta(i\omega_m)$ and $Z(i\omega_m)$; ii) the percentage difference between $\Delta(i\omega_M)$ and $a/M^2$ (and, likewise, $Z(i\omega_M)$ and $b/|M|$); and iii) the percentage difference between the numerically-found value of $a$ and the estimate for $a$ from weak Eliashberg theory. 
    \item If the results for i)-iii) are not below given thresholds, the value of $m'$ in step 1 of the above is increased, and step 2 is repeated. If steps 1-3 are repeated until $m'=M-200$, $M$ is increased to $M+50$, and step 1) is repeated. 
    \item Upon successfully finding a value of $m=M$ where the values computed in i)-iii) are below the given thresholds, the gap and renormalization functions are extended from $m=M$ to $m=M'$ using $a/m^2$ and $b/|m|$, where $M'\gg M$.
\end{enumerate}

The advantage of using the tail extension is that it enables us to extend our numerically computed values of $\Delta(i\omega_n)$ and $Z(i\omega_n)$ to larger Matsubara frequencies without the need for the iterative method. For the results in the main text, we performed the iterative method up to $M\approx 600$, and then extend with the tail approximation up to $M'\approx 10^6$.

In Fig.~\ref{Fig4}, we see numerical verification that the weak Eliashberg approximation to the tail remains reasonable for larger values of $\lambda$. In Fig. \ref{Fig5}, we see a plot of the gap and renormalization functions versus Matsubara frequencies, from $\lambda=0.3$ to $\lambda=100$. For all values of $\lambda$ considered, the Ouroboros algorithm with a tail extension is utilized to obtain the numerical values under consideration. Note that, for all coupling strengths considered, a Lorentzian structure is preserved in both the gap function and the renormalization function, as already seen in the weak Eliashberg ($\lambda \lessapprox 0.5$) regime. The values of the gap and renormalization functions shown in Fig.~\ref{Fig5} are the values we use in the calculation of the superfluid density reported in the main text.

\subsection{Numerically solving for the Homes slope}

In the previous subsection, we discussed the algorithm  that was implemented to find the gap and renormalization functions for arbitrary values of $\lambda$. From these quantities, the superfluid density and normal-state conductivity can be numerically determined for arbitrary $\lambda$ and $1/(\tau \omega_E)$. In this final subsection of the Supplemental Material, we will outline precisely how linearity in the Homes relation is determined.

One will notice that we have carefully defined the value of $\eta_H(\tau,\lambda)$ as a Homes "proportionality factor", as opposed to always defining it as a "Homes slope". This is because the Homes proportionality is only formally defined to be a "slope" when the $T=0$ superfluid density scales linearly with the product $\sigma(\tau,\lambda)\cdot T_c$. The proportionality factor may be well-defined, but linear scaling may not be present (for example, when $\eta_H(\tau, \lambda)=0$). For this reason, we briefly outline the algorithm that was implemented to determine whether linear Homes scaling was obeyed:

\begin{enumerate}
\item For a given value of $1/(\tau \omega_E)$, find the values of the superfluid density $n_s(\tau,\lambda)/n$ and $\sigma(\tau,\lambda)\cdot T_c$ for a given range of $\lambda$ values.
\item Consider a series of linear fits for small "patches" of the Homes plot; namely, small segments of the plot of $n_s(\tau,\lambda)/n$ and $\sigma(\tau,\lambda)\cdot T_c$ for various different sizes. For each "patch", do a line of best fit, and calculate the coefficient of determination between each segment of data and the corresponding fit. From these values, calculate the average coefficient of determination for a given position on the $x$-axis (i.e., a given $\sigma(\tau,\lambda)\cdot T_c$) {and the $y$-intercept of each path}. 
\item For a given $\lambda$, if the averaged coefficient of determination for the Homes slope "patches" is above a given threshold, we will say that this value of $\lambda$ and $1/(\tau \omega_E)$ supports linear Homes scaling, with a Homes slope given by the Homes proportionality factor $\eta_H(\tau, \lambda)$. If, however, the averaged coefficient of determination is below a given threshold, we say that linear Homes scaling is violated for this value of $\lambda$ and $1/(\tau \omega_E)$. {If the $y$-intercept of the patches is above a small threshold, then we will say the scaling relation is violated due to flattening (as in the clean weak-coupling scenario).}
\end{enumerate}
The regime of linear Pimenov scaling and linear Holstein scaling is determined in the same manner.

\bibliography{main}{}